%% file: draft.tex
\documentclass[a4paper,11pt]{article}
\usepackage{jheppub} 
\usepackage[T1]{fontenc}
\usepackage{multirow}
\usepackage{color}
\usepackage{ulem}
\usepackage{rotating}
\usepackage{booktabs}
\usepackage{dcolumn}
\usepackage{array} 
\usepackage{multirow}
\usepackage{lineno} 
\usepackage{siunitx}
\usepackage{amsmath}
\usepackage{bm}
\usepackage{mathrsfs}
\usepackage{indentfirst}
\usepackage{subcaption} 
\usepackage{pgffor}     
\usepackage[labelformat=simple]{caption} 
\usepackage{booktabs}
\usepackage{array}
\usepackage{longtable}
\usepackage{graphicx}

\DeclareSIUnit{\bmm}{\bm{m}}

\DeclareSIUnit{\clight}{\textnormal{\textit{c}}}
\DeclareSIUnit{\clight}{\textnormal{\textit{c}}}

\newcolumntype{d}{D{.}{.}{-1}}
\newcolumntype{e}{D{.}{.}{8}}
\newcolumntype{f}{D{.}{.}{18}}
\newcolumntype{h}{D{.}{.}{13}}
\newcolumntype{g}{D{.}{.}{12}}

\newcommand{\BESIIIorcid}[1]{\href{https://orcid.org/#1}{\hspace*{0.1em}\raisebox{-0.45ex}{\includegraphics[width=1em]{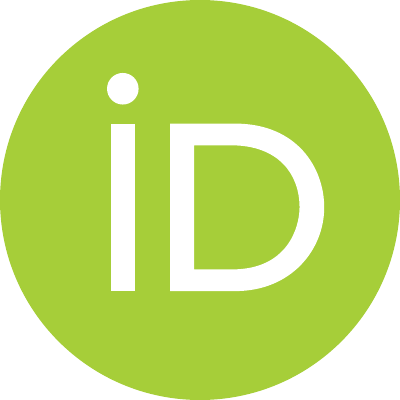}}}} 

\title{\protect\boldmath  
Measurement of Born cross sections of $e^+e^-\to pK^{-}\bar\Sigma^{0}$ + c.c. at $\sqrt{s}$ = 3.5-4.9~\textbf{GeV} and partial wave analysis at $\sqrt{s}$ = 3.773~\textbf{GeV} }

\collaborationImg{\includegraphics[width=0.25\textwidth]{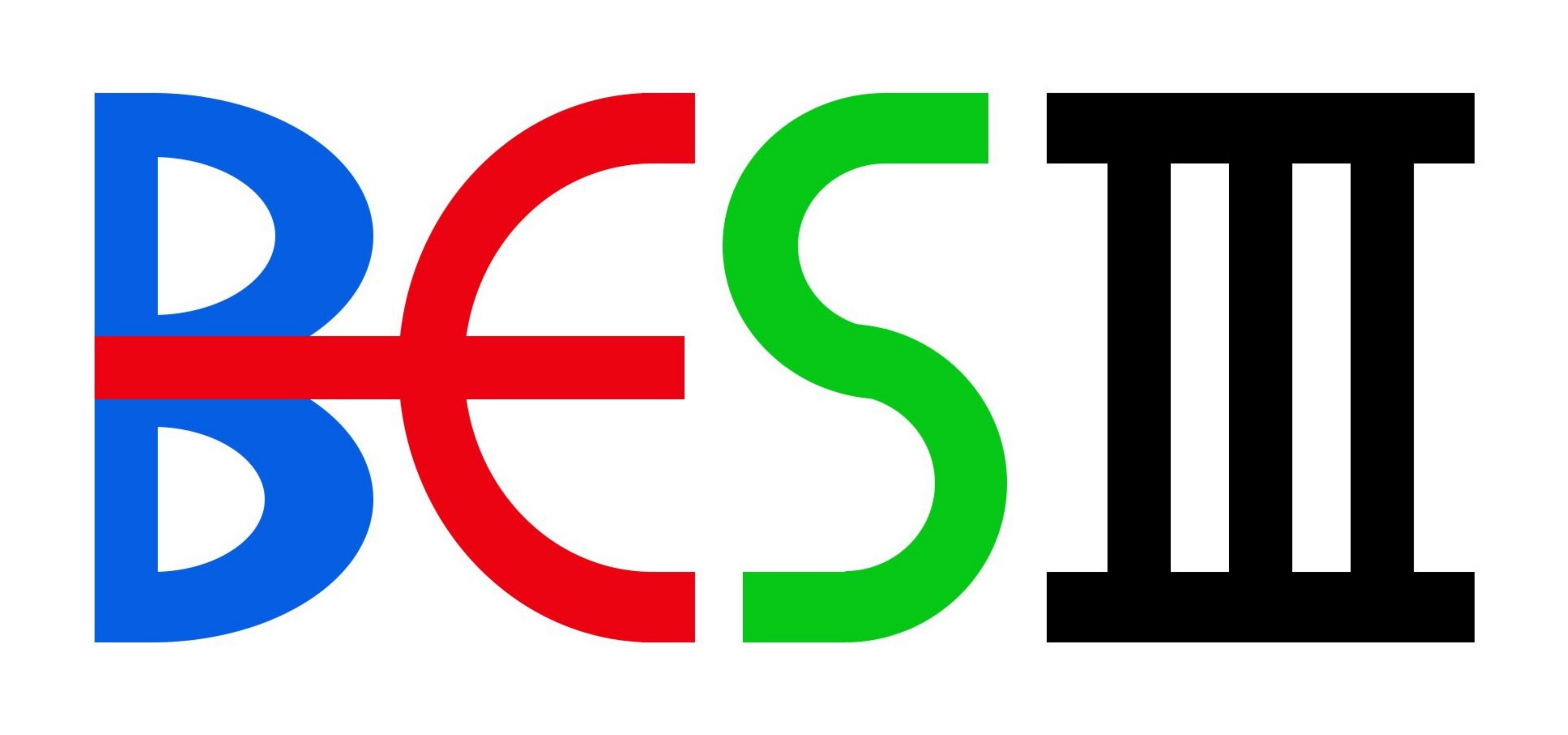}}
\collaboration{The BESIII Collaboration}
\emailAdd{besiii-publications@ihep.ac.cn}

\begin{document} 
\abstract{
Using $e^+e^-$ collision data collected by the BESIII detector at BEPCII corresponding to a total integrated luminosity of 44.6 $\rm fb^{-1}$, we present the first measurement of the Born cross sections for the process $e^+e^-\to pK^{-}\bar{\Sigma}^{0}$ + c.c. at fifty-six center-of-mass energies between 3.51 and 4.95~GeV. 
By fitting the dressed cross sections of $e^+e^-\to pK^{-}\bar{\Sigma}^{0}$ + c.c. with the assumption of a power-law function plus a charmonium(-like) resonance, i.e., $\psi(3770)$, $\psi(4040)$, $\psi(4160)$, $Y(4230)$, $Y(4360)$, $\psi(4415)$, $\psi(4500)$, $Y(4660)$ or $\psi(4710)$, no significant charmonium(-like) state decaying into $pK^{-}\bar{\Sigma}^{0}$ + c.c. is observed. The product of branching fractions and electronic partial widths including their upper limits at the 90\% confidence level are provided for each decay.
In addition, a partial wave analysis of the process $e^+e^-\to pK^{-}\bar{\Sigma}^{0}$ + c.c. at $\sqrt{s}$ = 3.773 GeV is performed to investigate $N^{*}$, $\Lambda^{*}$ and $\Sigma^{*}$ resonances in the $K^{-}\bar{\Sigma}^{0}$ and $pK^{-}$ invariant mass spectra. Significant contributions are found from the $\Lambda(1520)$, $\Sigma(1670)$, $\Sigma(1915)$, $N(1900)$, $N(2300)$ and $N(2570)$. The masses, widths and production fractions for each component are determined.
}

\maketitle
\flushbottom    
\section{Introduction}
\label{sec:intro}
\noindent 
The study of the production of charmonium(-like) states above the open-charm threshold in $e^+e^-$ annihilation and their subsequent hadronic decays can provide a test for quantum chromodynamics (QCD) calculations, and a valuable opportunity to probe $XY\!Z$ states~\cite{Brambilla:2010cs,Briceno:2015rlt,Wang:2025dur,Bai:2026atm}.
According to the potential model, there are five vector charmonium states between $3.7~\text{GeV}/c^2$ and $4.7~\text{GeV}/c^2$, namely, the $3S$, $2D$, $4S$, $3D$ and $5S$ states. 
In the past decades, an abundance of charmonium(-like) vector states has been observed with masses above the open-charm threshold at $e^+e^-$ colliders. Several states, such as the $\psi(4040)$, $\psi(4160)$ and $\psi(4415)$, are considered to be ordinary charmonium states and are directly visible as peaks in the $e^+e^- \to \text{hadrons}$ cross section~\cite{PDG}. Additionally, five charmonium-like states, $Y(4230)$, $Y(4360)$, $Y(4500)$, $Y(4660)$, and $Y(4710)$, have been observed via initial-state radiation (ISR) processes at BaBar~\cite{BaBar:2005hhc,BaBar:2006ait,BaBar:2012vyb,BaBar:2012hpr} and Belle~\cite{Belle:2007umv,Belle:2007dxy,Belle:2008xmh,Belle:2013yex,Belle:2014wyt} and in direct production processes at CLEO-c~\cite{CLEO:2006ike} and BESIII~\cite{BESIII:2014rja,BESIII:2022joj,BESIII:2023cmv,BESIII:2023wqy}.
These new states may not be regarded as resonances consisting solely of a simple and pure $c\bar{c}$ quark component. Various scenarios~\cite{Close:2005iz,Chen:2016qju,Wang:2019mhs,Qian:2021neg,Dai:2023vsw,Wang:2023zxj},
including glueballs composed of pure gluons, multiquark states such as tetraquark states and molecular states, and hybrids  with a mixture of quark pairs and valence gluons have been proposed. These hadrons that go beyond the conventional quark model are collectively referred to as charmonium-like states or exotic states. 

The precise measurement of the exclusive production cross section for baryonic final states offers a novel approach to enrich and improve our understanding of the non-perturbative QCD problem of strong interactions associated with charmonium(-like) states. 
Although a series of investigations have been conducted by BESIII experiment~\cite{BESIII:2019cuv,BESIII:2021aer,BESIII:2021cvv,Wang:2022zyc,Liu:2023xhg,BESIII:2023rwv,BESIII:2023rse,BESIII:2024umc,BESIII:2024ogz,BESIII:2024ths,BESIII:2024ues,BESIII:2024gql,BESIII:2025lbj,Zhang:2026qjt,Zhang:2025qmo,BESIII:2026hgj,Wang:2026daa,BESIII:2026hie,BESIII:2026oyx,BESIII:2026lcb,BESIII:2026ont,BESIII:2026fcp},
only four pieces of evidence for the decays of $\psi(3770)\to\Lambda\bar\Lambda$, $\Xi^-\bar\Xi^+$, $\psi(4160)\to K^-\bar\Xi^+\Lambda$ and $Y(4230)\to p\bar{p}$~\cite{BESIII:2021cvv,BESIII:2023rse,BESIII:2024ogz,BESIII:2026lcb} 
and three discoveries for the decays of $\psi(3770)\to\Sigma^-\bar\Sigma^+$, $K^-\bar\Xi^0\Sigma^+$ and $p\bar{p}$~\cite{BESIII:2026hie,BESIII:2026oyx,BESIII:2026lcb} have been reported to data. No significant baryonic decays have been observed for any other vector charmonium(-like) state. 
To gain a deeper understanding, more experimental studies related to charmonium(-like) states with different final states are highly desirable.  

Moreover, compared to mesons, the baryons offer the simplest system in which the three colors of QCD neutralize into colorless objects. Benefiting from a clean and complementary environment with low background and excellent energy resolution, baryon resonance studies serve as important experimental probes for understanding QCD.
In recent years, numerous studies have focused on the properties of excited states of nucleons, yet only a few results are available for baryons containing a strange quark, namely the $\Lambda$, $\Sigma$ and $\Xi$ hyperons~\cite{PDG,Wang:2026daa}. The large data samples collected from $e^+e^-$ annihilation at BESIII provide an excellent opportunity to study the baryon spectroscopy. 
In particular, the data sample collected at the center-of-mass (CM) energy ($\sqrt{s}$) of 3.773 GeV corresponds to a higher energy than the $J/\psi$ and $\psi(3686)$ resonances. The resulting larger decay phase space (PHSP) makes it possible to access higher-mass resonances.

In this article, we present measurements of the Born cross sections for the $e^+e^- \to pK^{-}\bar{\Sigma}^{0}$ + c.c. reaction, in the range of CM energies between 3.51 and 4.95 GeV~\cite{BESIII:2015zbz,BESIII:2020eyu}. These measurements are based on $e^+e^-$ collision data corresponding to a total integrated luminosity of 44.6~${\rm fb}^{-1}$~\cite{Ablikim:2013ntc,BESIII:2015qfd,BESIII:2022dxl,BESIII:2022ulv,BESIII:2024lks}, collected by the BESIII detector~\cite{BESIII:2009fln} at BEPCII~\cite{ref::collider}. Possible resonances are searched for by analyzing the line shape of the dressed cross sections of the $e^+e^- \to pK^{-}\bar{\Sigma}^{0}$ + c.c. reaction. The product of branching fractions and electronic partial widths for charmonium(-like) states decaying into $pK^{-}\bar{\Sigma}^{0}$ + c.c. as well as their upper limits at the 90\% confidence level (C.L.) are reported. In addition, a partial wave analysis of the process $e^+e^-\to pK^{-}\bar{\Sigma}^{0}$ + c.c. at 3.773 GeV is performed to investigate $N^{*}$, $\Lambda^{*}$ and $\Sigma^{*}$ resonances in the $K^{-}\bar{\Sigma}^{0}$ and $pK^{-}$ invariant mass distributions.

\section{BESIII Detector and Monte Carlo simulation}
\noindent 
The BESIII detector~\cite{BESIII:2009fln} records symmetric $e^+e^-$ collisions provided by the BEPCII storage ring in the CM energy range from 1.84~GeV to 4.95~GeV, with a peak luminosity of $1.1 \times 10^{33}~\text{cm}^{-2}\text{s}^{-1}$ achieved at $\sqrt{s} = 3.773$~GeV. BESIII has collected large data samples in this energy region. The cylindrical core of the BESIII detector covers 93\% of the full solid angle and consists of a helium-based multilayer drift chamber (MDC), a time-of-flight system (TOF), and a CsI (Tl) electromagnetic calorimeter (EMC), which are all enclosed in a superconducting solenoidal magnet providing a 1.0~T magnetic field. The solenoid is supported by an octagonal flux-return yoke with resistive plate counter muon identification modules interleaved with steel. The momentum resolution for charged particles at 1 GeV/$c$ is 0.5\%, while the $\mathrm{d}E/\mathrm{d}x$ resolution for electrons from Bhabha scattering is 6\%. The EMC measures photon energies with a resolution of 2.5\% (5\%) at 1~GeV in the barrel (end cap) region. The time resolution in the plastic scintillator TOF barrel region was 68~ps, while that in the end cap region was 110~ps. The end cap TOF system was upgraded in 2015 using multigap resistive plate chamber technology providing a time resolution of 60~ps, which benefits 64\% of the data used in this analysis~\cite{Li:2017jpg,Guo:2017sjt,Cao:2020ibk}.

Monte Carlo (MC) simulated data samples produced with a \textsc{geant4}-based~\cite{GEANT4:2002zbu} package, which includes the geometric description of the BESIII detector and the detector response, are used to determine detection efficiencies and to estimate backgrounds. The simulation models the beam energy spread and ISR in the $e^+e^-$ annihilations with the generator \textsc{kkmc}~\cite{JADACH2000260,Jadach:2000ir}. The inclusive MC sample includes the production of open charm processes, the ISR production of vector charmonium(-like) states, and the continuum processes incorporated in \textsc{kkmc}. All particle decays are modeled with \textsc{evtgen}~\cite{Ping:2008zz,Lange:2001uf} using branching fractions either taken from the Particle Data Group (PDG)~\cite{PDG}, when available, or otherwise estimated with \textsc{lundcharm}~\cite{Chen:2000tv}. To determine the detection efficiencies and evaluate the ISR factors in the $e^+e^- \to pK^{-}\bar{\Sigma}^{0}$ + c.c. process, exclusive MC samples with 100,000 signal events are generated for each of the 56 CM energies from 3.5~GeV to 4.9~GeV using a PHSP model and incorporating ISR effects.

\section{Event selection}
\noindent 
A full-reconstruction technique is employed to select the $e^+e^- \to pK^{-}\bar{\Sigma}^{0}$ + c.c. candidate events. The $\Sigma^{0}$ and the $\bar{\Sigma}^0$ are reconstructed through the $\Sigma^0 \to \Lambda \gamma$ and $\bar{\Sigma}^0 \to \bar{\Lambda} \gamma$ decays, respectively. Here, $\Lambda \to p \pi^-$ and $\bar{\Lambda} \to \bar{p} \pi^+$.

Charged tracks must be reconstructed in the MDC within the angular region $|\cos\theta| < 0.93$, where $\theta$ is the polar angle with respect to the $z$-axis in the laboratory frame, which is the symmetry axis of the MDC. The closest approach distance to the interaction point~(IP) must be less than 30 cm along the $z$ axis ($|V_z|<30$ cm) and less than 15 cm in the transverse plane ($|V_{xy}|<15$ cm) for all the tracks. The number of positive and negative charged tracks is required to be equal to 2.
Particle identification for charged tracks combines the measurements of the specific ionization energy loss in the MDC ($\mathrm{d}E/\mathrm{d}x$) and the time of flight in TOF to form likelihoods $\mathcal{L}(h)$ ($h = p, K, \pi$) for each hadron $h$ hypothesis. Each track is assigned a particle type corresponding to the hypothesis with the highest confidence level. For this analysis, four tracks identified as $p$, $\bar{p}$, $K^-$ and $\pi^+$ are required. 

Photons are reconstructed from isolated showers in the EMC. The energy deposited in the nearby TOF detector is incorporated to enhance the reconstruction efficiency and energy resolution. The deposited energy of each photon must be more than 25 MeV in the barrel region ($|\cos\theta| <0.8$) and more than 50~MeV in the end cap region ($0.86< |\cos\theta| <0.92$). Moreover, to suppress electronic noise and showers not related to the event, the difference between the EMC time and the event start time must fall within the range (0, 700) ns. To exclude showers that originate from charged tracks, the angle between the direction of each shower in the EMC and the direction of the closest extrapolated charged track must be greater than 10 degrees. Events with at least one photon are kept for further analysis.

To suppress backgrounds from fake tracks, the $p$ and $K^-$ candidates are constrained to originate from a common vertex using a vertex fit, and are required to satisfy $|V_z|<10$ cm and $|V_{xy}|<1$ cm in the case of the $\gamma pK^-\bar{\Lambda}$ modes. The same procedure is applied for the respective antiparticle combinations in the charge-conjugate mode. 
$\bar{\Lambda}$ candidates are reconstructed from $\bar{p}\pi^+$ combinations by performing both a primary and a secondary vertex fit. In the primary vertex fit, the tracks are constrained to the $e^{+}e^{-}$ interaction point, while in the secondary vertex fit they are constrained to a common decay vertex. Both fits are required to converge.
To suppress background, the $\bar{p}\pi^+$ invariant mass is required to lie within 8.5 MeV/$c^2$ of the $\bar{\Lambda}$ nominal mass. The decay length of the $\bar{\Lambda}$ candidate is required to be greater than zero to suppress background from non-$\bar{\Lambda}$ events~\cite{BESIII:2024umc}. 
To improve the momentum and energy resolution and suppress background events, a four-constraint (4C) kinematic fit is applied under the hypothesis of $e^+e^- \to pK^-\bar{p}\pi^{+}\gamma$, enforcing energy-momentum conservation from the initial $e^+e^-$ to the final states. For events with more than one photon candidate, the combination with the least $\chi^2_{\rm{4C}}$ is selected from all possible combinations. The requirement $\chi^2_{\rm{4C}}<60$ is imposed after selecting the best candidate and optimizing the figure of merit (FOM), defined as $S/\sqrt{S+B}$, where $S$ is the number of signal MC events normalized to the data, and $B$ is the number of background events estimated from the inclusive MC sample. To reject possible background events from $e^+e^- \to pK^-\bar{\Lambda}$ and $pK^-\bar{\Lambda}\gamma\gamma$ processes, we further require the $\chi^2_{\rm{4C}}$ value for the $e^+e^- \to pK^-\bar{\Lambda}\gamma$ hypothesis to be smaller than those obtained under either background hypothesis. 
After applying the aforementioned event selection criteria and analyzing the inclusive MC sample at the CM energy of $\sqrt{s}$ = 3.773 GeV, the surviving backgrounds are found to be smoothly distributed in the $\bar{\Sigma}^{0}$ signal region on the mass spectrum of $\gamma\bar{\Lambda}$ ($M_{\bar{\Sigma}^{0}}$).

\section{Born cross-section measurement}
\subsection{Determination of the signal yields}

\begin{figure}[htbp]
  \centering
\begin{minipage}[b]{0.24\linewidth}
    \begin{picture}(50,80)
      \put(0,0){\includegraphics[width=\linewidth]{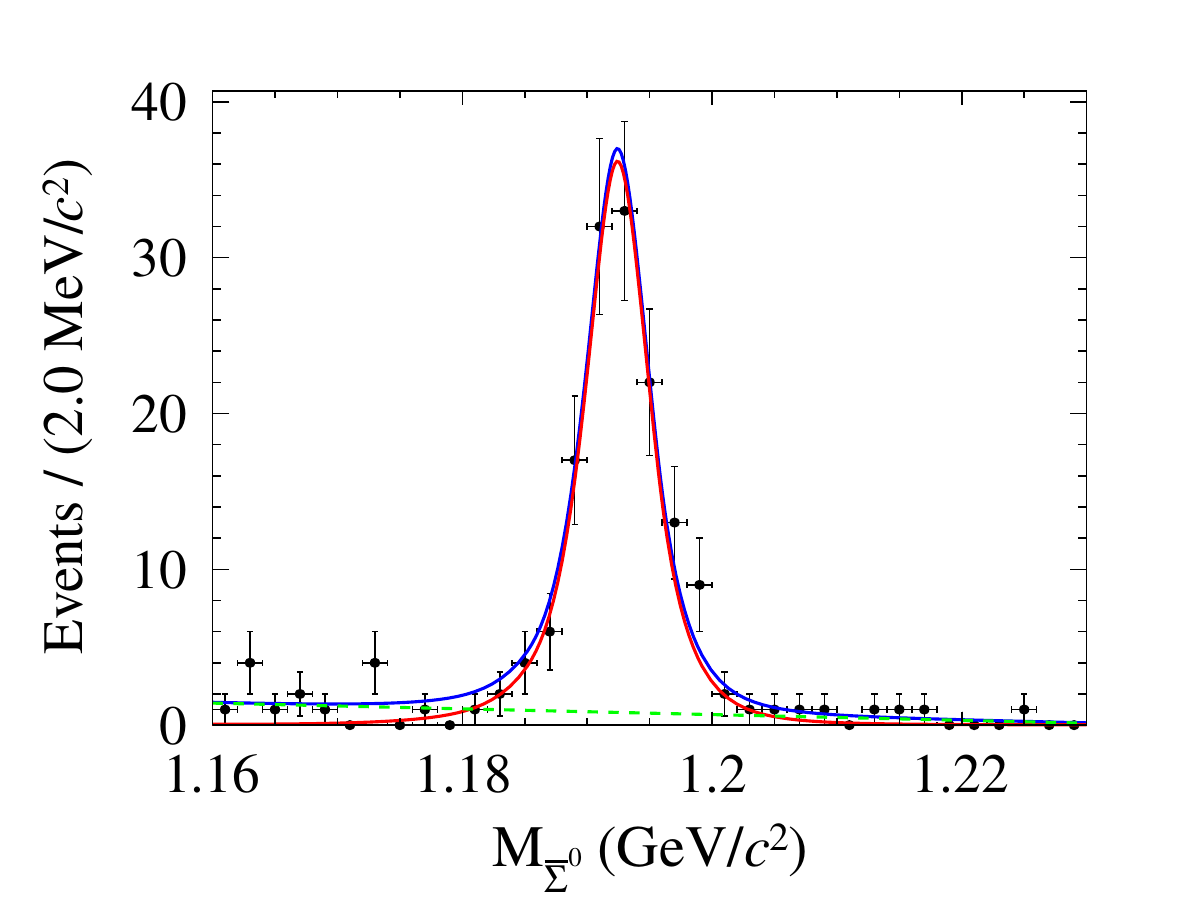}}
      \put(20,65){\color{blue}\scalebox{0.4}{$\sqrt{s}=3.508$ GeV}}
    \end{picture}
  \end{minipage}
  \begin{minipage}[b]{0.24\linewidth}
    \begin{picture}(50,80)
      \put(0,0){\includegraphics[width=\linewidth]{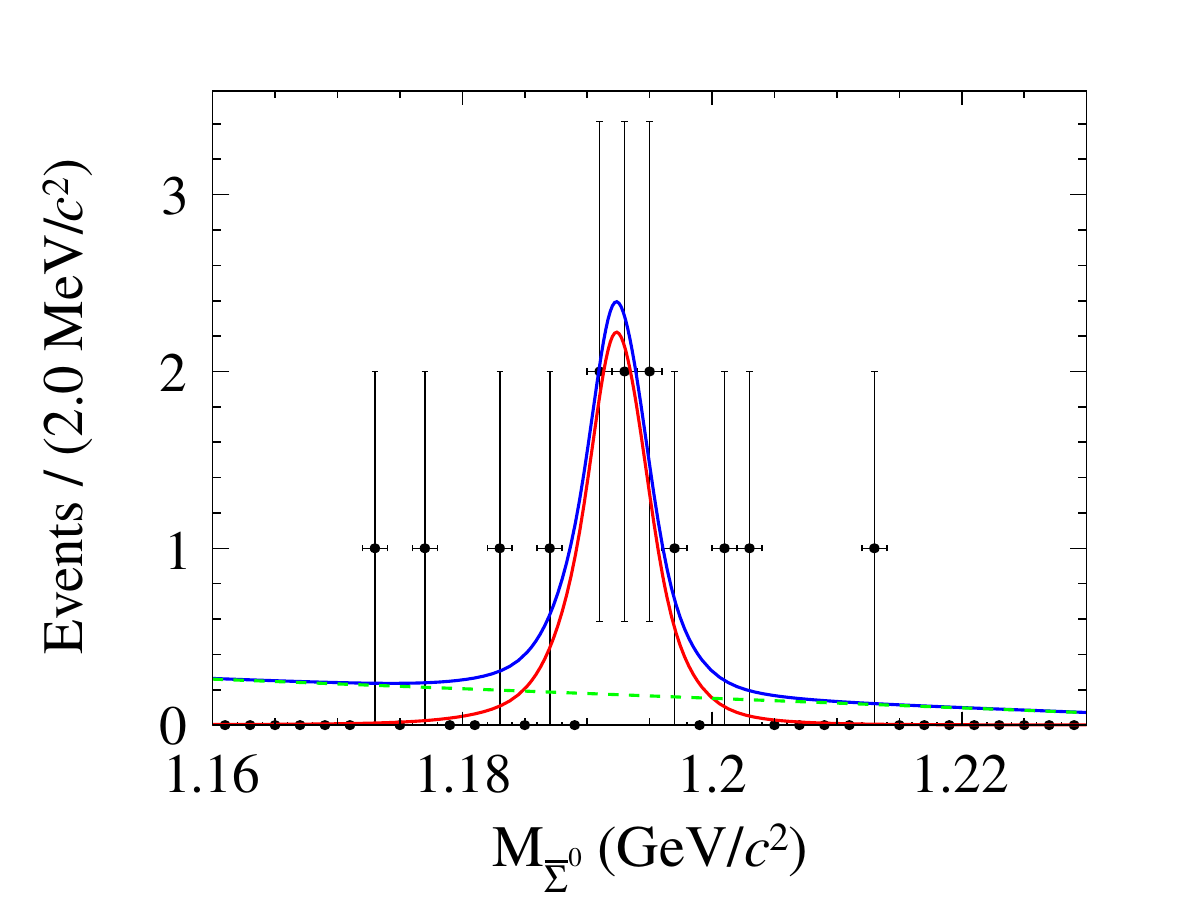}}
      \put(20,65){\color{blue}\scalebox{0.4}{$\sqrt{s}=3.514$ GeV}}
    \end{picture}
  \end{minipage}
  \begin{minipage}[b]{0.24\linewidth}
    \begin{picture}(50,80)
      \put(0,0){\includegraphics[width=\linewidth]{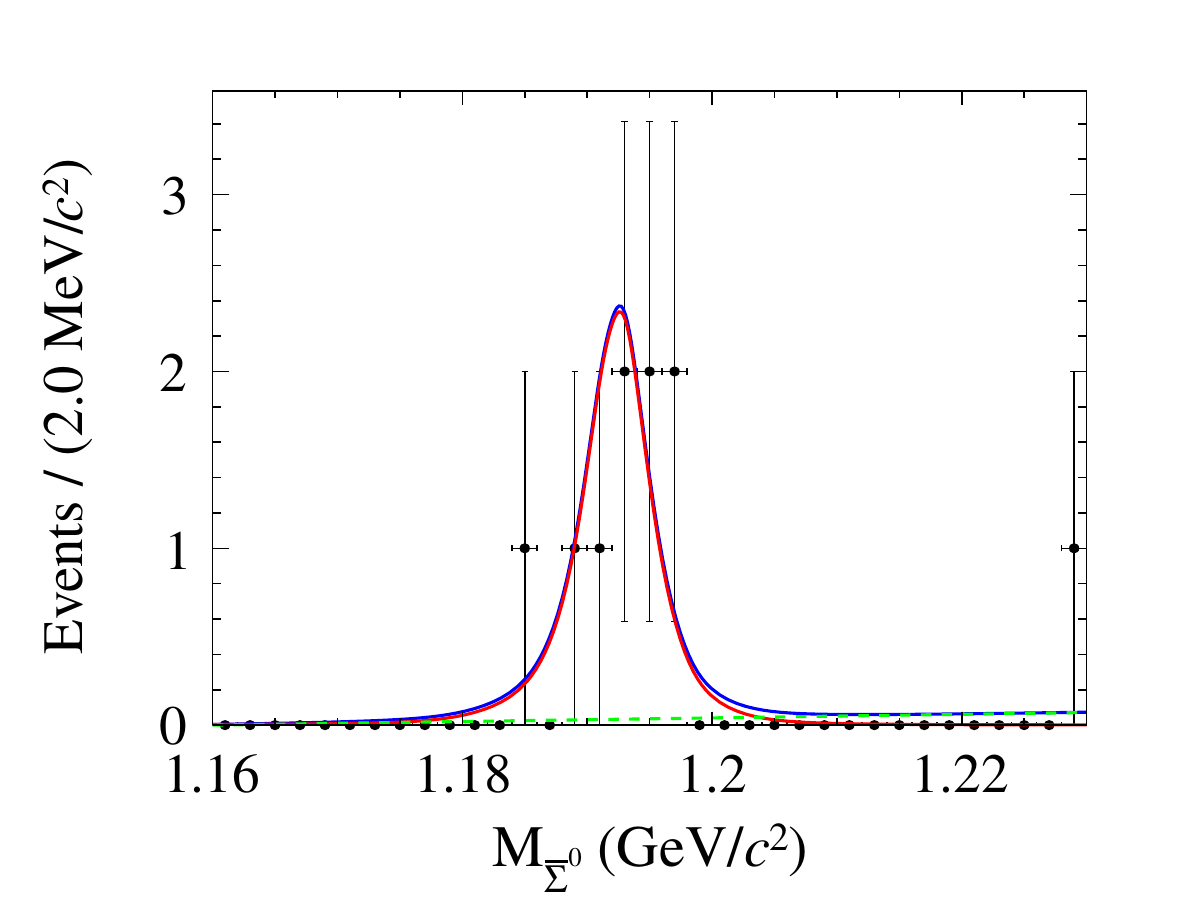}}
      \put(20,65){\color{blue}\scalebox{0.4}{$\sqrt{s}=3.553$ GeV}}
    \end{picture}
  \end{minipage}
  \begin{minipage}[b]{0.24\linewidth}
    \begin{picture}(50,80)
      \put(0,0){\includegraphics[width=\linewidth]{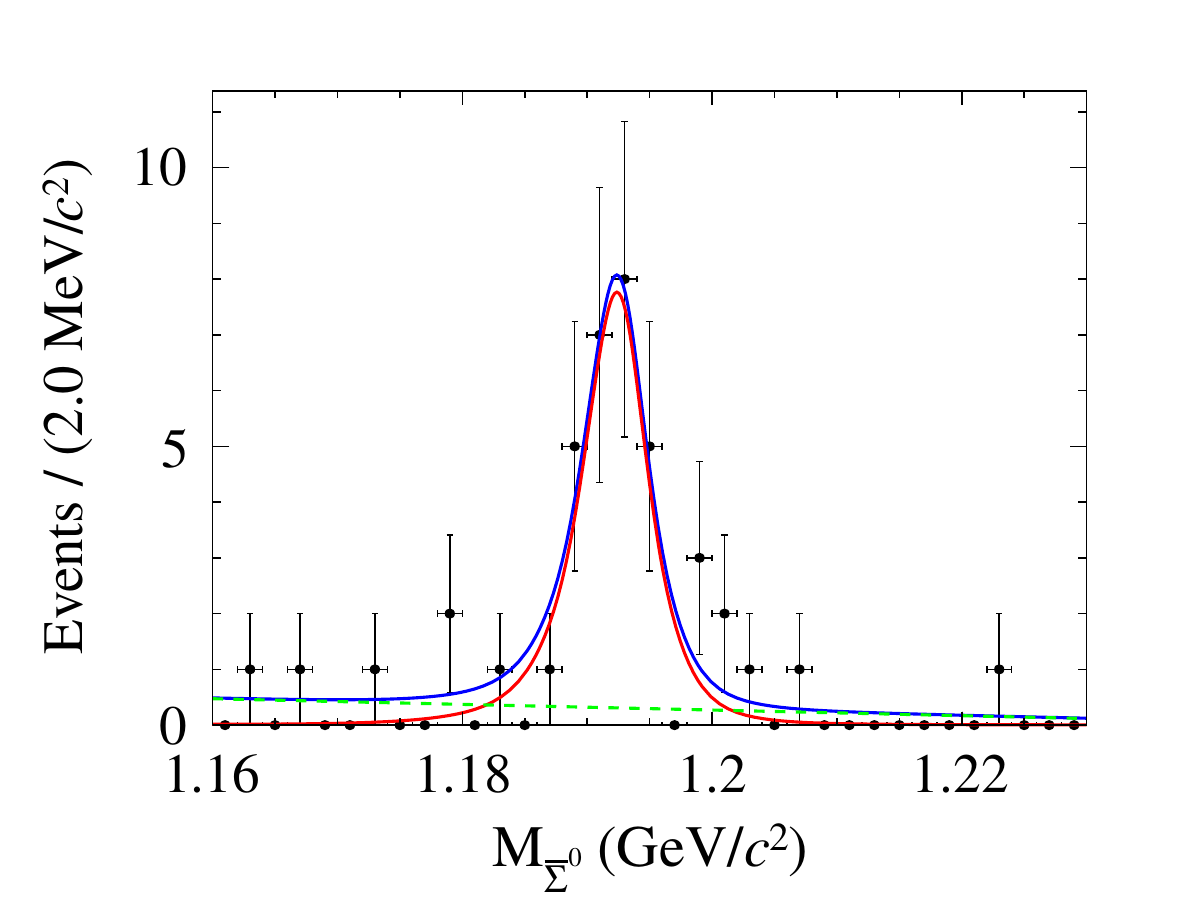}}
      \put(20,65){\color{blue}\scalebox{0.4}{$\sqrt{s}=3.554$ GeV}}
    \end{picture}
  \end{minipage}

  \begin{minipage}[b]{0.24\linewidth}
    \begin{picture}(50,80)
      \put(0,0){\includegraphics[width=\linewidth]{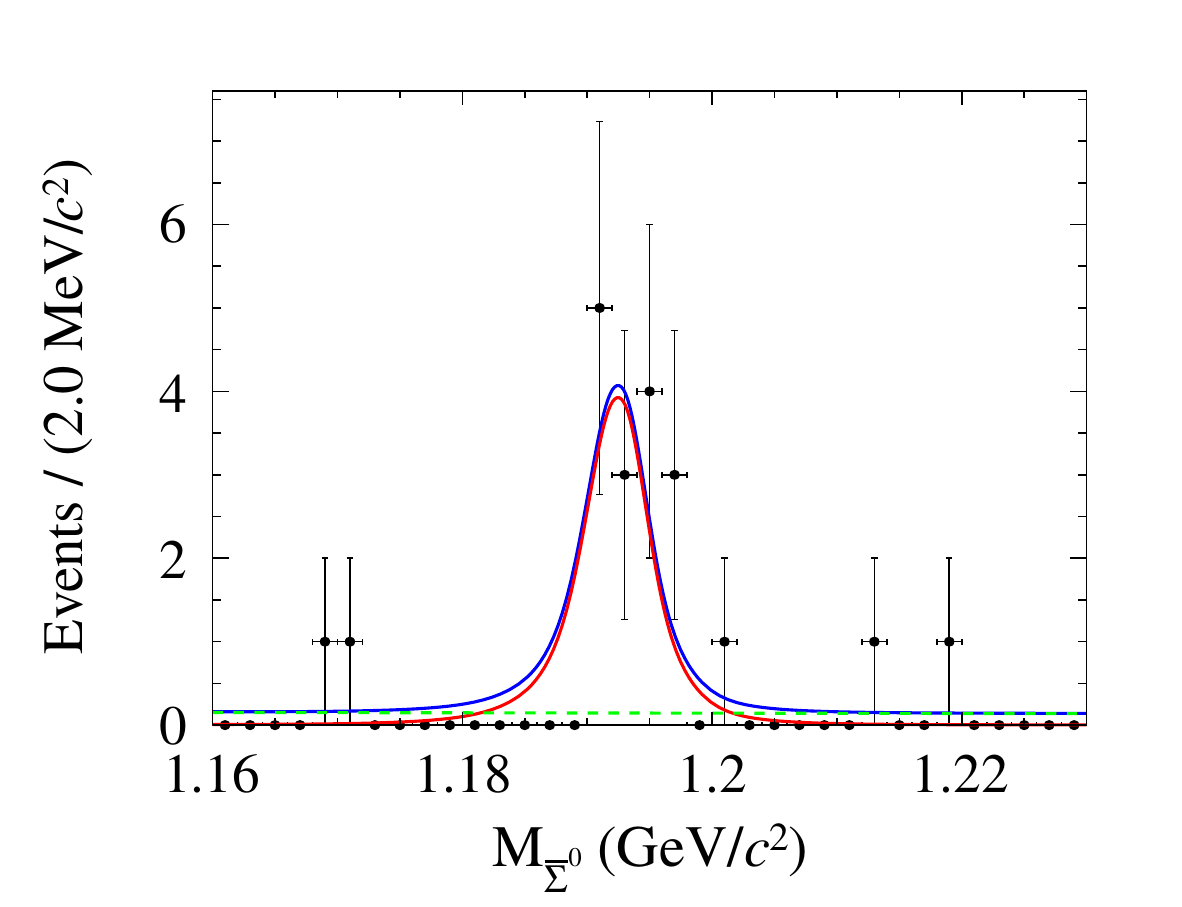}}
      \put(20,65){\color{blue}\scalebox{0.4}{$\sqrt{s}=3.582$ GeV}}
    \end{picture}
  \end{minipage}
  \begin{minipage}[b]{0.24\linewidth}
    \begin{picture}(50,80)
      \put(0,0){\includegraphics[width=\linewidth]{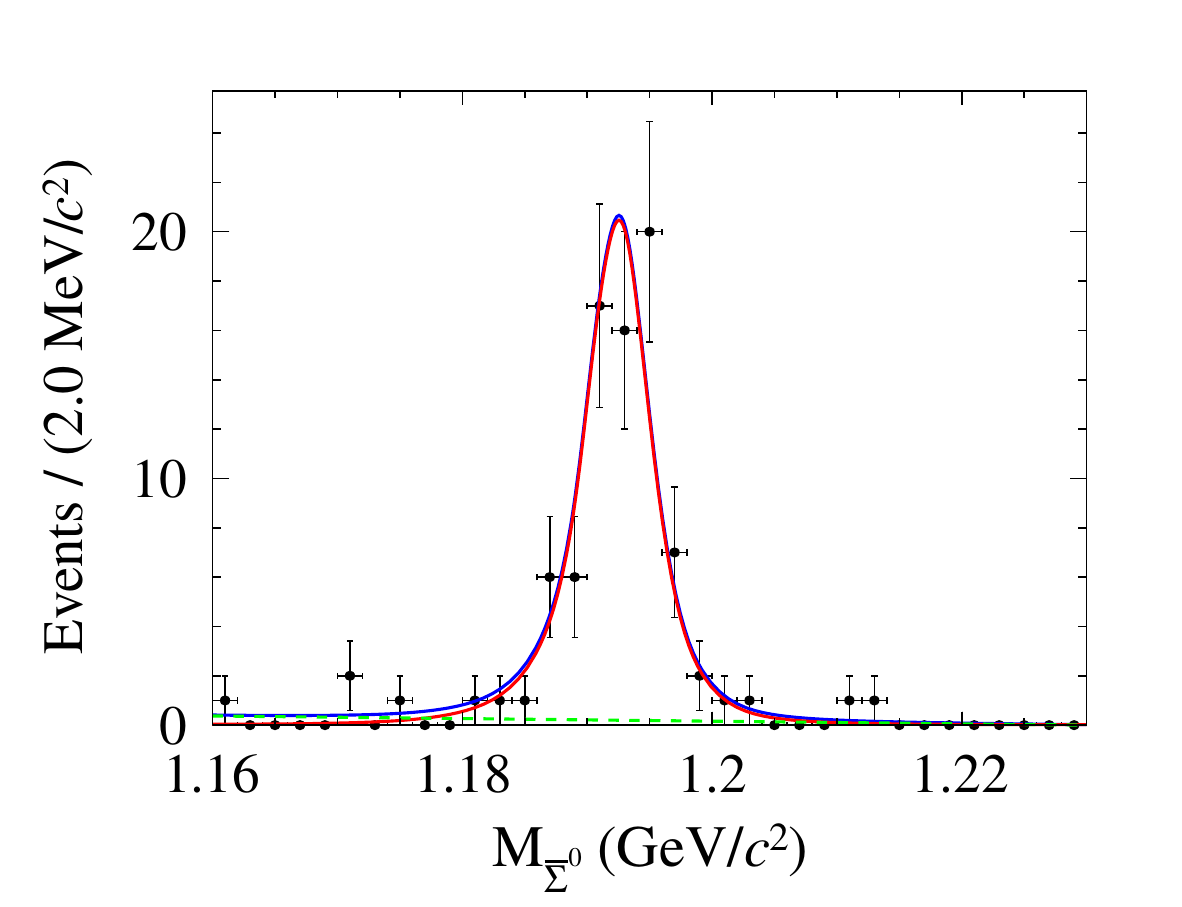}}
      \put(20,65){\color{blue}\scalebox{0.4}{$\sqrt{s}=3.650$ GeV}}
    \end{picture}
  \end{minipage}
  \begin{minipage}[b]{0.24\linewidth}
    \begin{picture}(50,80)
      \put(0,0){\includegraphics[width=\linewidth]{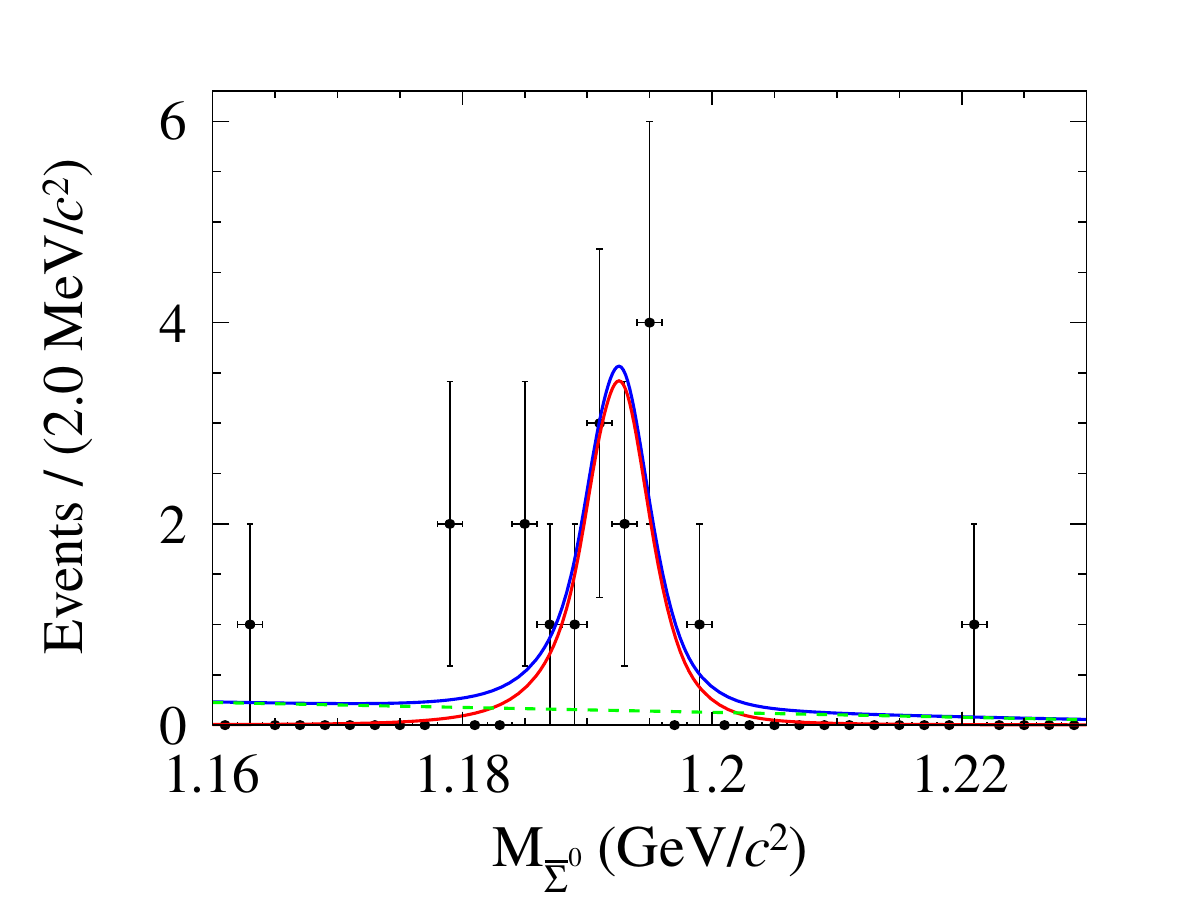}}
      \put(20,65){\color{blue}\scalebox{0.4}{$\sqrt{s}=3.670$ GeV}}
    \end{picture}
  \end{minipage}
  \begin{minipage}[b]{0.24\linewidth}
    \begin{picture}(50,80)
      \put(0,0){\includegraphics[width=\linewidth]{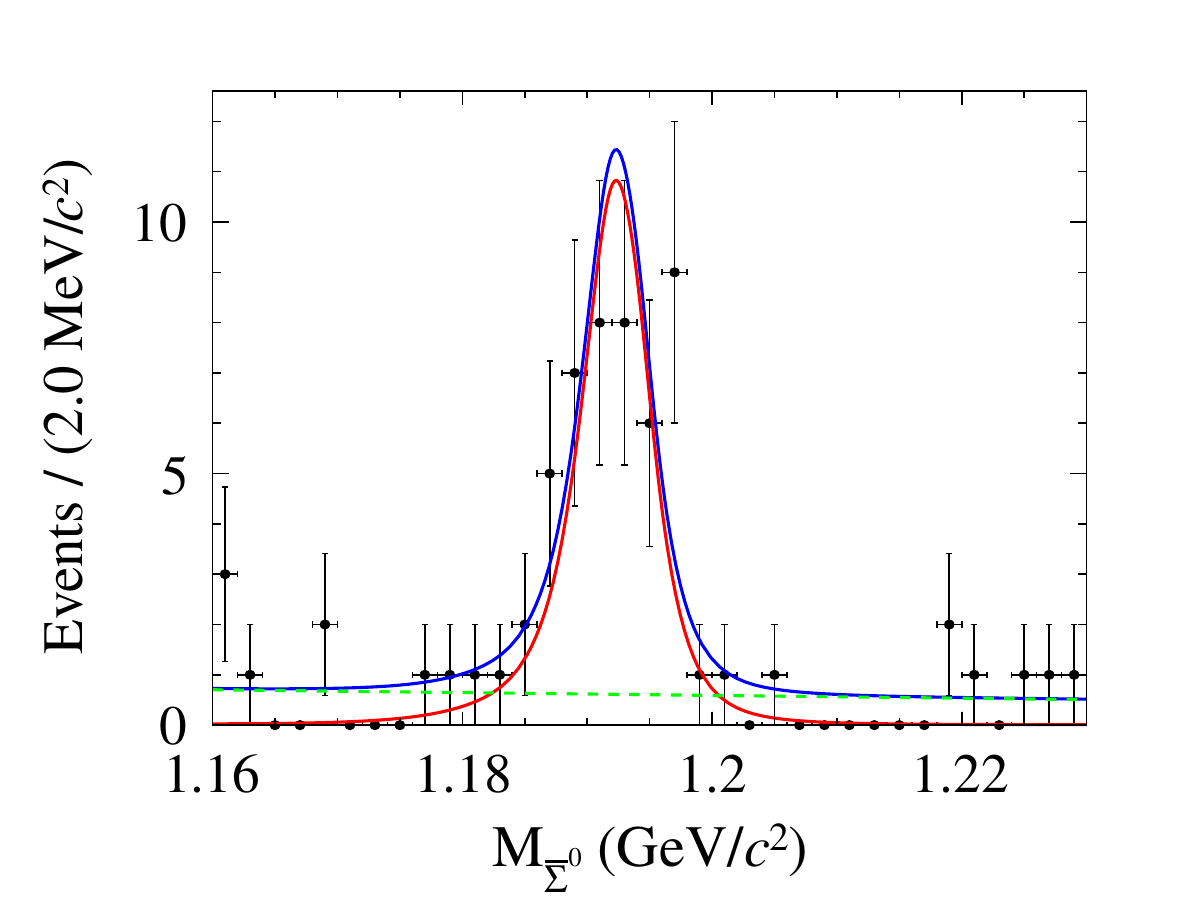}}
      \put(20,65){\color{blue}\scalebox{0.4}{$\sqrt{s}=3.768$ GeV}}
    \end{picture}
  \end{minipage}

  \begin{minipage}[b]{0.24\linewidth}
    \begin{picture}(50,80)
      \put(0,0){\includegraphics[width=\linewidth]{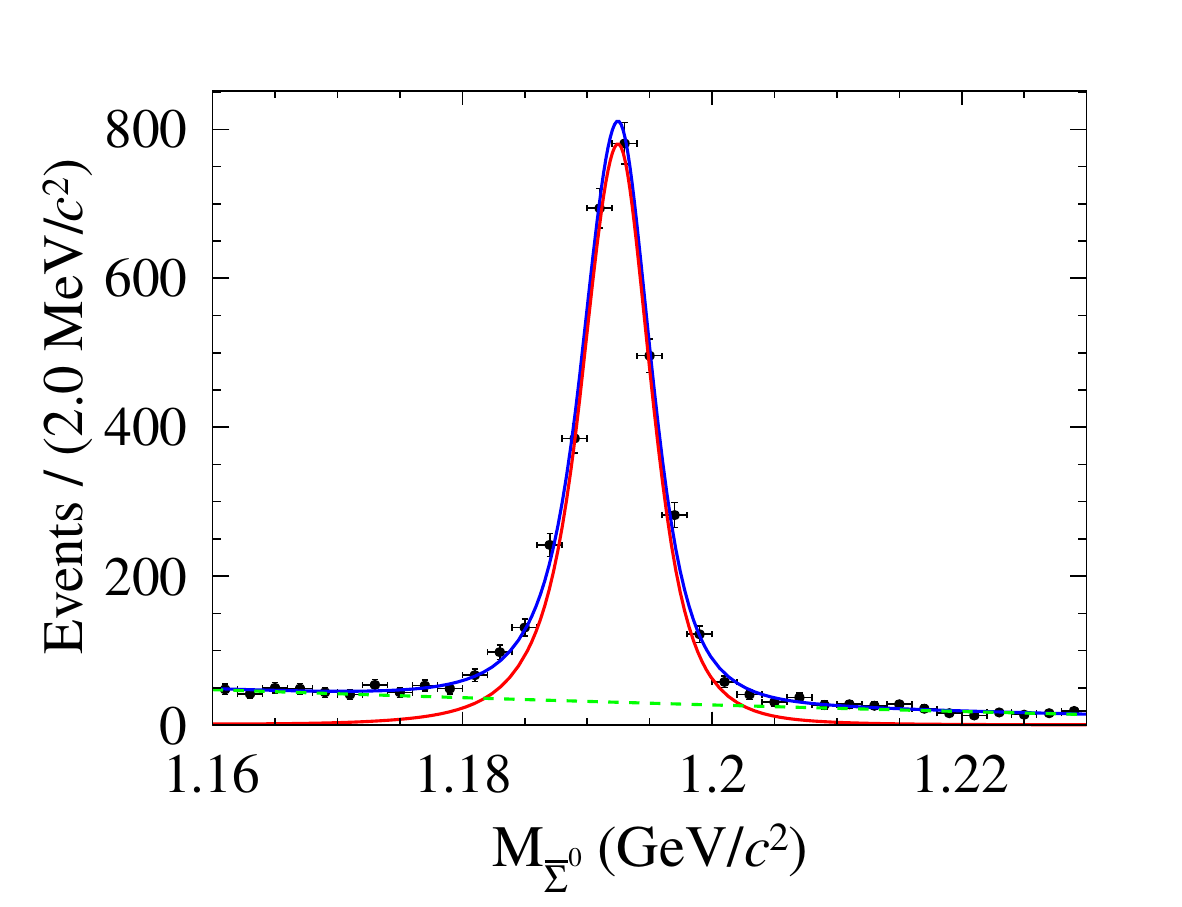}}
      \put(20,65){\color{blue}\scalebox{0.4}{$\sqrt{s}=3.773$ GeV}}
    \end{picture}
  \end{minipage}
  \begin{minipage}[b]{0.24\linewidth}
    \begin{picture}(50,80)
      \put(0,0){\includegraphics[width=\linewidth]{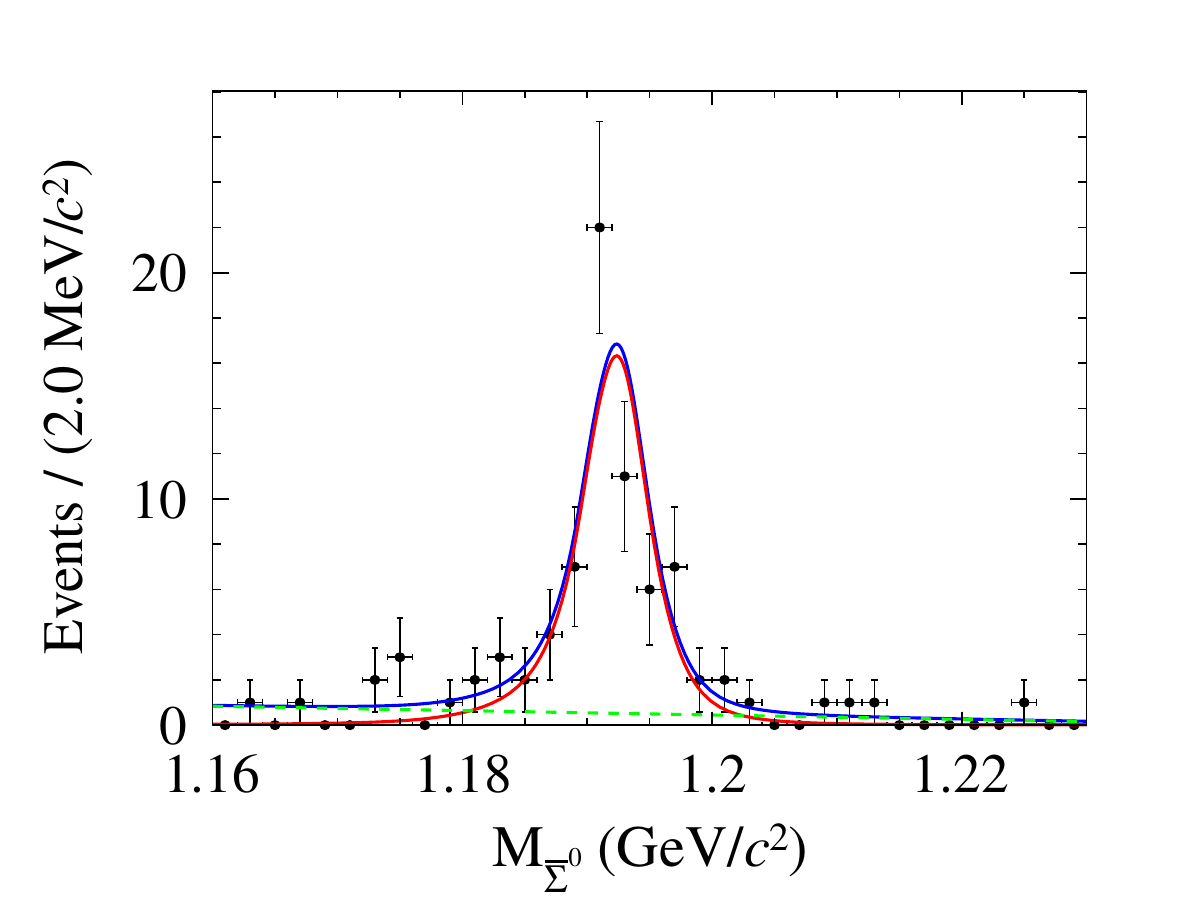}}
      \put(20,65){\color{blue}\scalebox{0.4}{$\sqrt{s}=3.780$ GeV}}
    \end{picture}
  \end{minipage}
  \begin{minipage}[b]{0.24\linewidth}
    \begin{picture}(50,80)
      \put(0,0){\includegraphics[width=\linewidth]{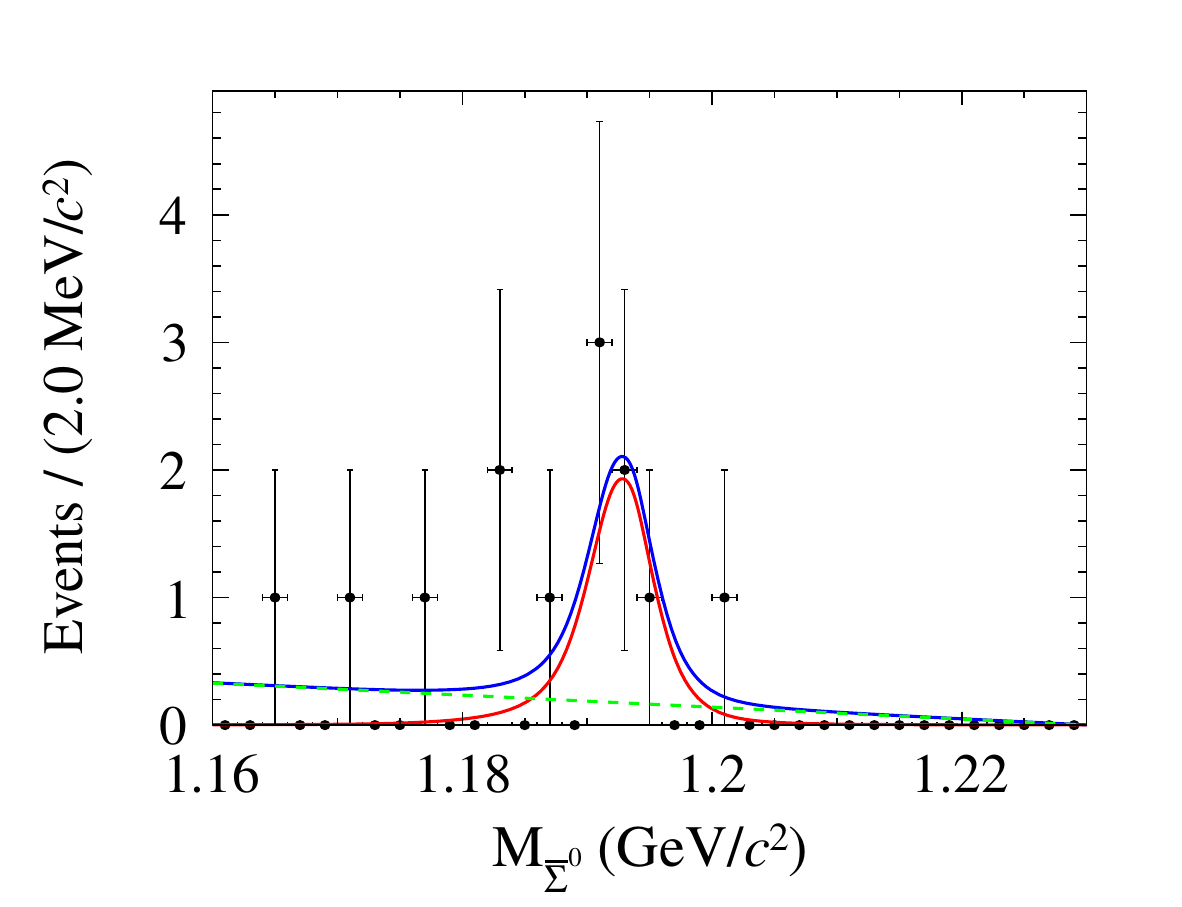}}
      \put(20,65){\color{blue}\scalebox{0.4}{$\sqrt{s}=3.808$ GeV}}
    \end{picture}
  \end{minipage}
  \begin{minipage}[b]{0.24\linewidth}
    \begin{picture}(50,80)
      \put(0,0){\includegraphics[width=\linewidth]{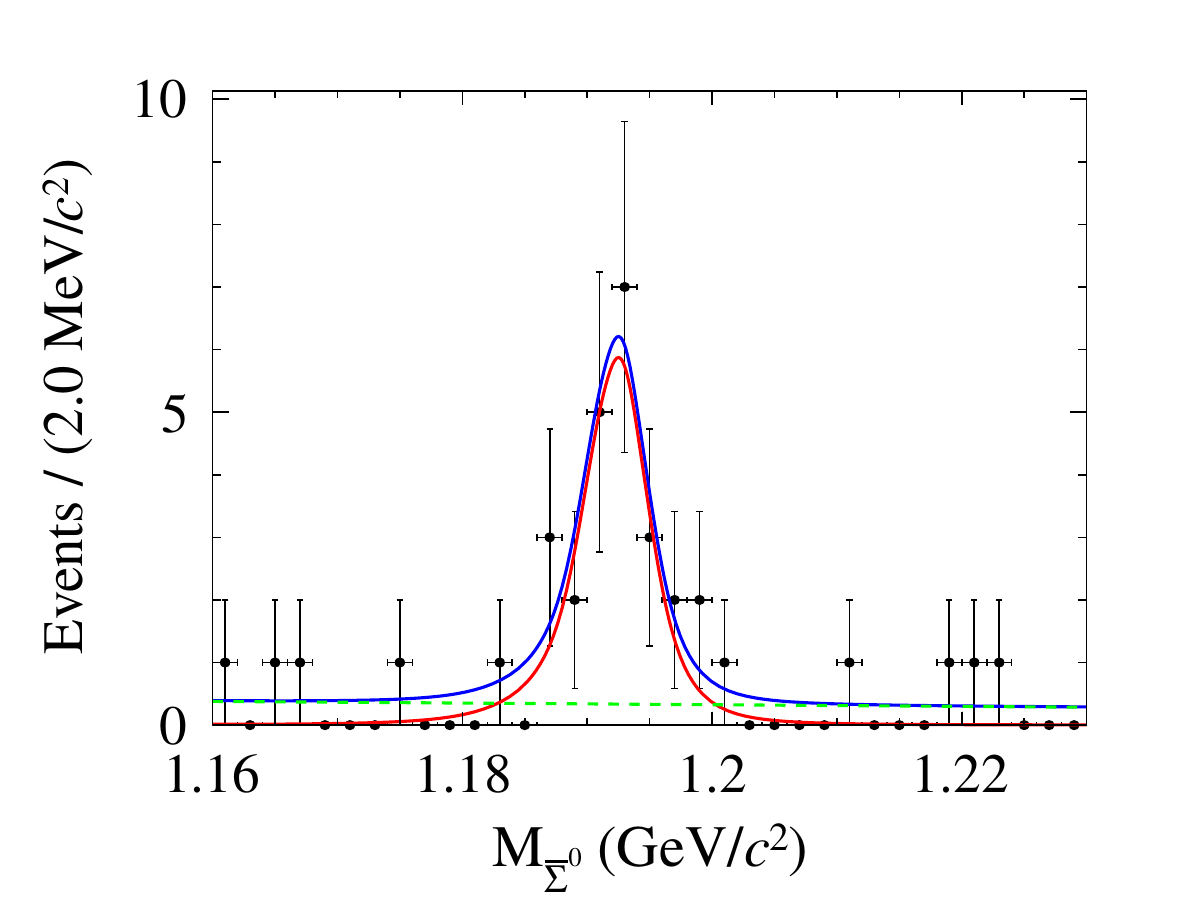}}
      \put(20,65){\color{blue}\scalebox{0.4}{$\sqrt{s}=3.867$ GeV}}
    \end{picture}
  \end{minipage}

  \begin{minipage}[b]{0.24\linewidth}
    \begin{picture}(50,80)
      \put(0,0){\includegraphics[width=\linewidth]{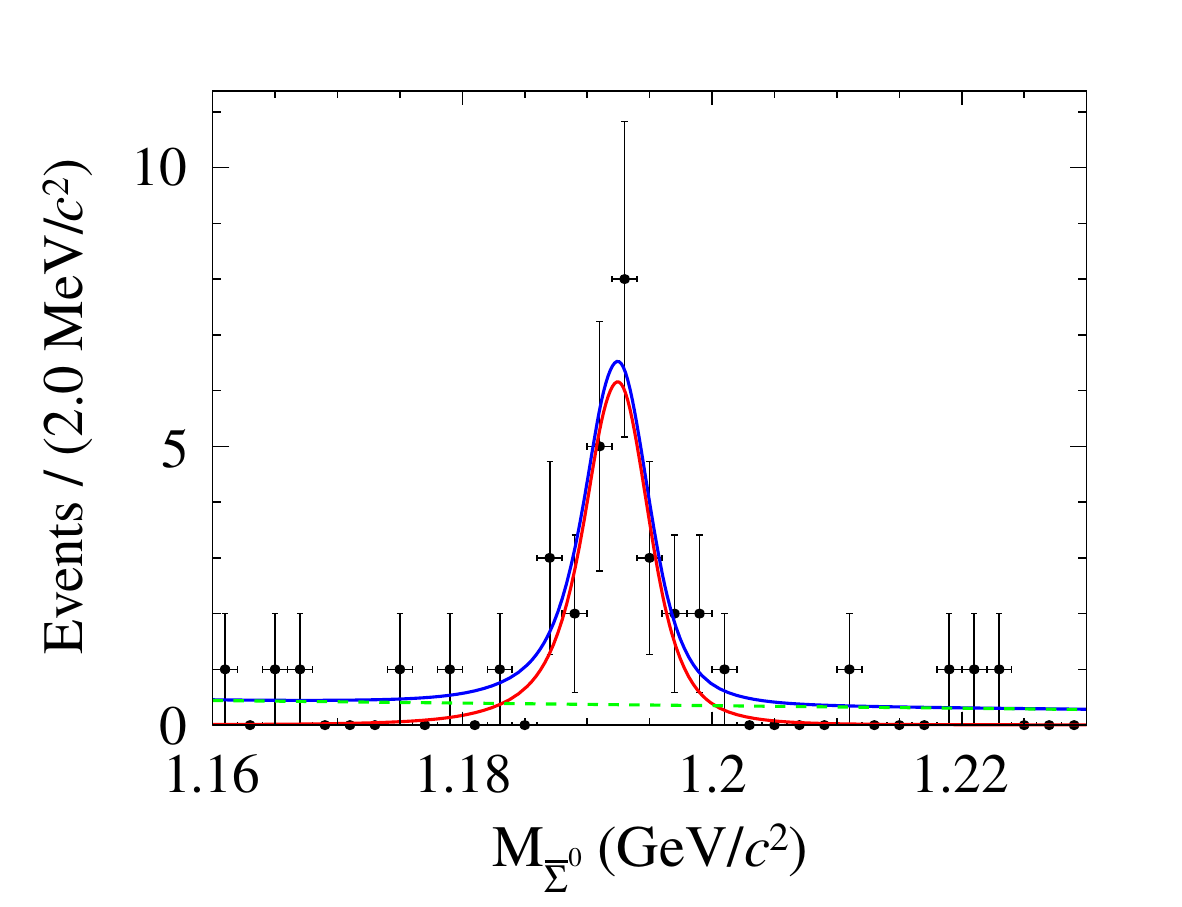}}
      \put(20,65){\color{blue}\scalebox{0.4}{$\sqrt{s}=3.871$ GeV}}
    \end{picture}
  \end{minipage}
  \begin{minipage}[b]{0.24\linewidth}
    \begin{picture}(50,80)
      \put(0,0){\includegraphics[width=\linewidth]{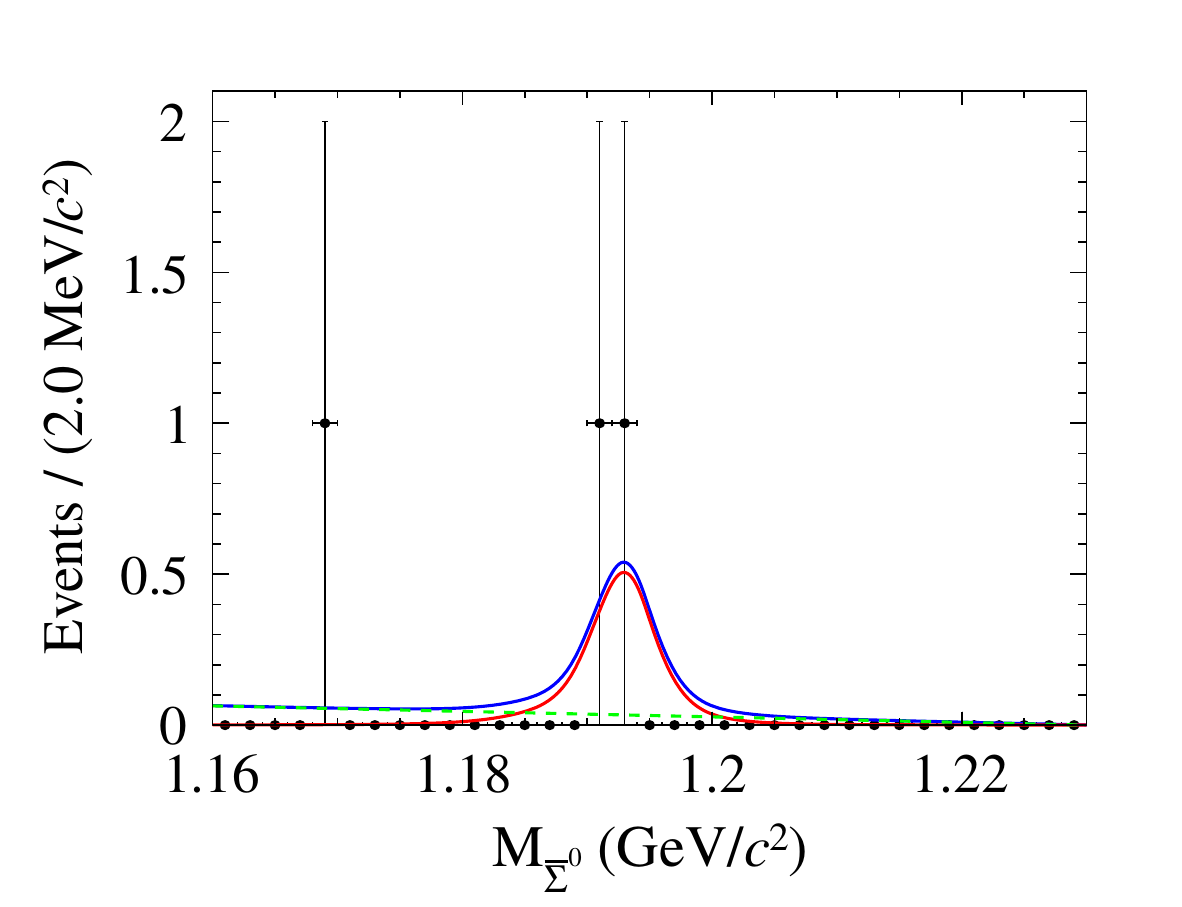}}
      \put(20,65){\color{blue}\scalebox{0.4}{$\sqrt{s}=3.896$ GeV}}
    \end{picture}
  \end{minipage}
  \begin{minipage}[b]{0.24\linewidth}
    \begin{picture}(50,80)
      \put(0,0){\includegraphics[width=\linewidth]{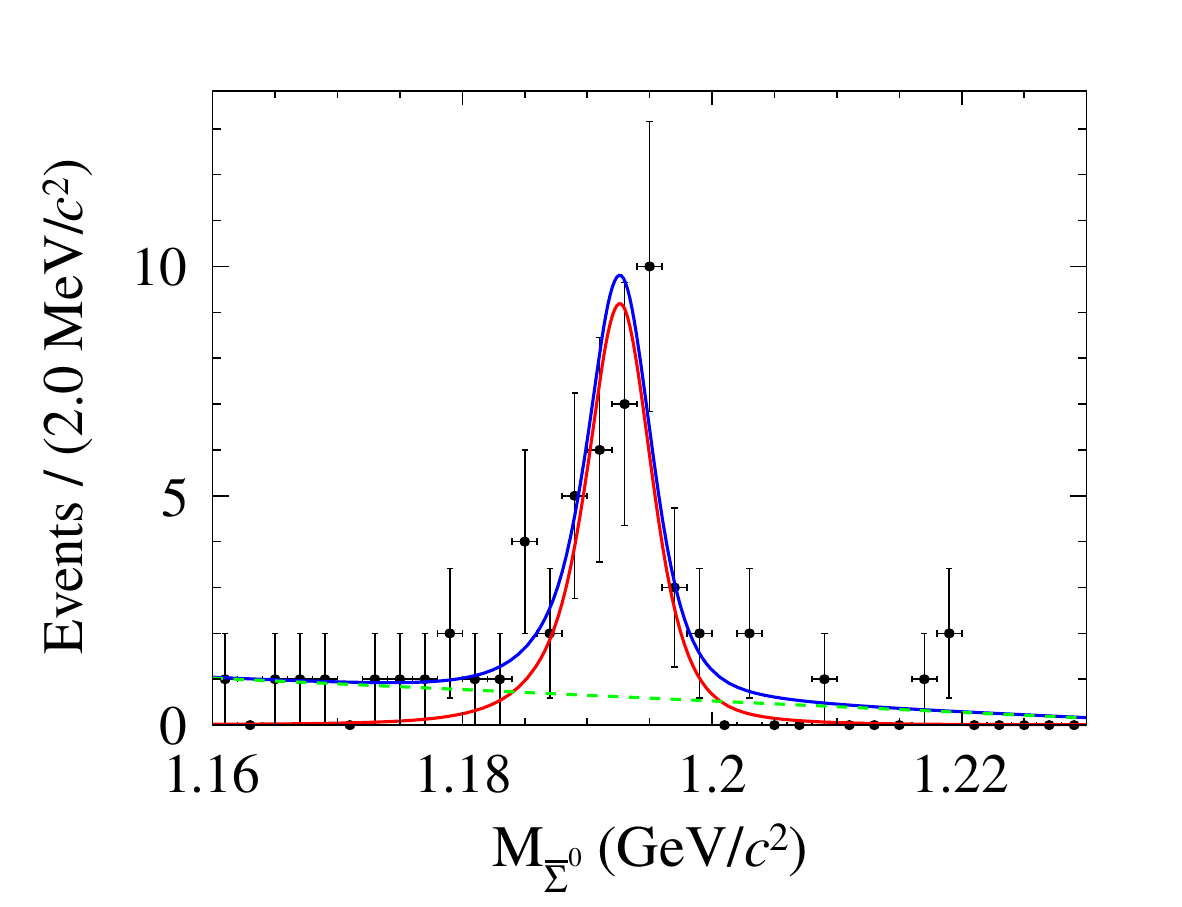}}
      \put(20,65){\color{blue}\scalebox{0.4}{$\sqrt{s}=4.008$ GeV}}
    \end{picture}
  \end{minipage}
  \begin{minipage}[b]{0.24\linewidth}
    \begin{picture}(50,80)
      \put(0,0){\includegraphics[width=\linewidth]{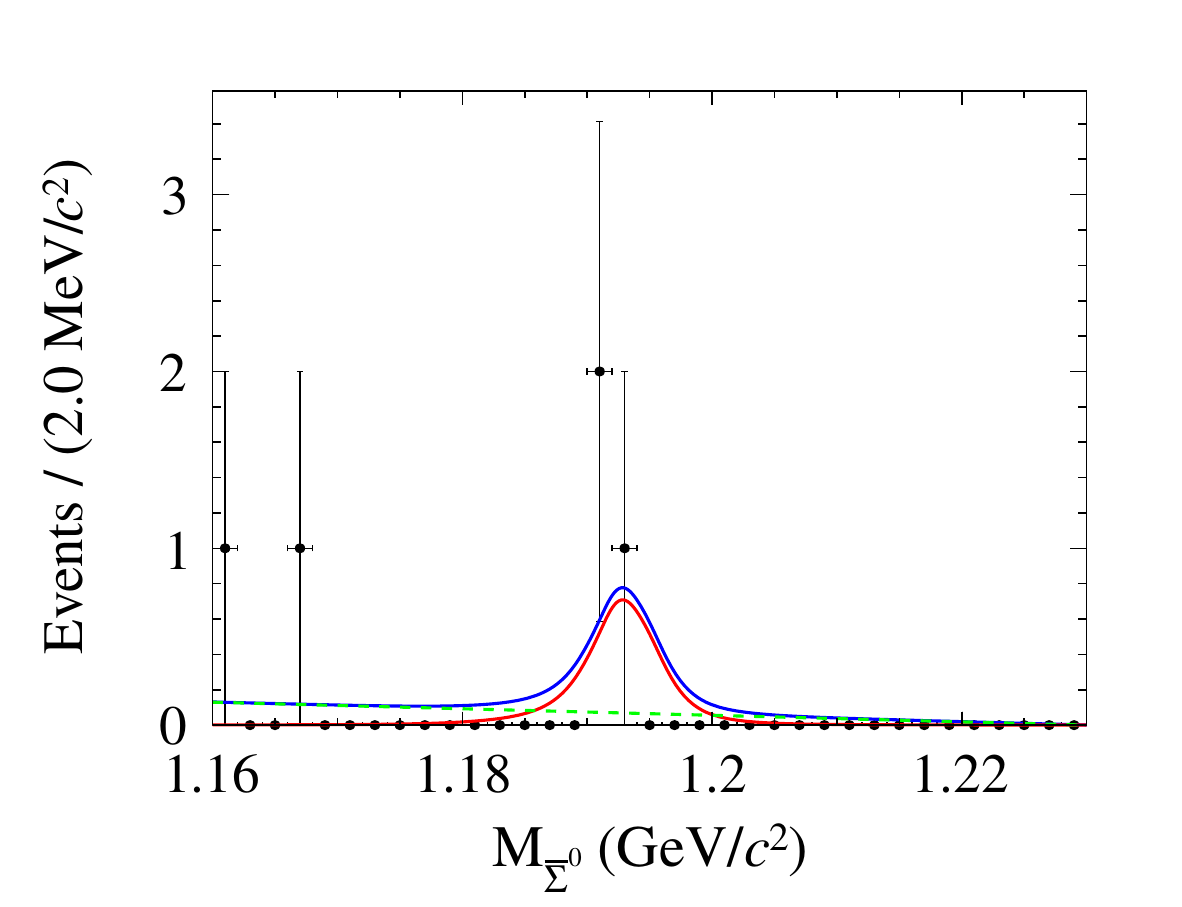}}
      \put(20,65){\color{blue}\scalebox{0.4}{$\sqrt{s}=4.085$ GeV}}
    \end{picture}
  \end{minipage}

  \begin{minipage}[b]{0.24\linewidth}
    \begin{picture}(50,80)
      \put(0,0){\includegraphics[width=\linewidth]{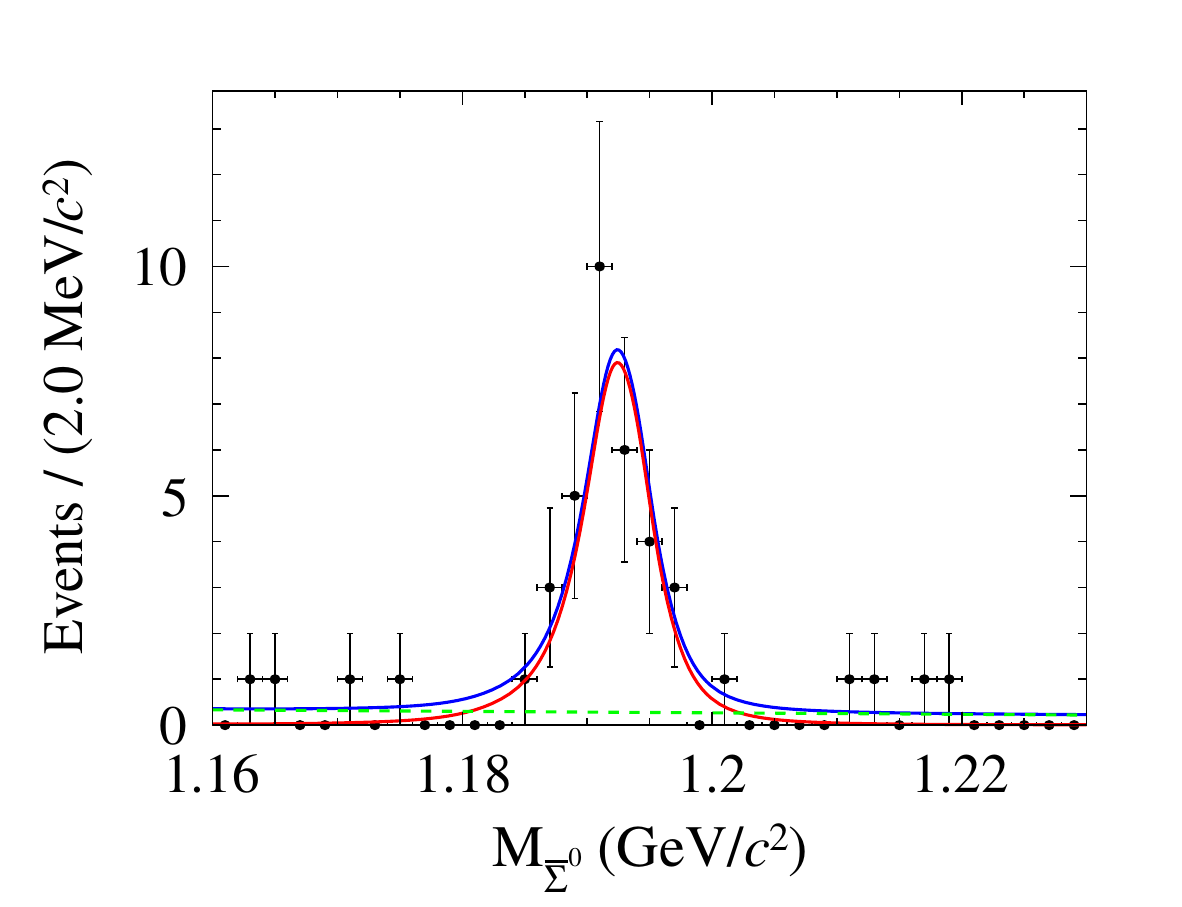}}
      \put(20,65){\color{blue}\scalebox{0.4}{$\sqrt{s}=4.128$ GeV}}
    \end{picture}
  \end{minipage}
  \begin{minipage}[b]{0.24\linewidth}
    \begin{picture}(50,80)
      \put(0,0){\includegraphics[width=\linewidth]{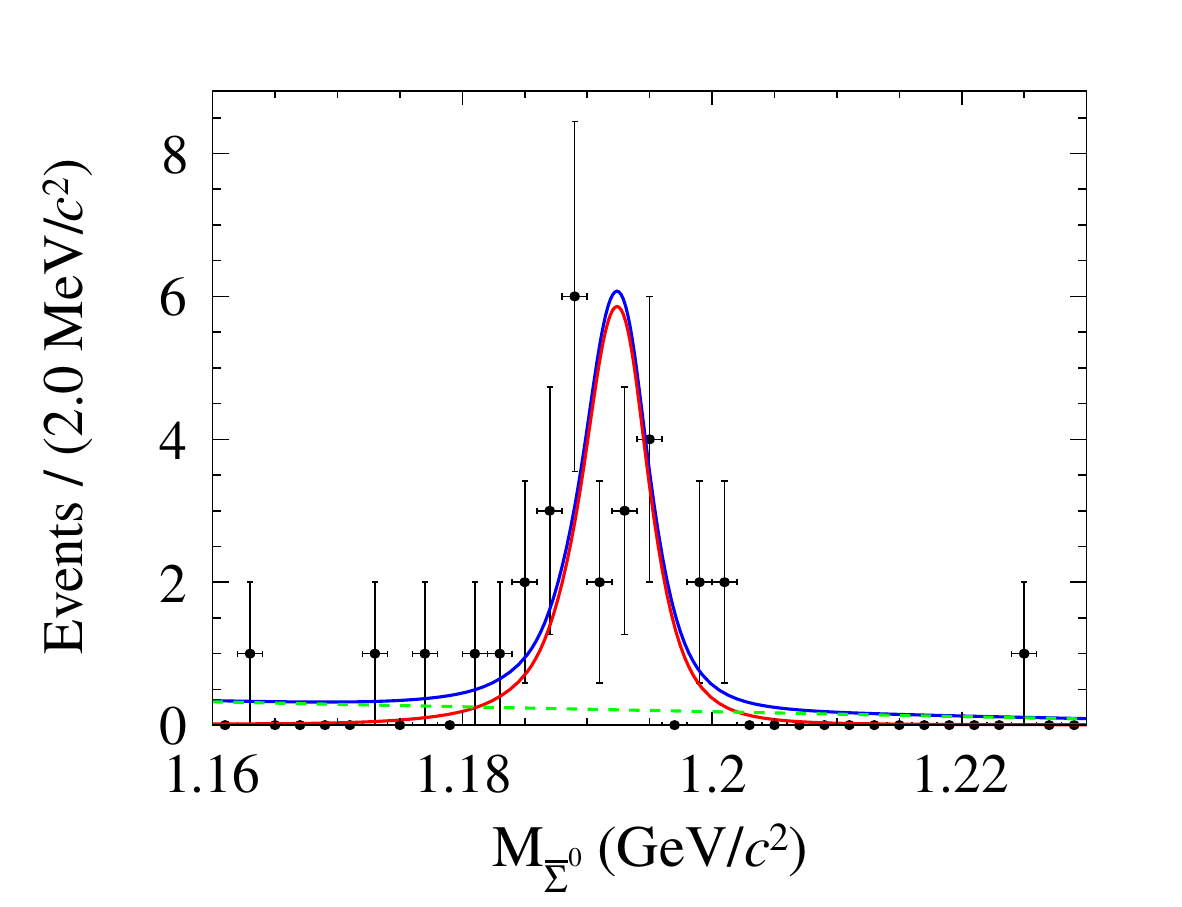}}
      \put(20,65){\color{blue}\scalebox{0.4}{$\sqrt{s}=4.157$ GeV}}
    \end{picture}
  \end{minipage}
  \begin{minipage}[b]{0.24\linewidth}
    \begin{picture}(50,80)
      \put(0,0){\includegraphics[width=\linewidth]{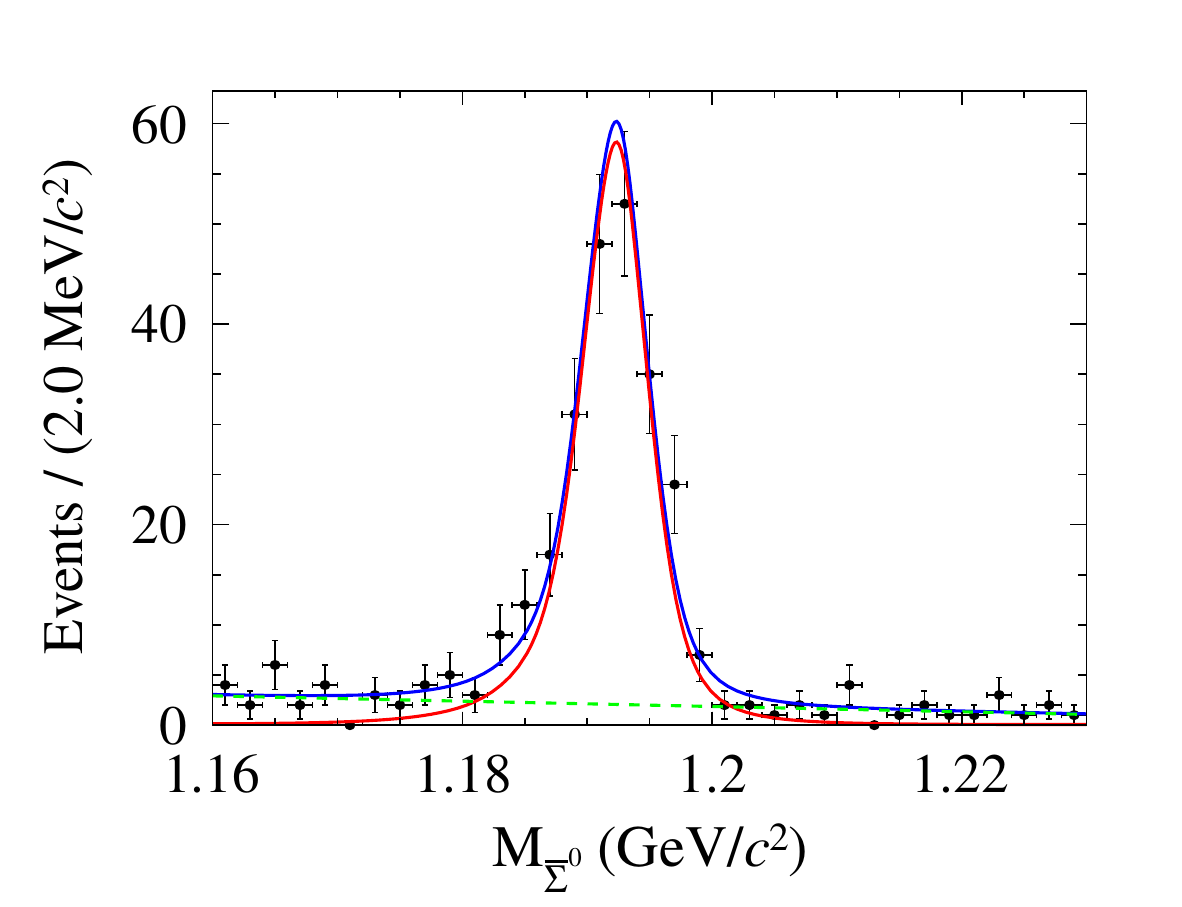}}
      \put(20,65){\color{blue}\scalebox{0.4}{$\sqrt{s}=4.178$ GeV}}
    \end{picture}
  \end{minipage}
  \begin{minipage}[b]{0.24\linewidth}
    \begin{picture}(50,80)
      \put(0,0){\includegraphics[width=\linewidth]{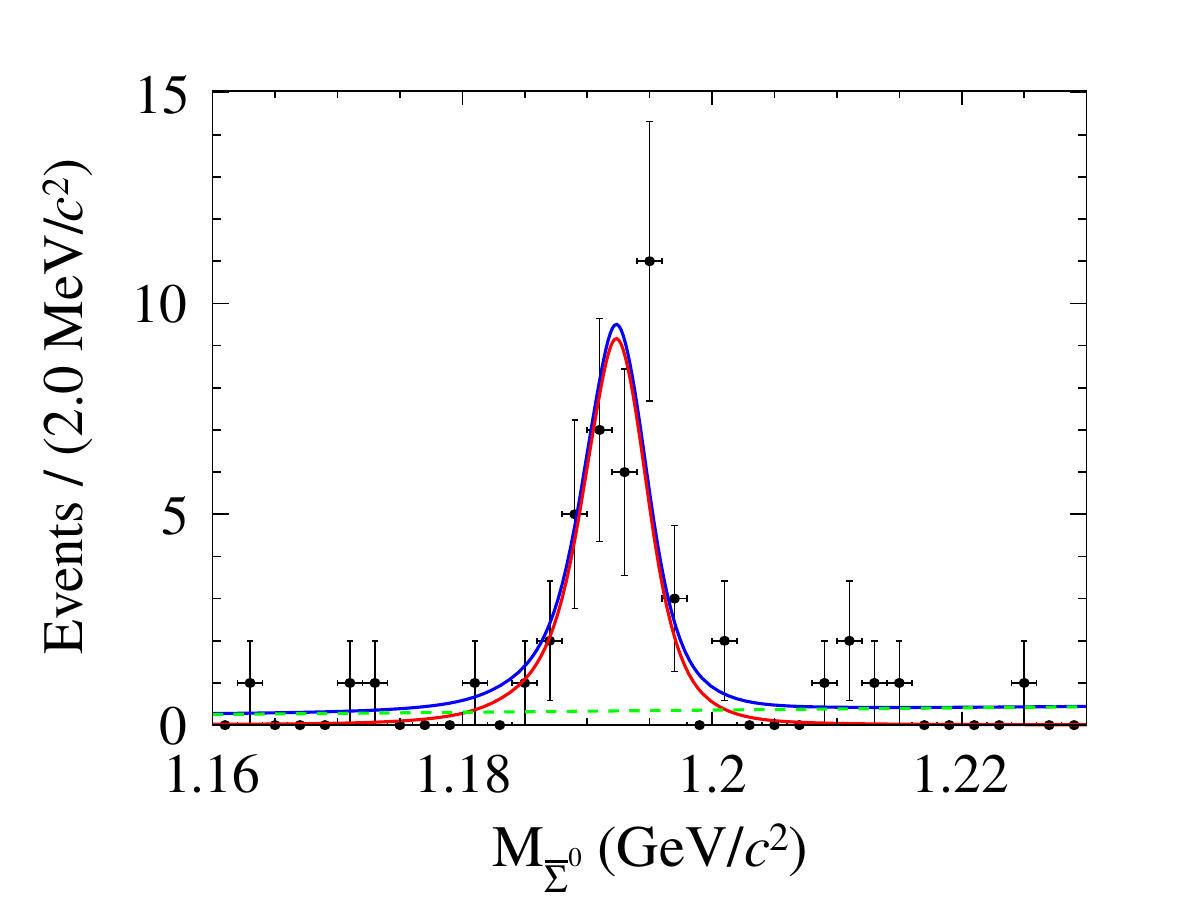}}
      \put(20,65){\color{blue}\scalebox{0.4}{$\sqrt{s}=4.189$ GeV}}
    \end{picture}
  \end{minipage}

  \begin{minipage}[b]{0.24\linewidth}
    \begin{picture}(50,80)
      \put(0,0){\includegraphics[width=\linewidth]{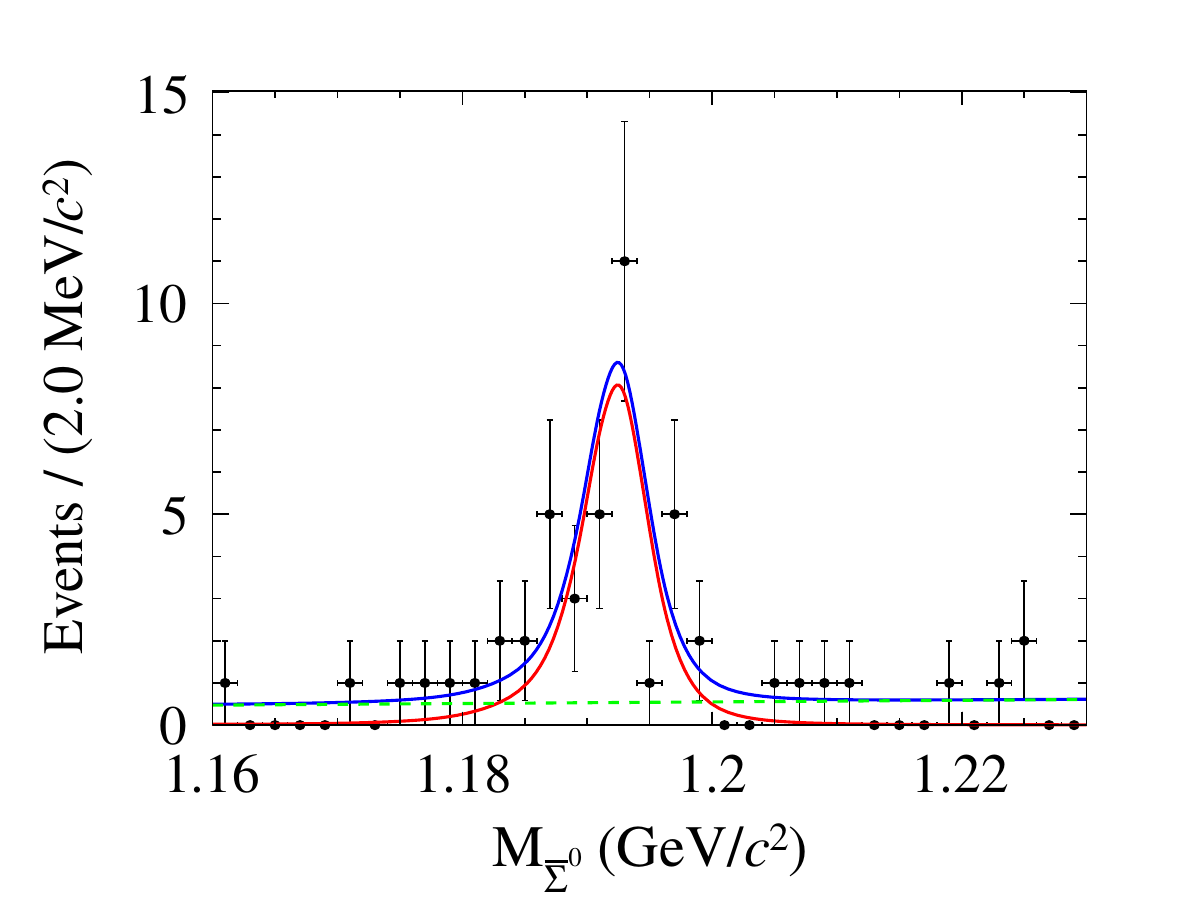}}
      \put(20,65){\color{blue}\scalebox{0.4}{$\sqrt{s}=4.199$ GeV}}
    \end{picture}
  \end{minipage}
  \begin{minipage}[b]{0.24\linewidth}
    \begin{picture}(50,80)
      \put(0,0){\includegraphics[width=\linewidth]{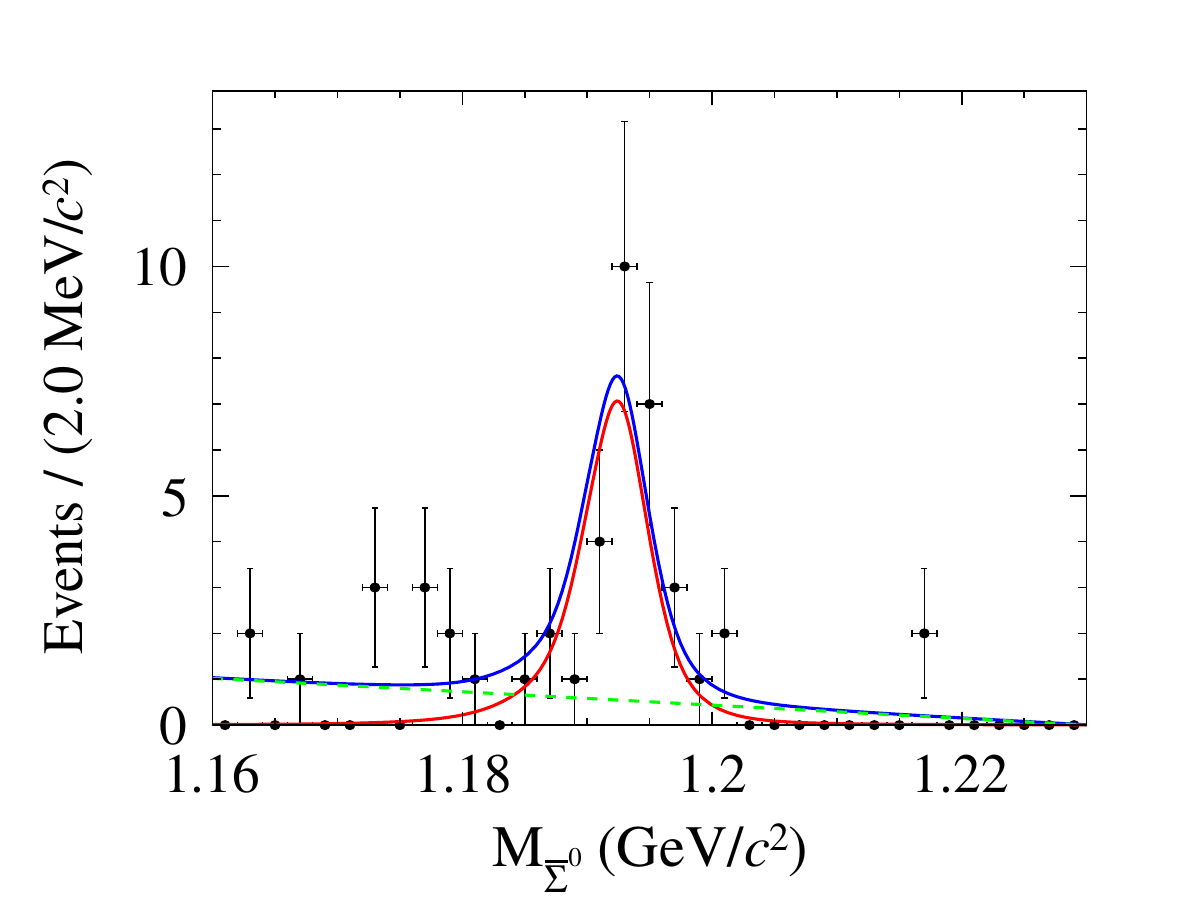}}
      \put(20,65){\color{blue}\scalebox{0.4}{$\sqrt{s}=4.208$ GeV}}
    \end{picture}
  \end{minipage}
  \begin{minipage}[b]{0.24\linewidth}
    \begin{picture}(50,80)
      \put(0,0){\includegraphics[width=\linewidth]{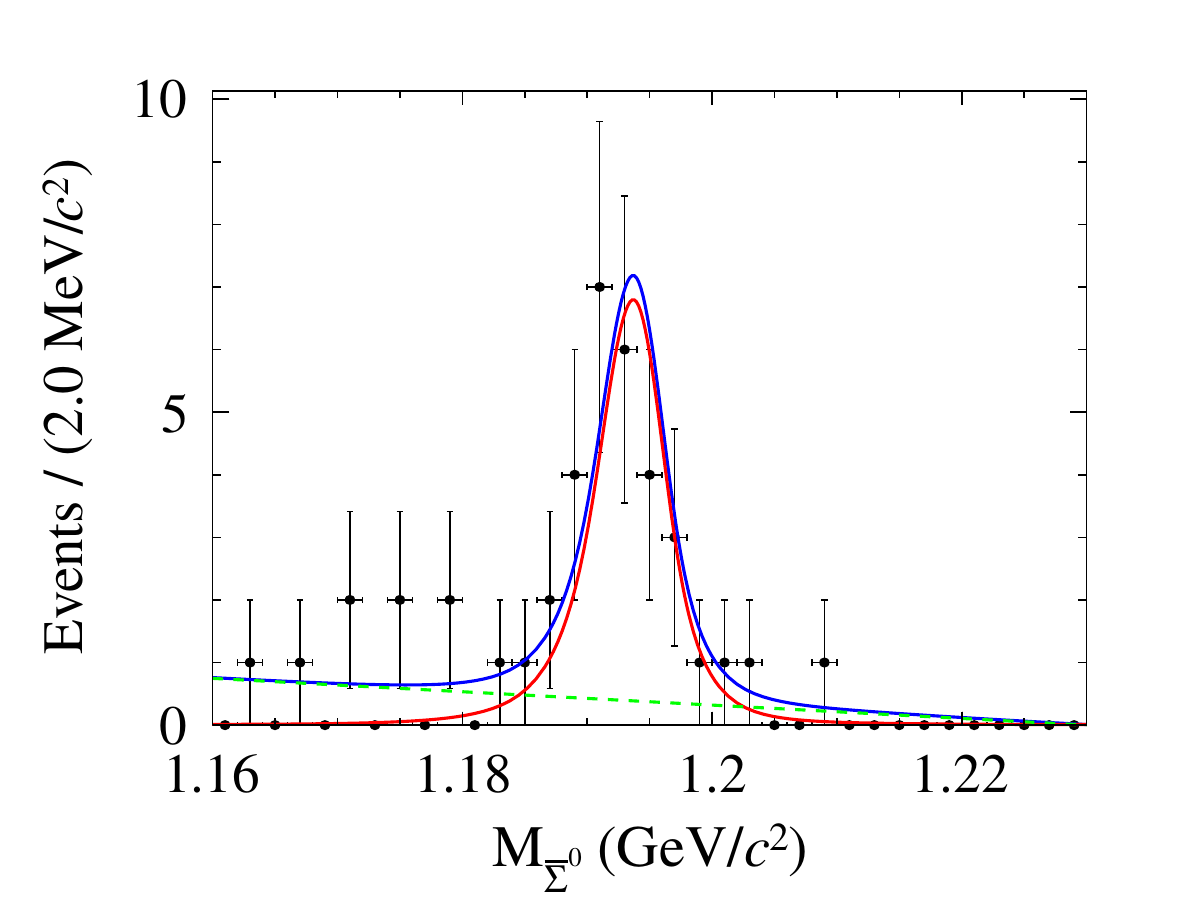}}
      \put(20,65){\color{blue}\scalebox{0.4}{$\sqrt{s}=4.217$ GeV}}
    \end{picture}
  \end{minipage}
  \begin{minipage}[b]{0.24\linewidth}
    \begin{picture}(50,80)
      \put(0,0){\includegraphics[width=\linewidth]{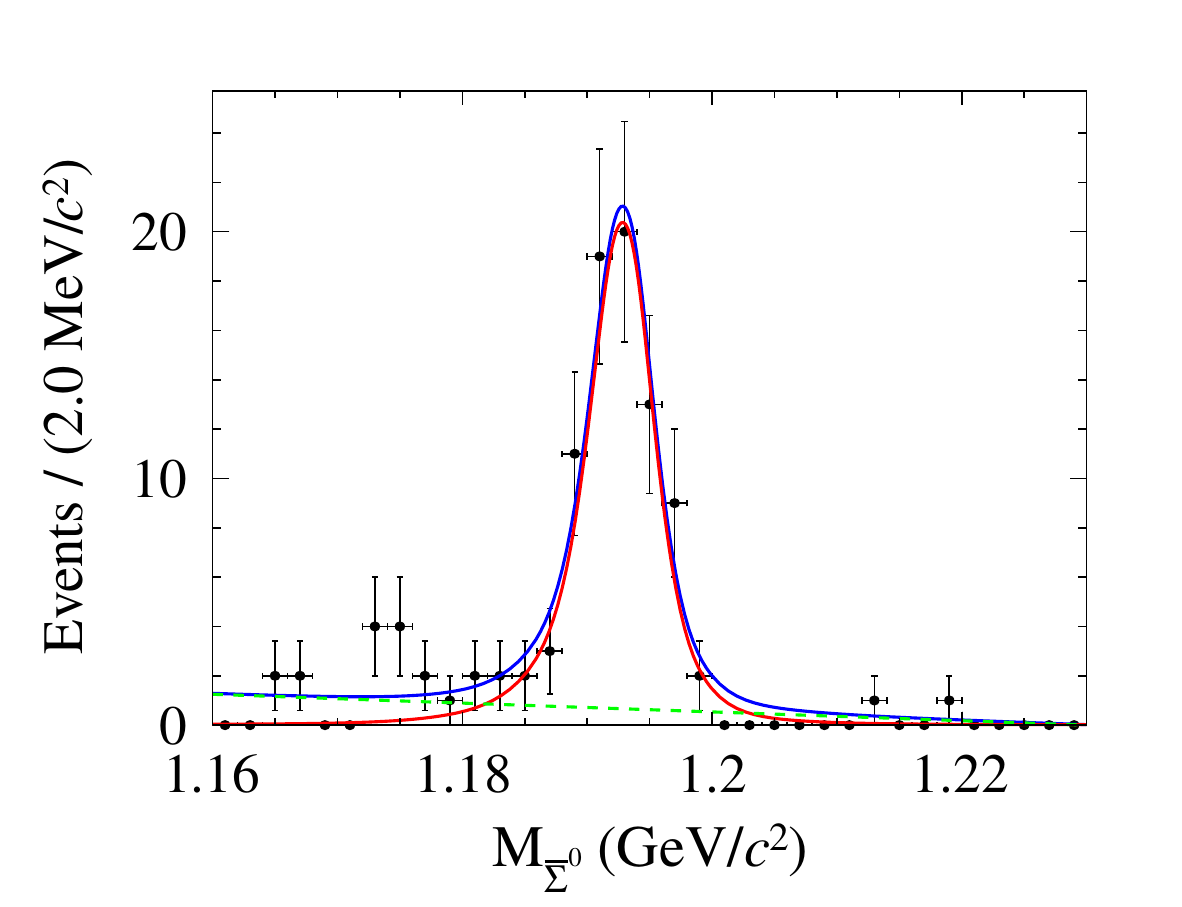}}
      \put(20,65){\color{blue}\scalebox{0.4}{$\sqrt{s}=4.226$ GeV}}
    \end{picture}
  \end{minipage}

  \begin{minipage}[b]{0.24\linewidth}
    \begin{picture}(50,80)
      \put(0,0){\includegraphics[width=\linewidth]{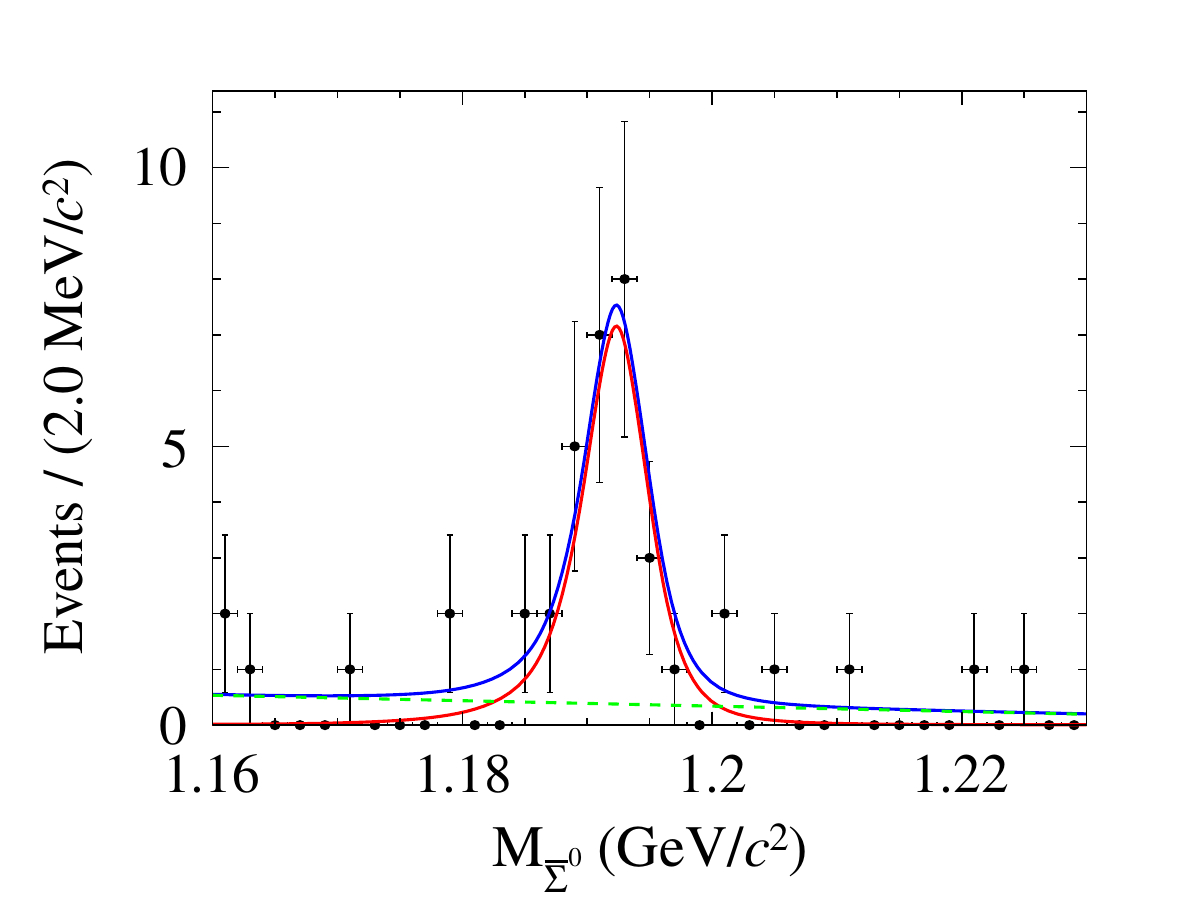}}
      \put(20,65){\color{blue}\scalebox{0.4}{$\sqrt{s}=4.236$ GeV}}
    \end{picture}
  \end{minipage}
  \begin{minipage}[b]{0.24\linewidth}
    \begin{picture}(50,80)
      \put(0,0){\includegraphics[width=\linewidth]{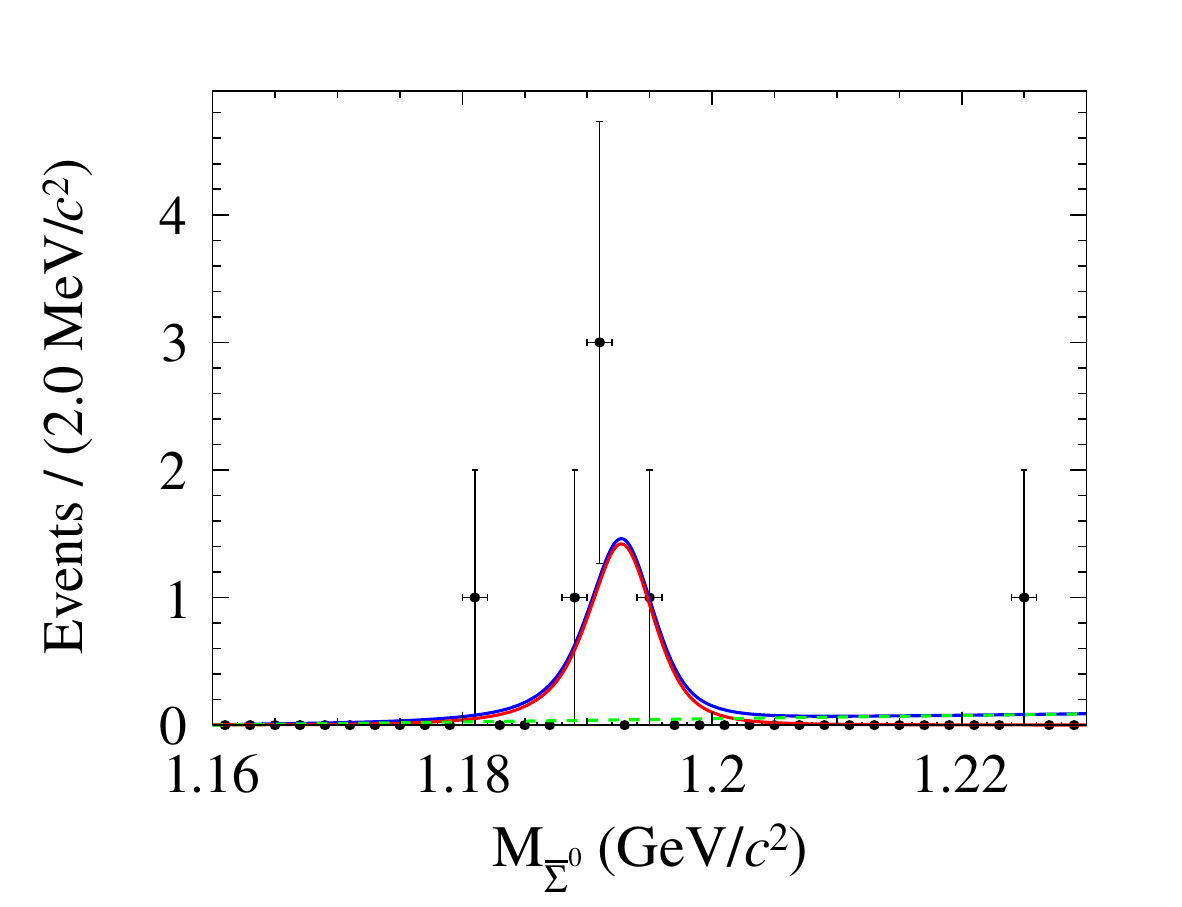}}
      \put(20,65){\color{blue}\scalebox{0.4}{$\sqrt{s}=4.242$ GeV}}
    \end{picture}
  \end{minipage}
  \begin{minipage}[b]{0.24\linewidth}
    \begin{picture}(50,80)
      \put(0,0){\includegraphics[width=\linewidth]{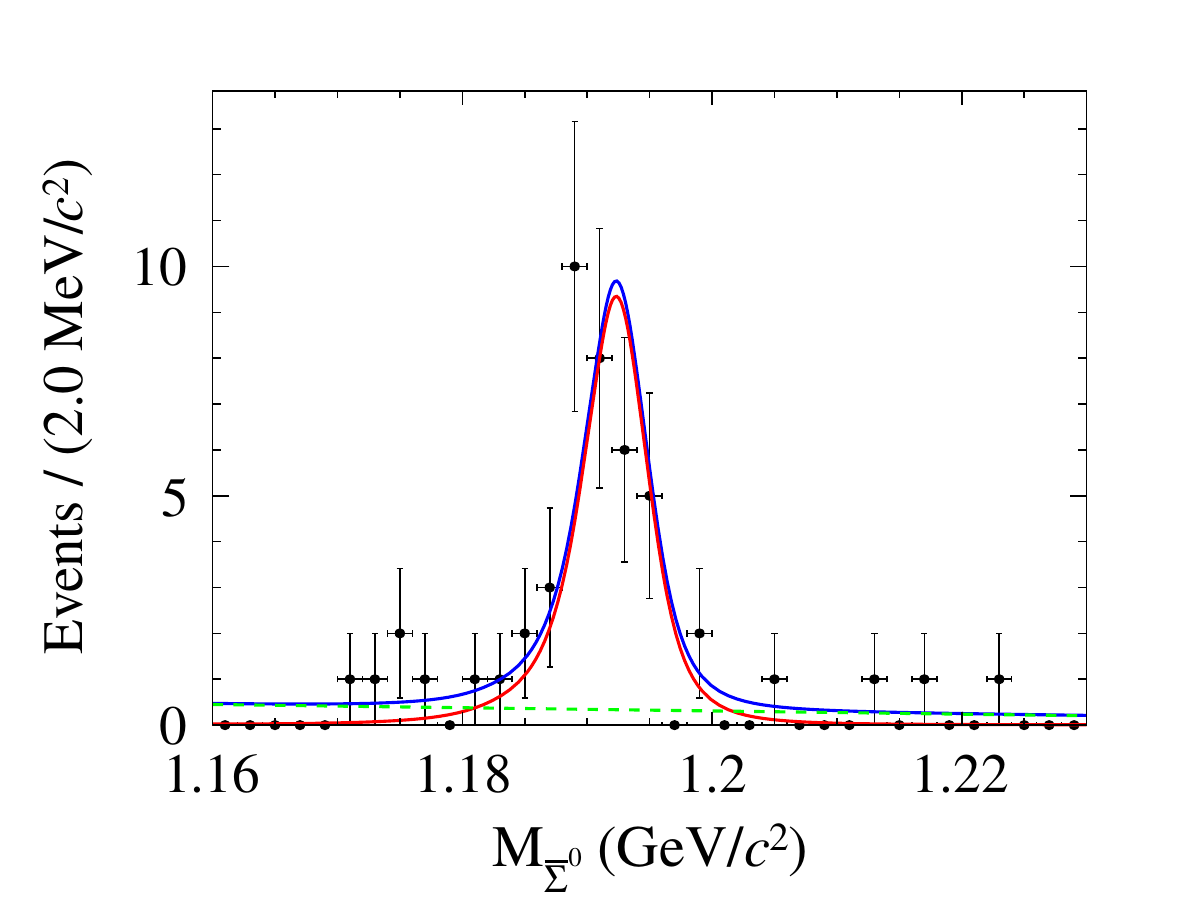}}
      \put(20,65){\color{blue}\scalebox{0.4}{$\sqrt{s}=4.244$ GeV}}
    \end{picture}
  \end{minipage}
  \begin{minipage}[b]{0.24\linewidth}
    \begin{picture}(50,80)
      \put(0,0){\includegraphics[width=\linewidth]{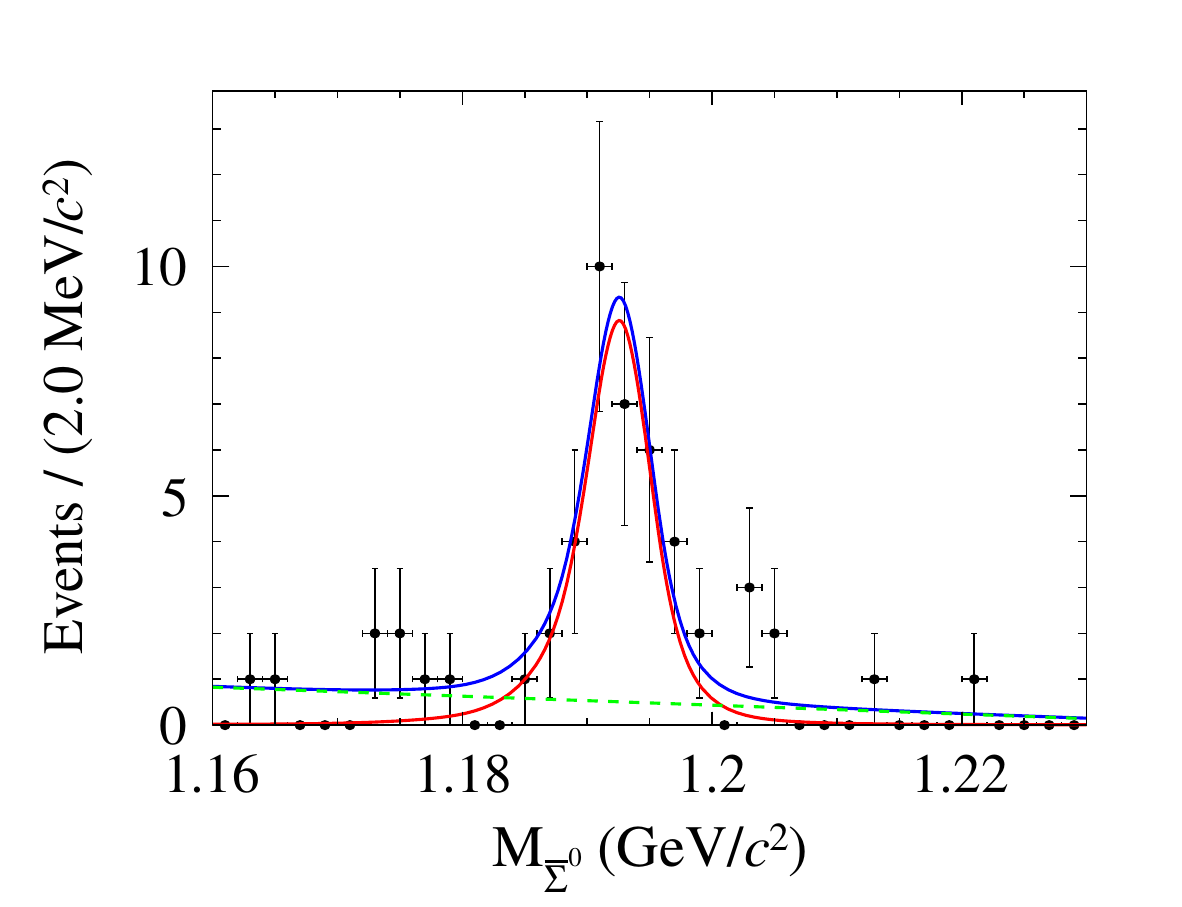}}
      \put(20,65){\color{blue}\scalebox{0.4}{$\sqrt{s}=4.258$ GeV}}
    \end{picture}
  \end{minipage}
  \caption{Fits to the $M_{\bar{\Sigma}^{0}}$ distributions at $\sqrt{s}$ between 3.51 and 4.26 GeV. The dots with error bars represent the data, the blue solid lines show the fit results, the red solid lines correspond to the signal shapes, and the green dashed lines indicate the background contributions.}
  \label{fitMsigma1}
\end{figure}

\begin{figure}[htbp]
  \centering
\begin{minipage}[b]{0.24\linewidth}
    \begin{picture}(50,80)
      \put(0,0){\includegraphics[width=\linewidth]{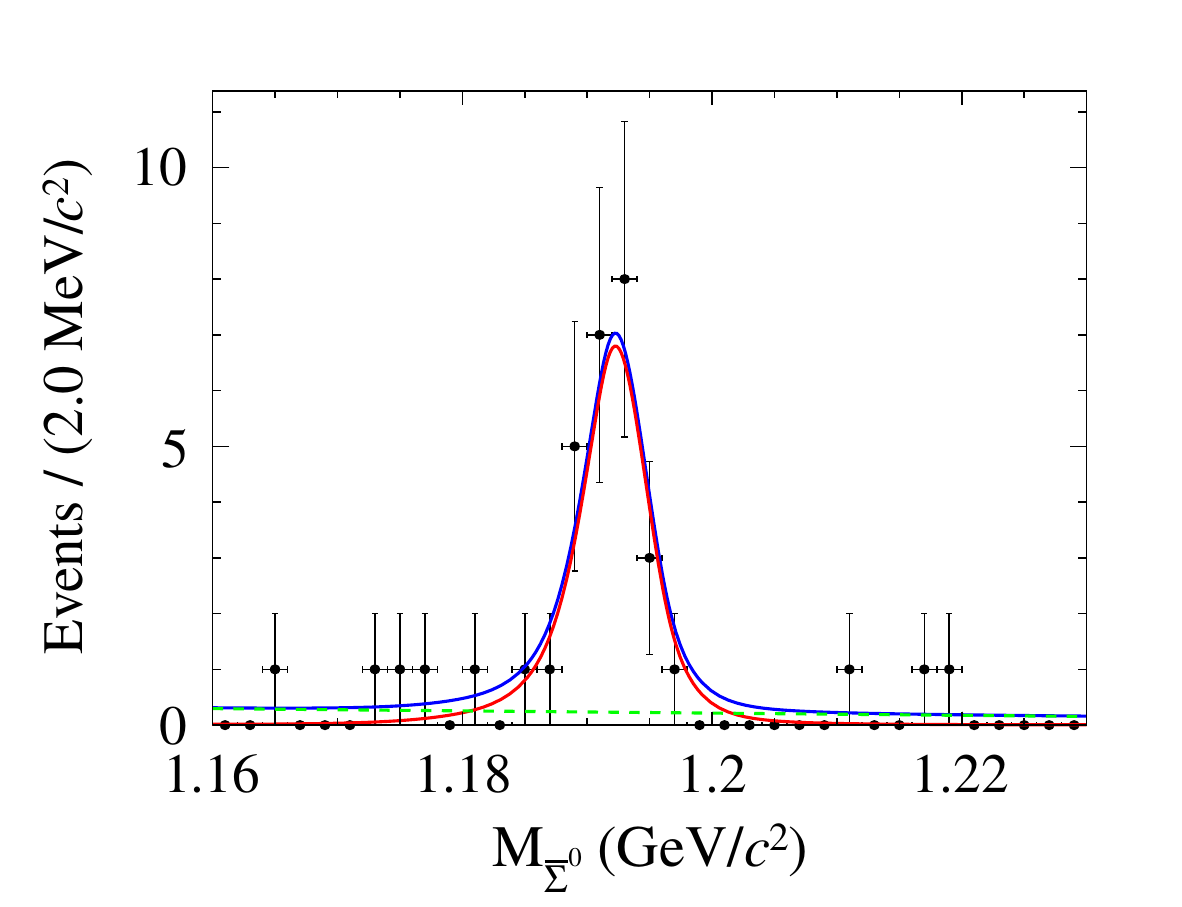}}
      \put(20,65){\color{blue}\scalebox{0.4}{$\sqrt{s}=4.267$ GeV}}
    \end{picture}
  \end{minipage}
  \begin{minipage}[b]{0.24\linewidth}
    \begin{picture}(50,80)
      \put(0,0){\includegraphics[width=\linewidth]{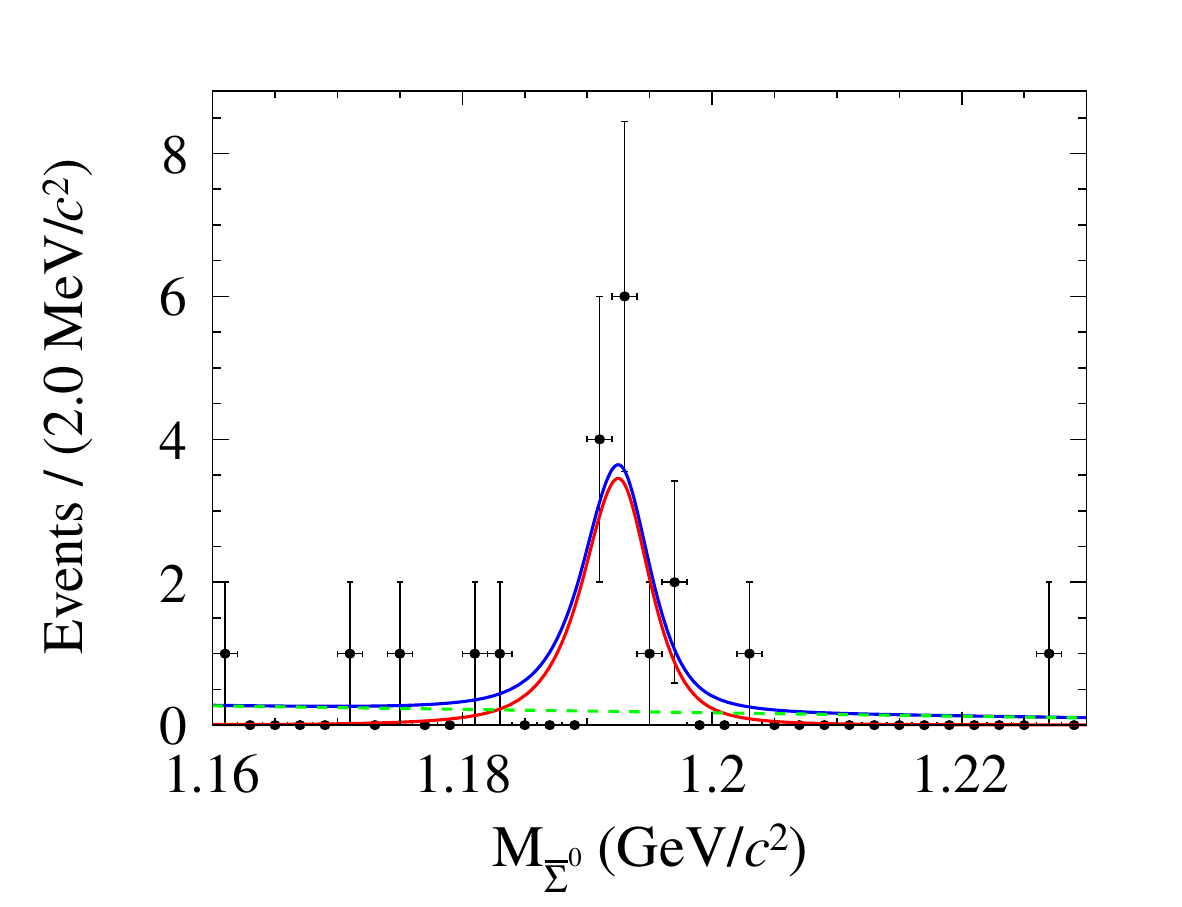}}
      \put(20,65){\color{blue}\scalebox{0.4}{$\sqrt{s}=4.278$ GeV}}
    \end{picture}
  \end{minipage}
  \begin{minipage}[b]{0.24\linewidth}
    \begin{picture}(50,80)
      \put(0,0){\includegraphics[width=\linewidth]{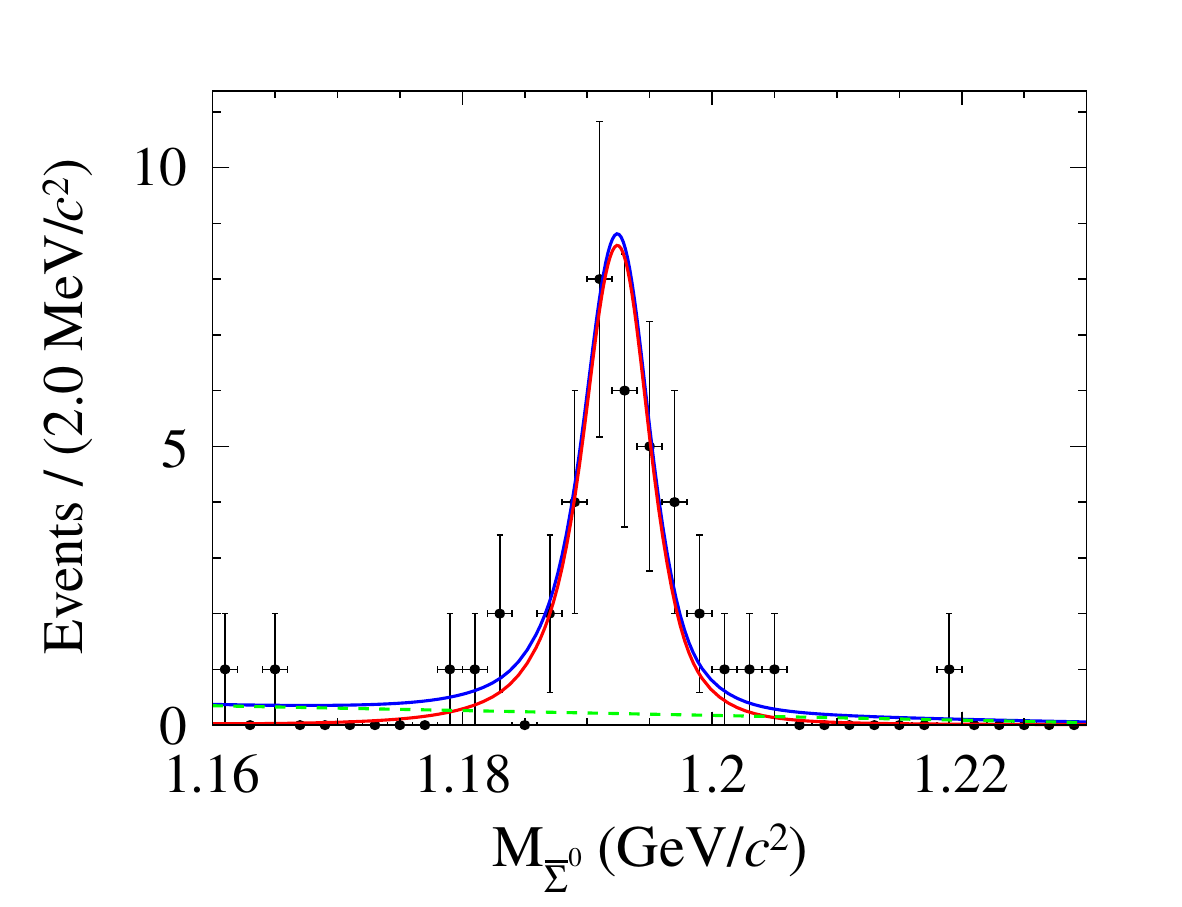}}
      \put(20,65){\color{blue}\scalebox{0.4}{$\sqrt{s}=4.288$ GeV}}
    \end{picture}
  \end{minipage}
  \begin{minipage}[b]{0.24\linewidth}
    \begin{picture}(50,80)
      \put(0,0){\includegraphics[width=\linewidth]{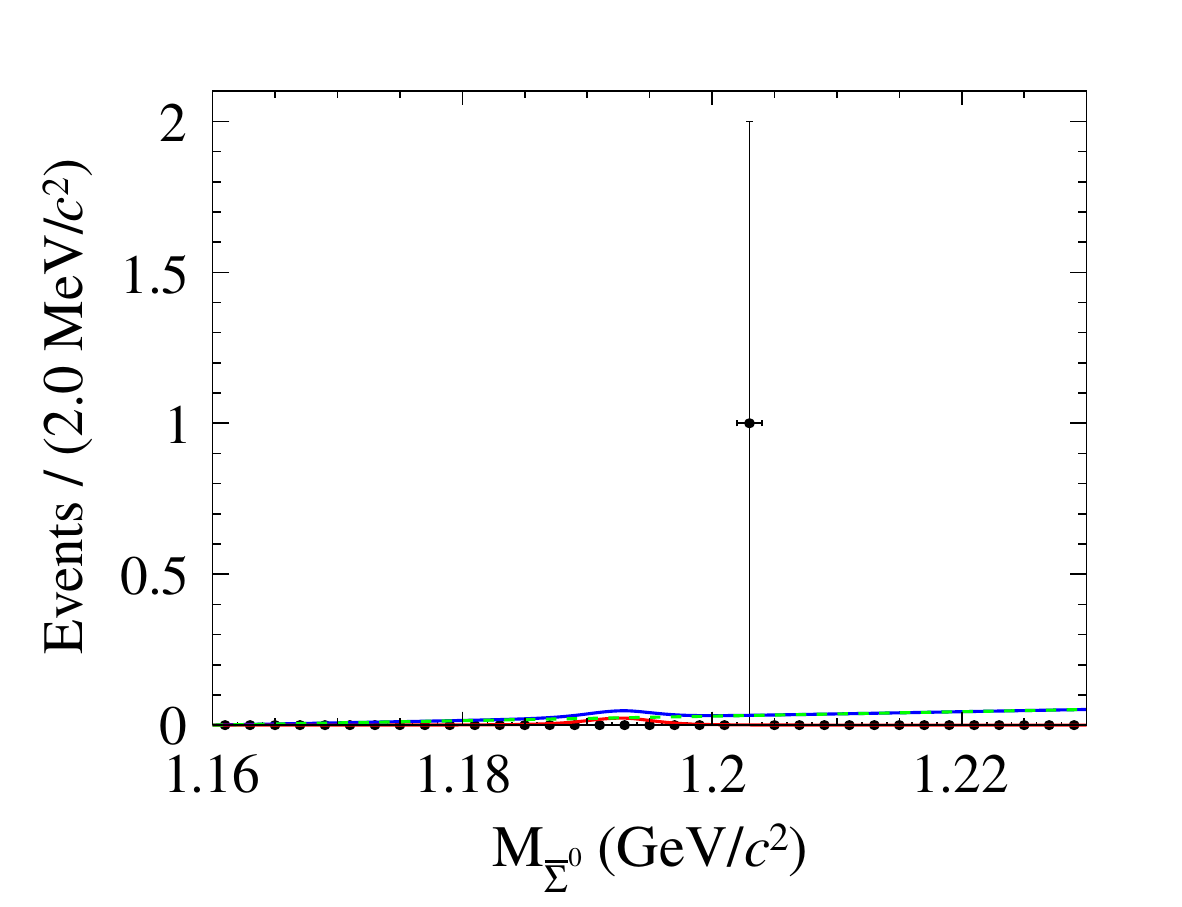}}
      \put(20,65){\color{blue}\scalebox{0.4}{$\sqrt{s}=4.308$ GeV}}
    \end{picture}
  \end{minipage}

  \begin{minipage}[b]{0.24\linewidth}
    \begin{picture}(50,80)
      \put(0,0){\includegraphics[width=\linewidth]{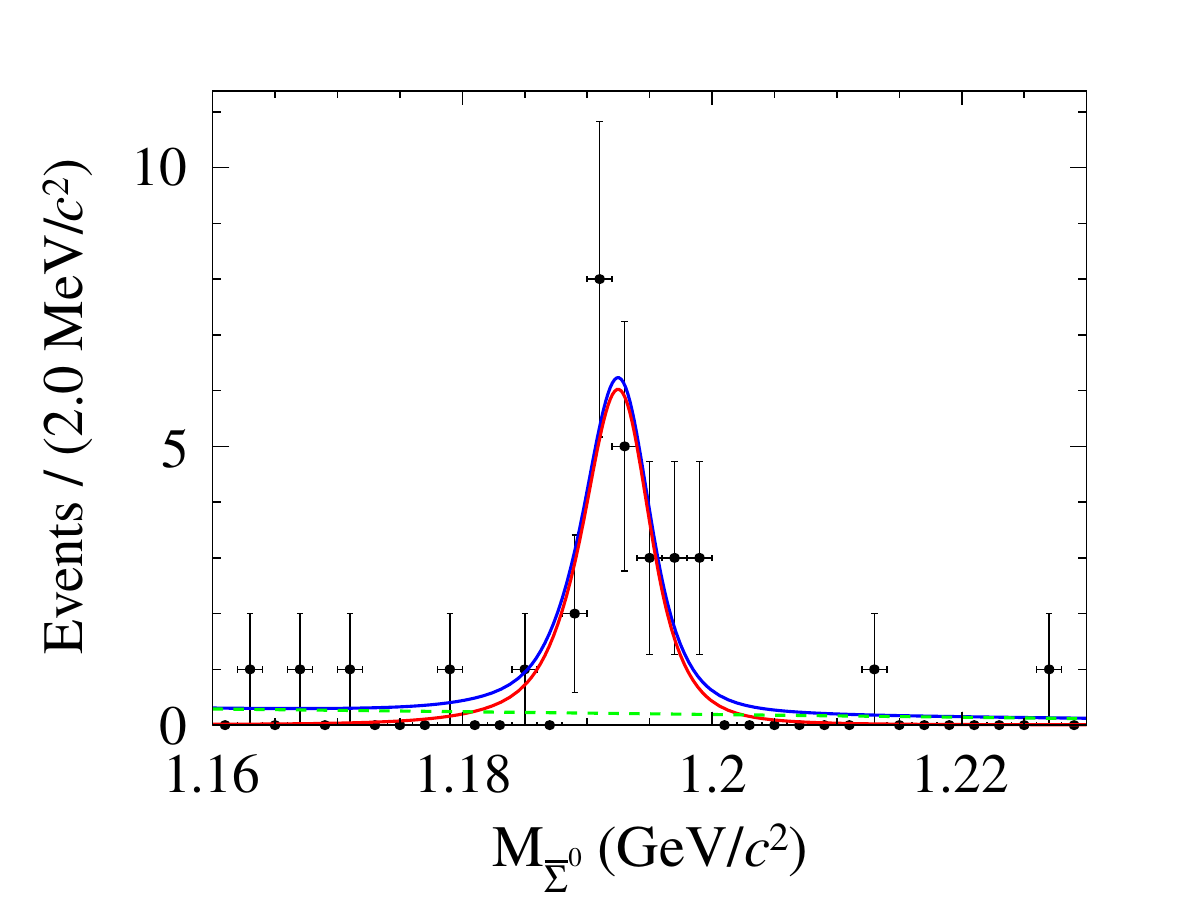}}
      \put(20,65){\color{blue}\scalebox{0.4}{$\sqrt{s}=4.312$ GeV}}
    \end{picture}
  \end{minipage}
  \begin{minipage}[b]{0.24\linewidth}
    \begin{picture}(50,80)
      \put(0,0){\includegraphics[width=\linewidth]{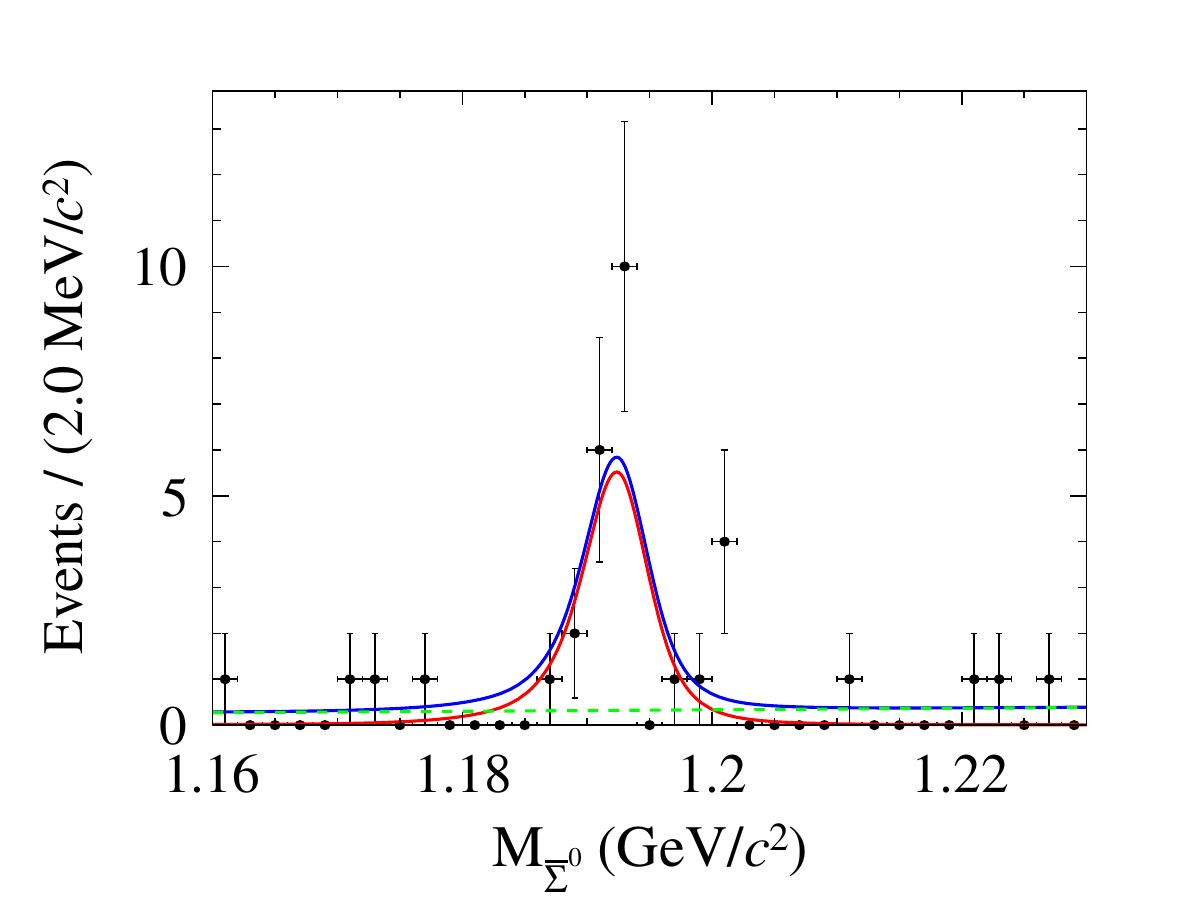}}
      \put(20,65){\color{blue}\scalebox{0.4}{$\sqrt{s}=4.337$ GeV}}
    \end{picture}
  \end{minipage}
  \begin{minipage}[b]{0.24\linewidth}
    \begin{picture}(50,80)
      \put(0,0){\includegraphics[width=\linewidth]{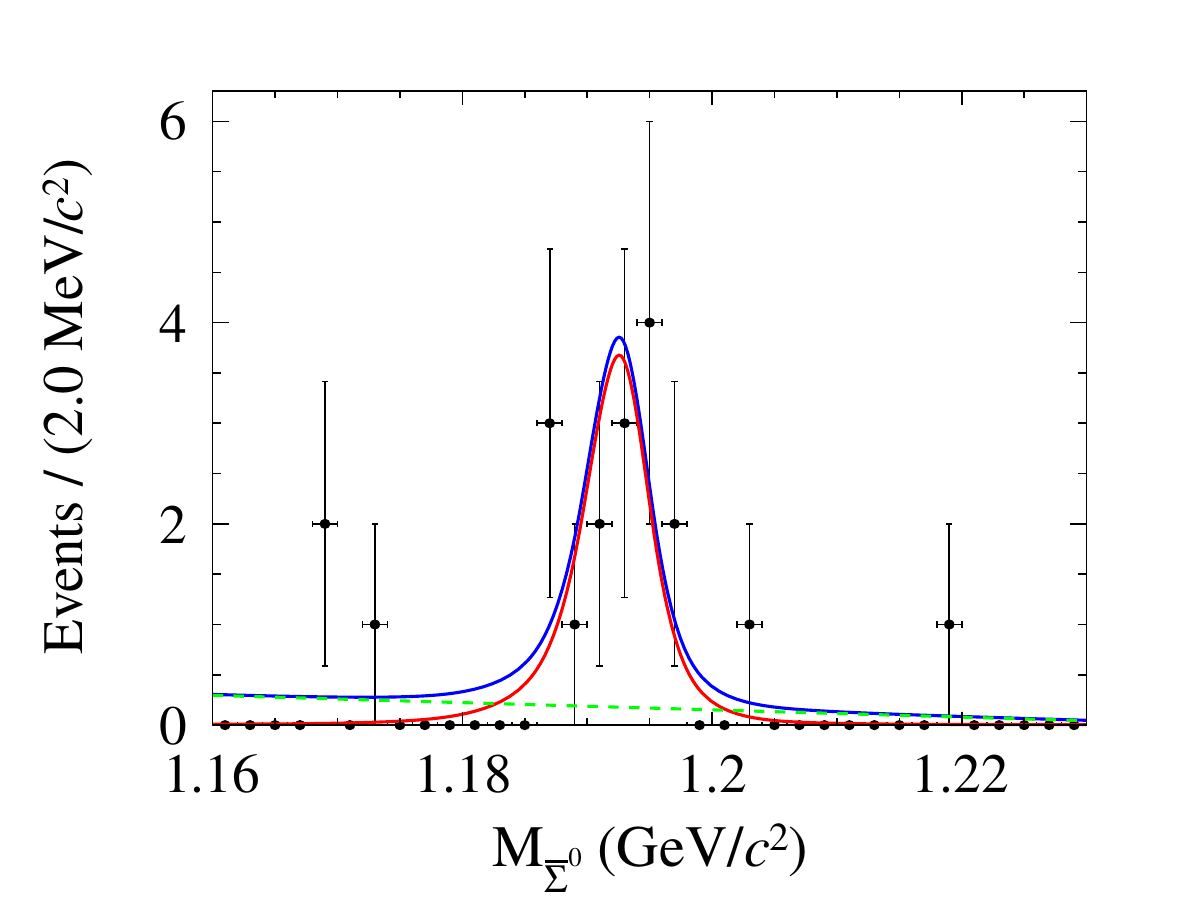}}
      \put(20,65){\color{blue}\scalebox{0.4}{$\sqrt{s}=4.358$ GeV}}
    \end{picture}
  \end{minipage}
  \begin{minipage}[b]{0.24\linewidth}
    \begin{picture}(50,80)
      \put(0,0){\includegraphics[width=\linewidth]{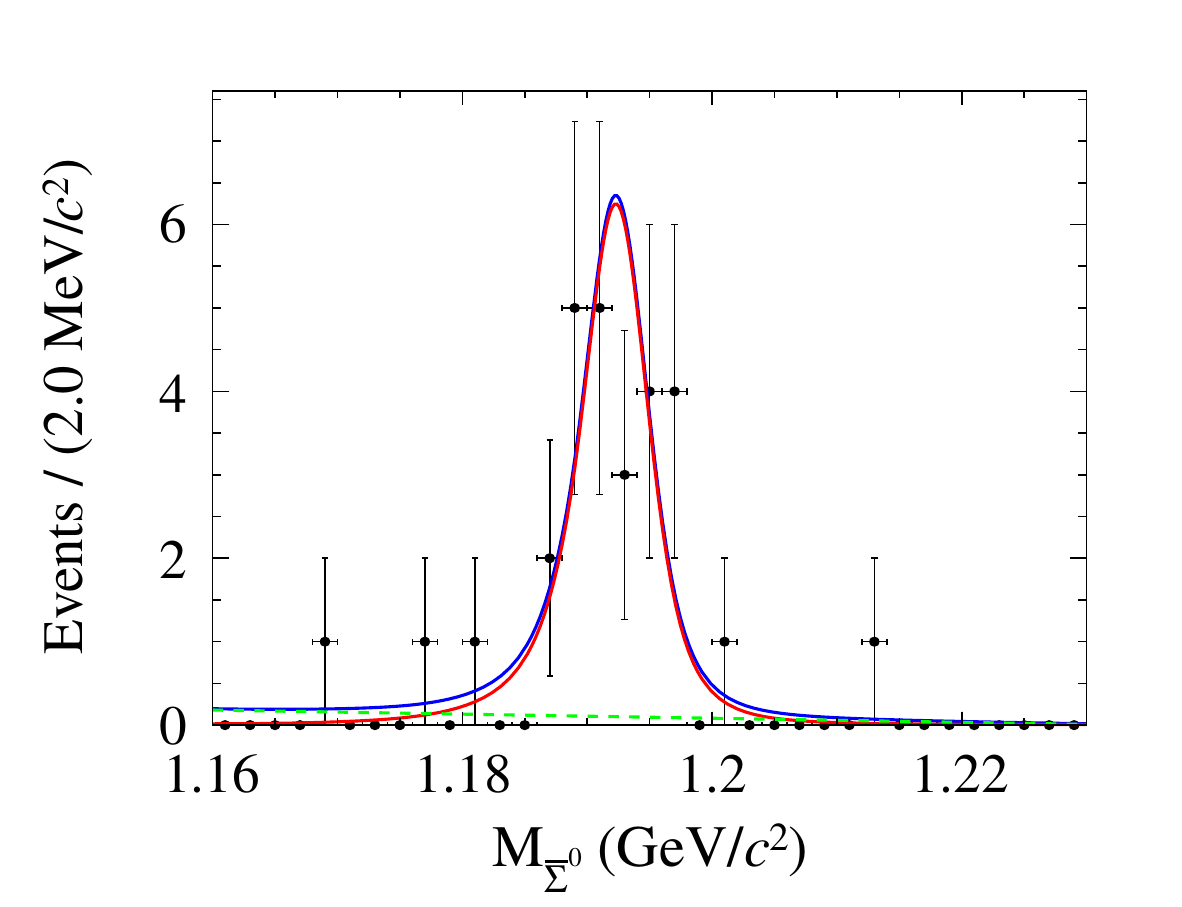}}
      \put(20,65){\color{blue}\scalebox{0.4}{$\sqrt{s}=4.377$ GeV}}
    \end{picture}
  \end{minipage}

  \begin{minipage}[b]{0.24\linewidth}
    \begin{picture}(50,80)
      \put(0,0){\includegraphics[width=\linewidth]{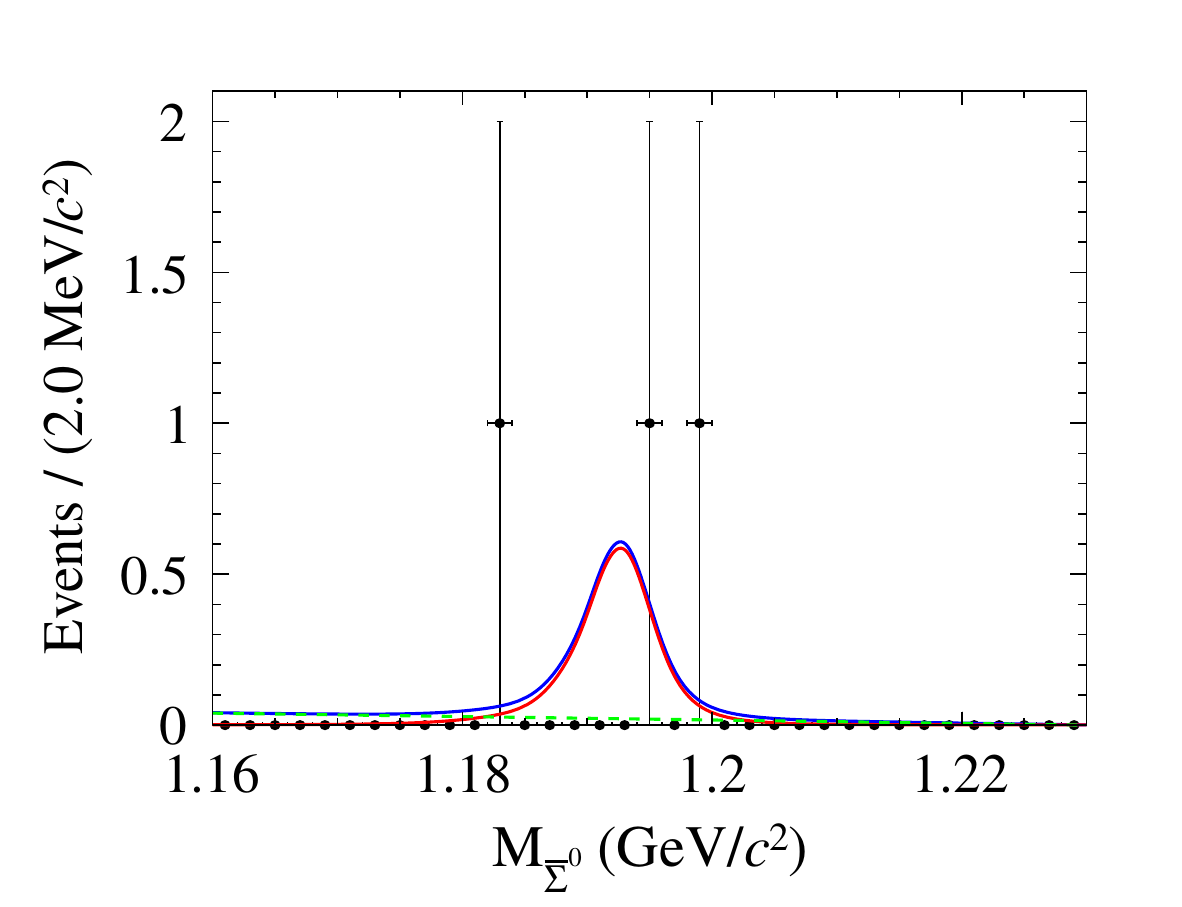}}
      \put(20,65){\color{blue}\scalebox{0.4}{$\sqrt{s}=4.387$ GeV}}
    \end{picture}
  \end{minipage}
  \begin{minipage}[b]{0.24\linewidth}
    \begin{picture}(50,80)
      \put(0,0){\includegraphics[width=\linewidth]{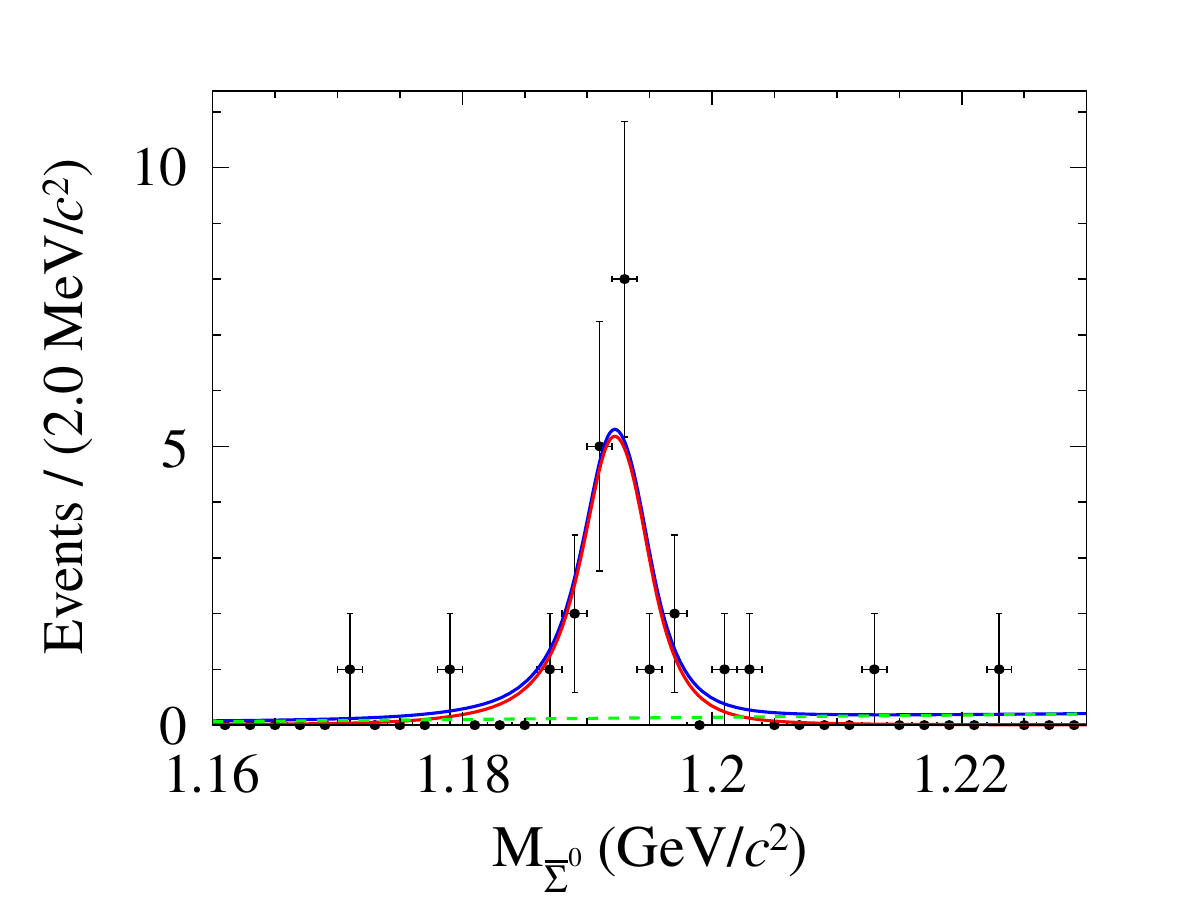}}
      \put(20,65){\color{blue}\scalebox{0.4}{$\sqrt{s}=4.396$ GeV}}
    \end{picture}
  \end{minipage}
  \begin{minipage}[b]{0.24\linewidth}
    \begin{picture}(50,80)
      \put(0,0){\includegraphics[width=\linewidth]{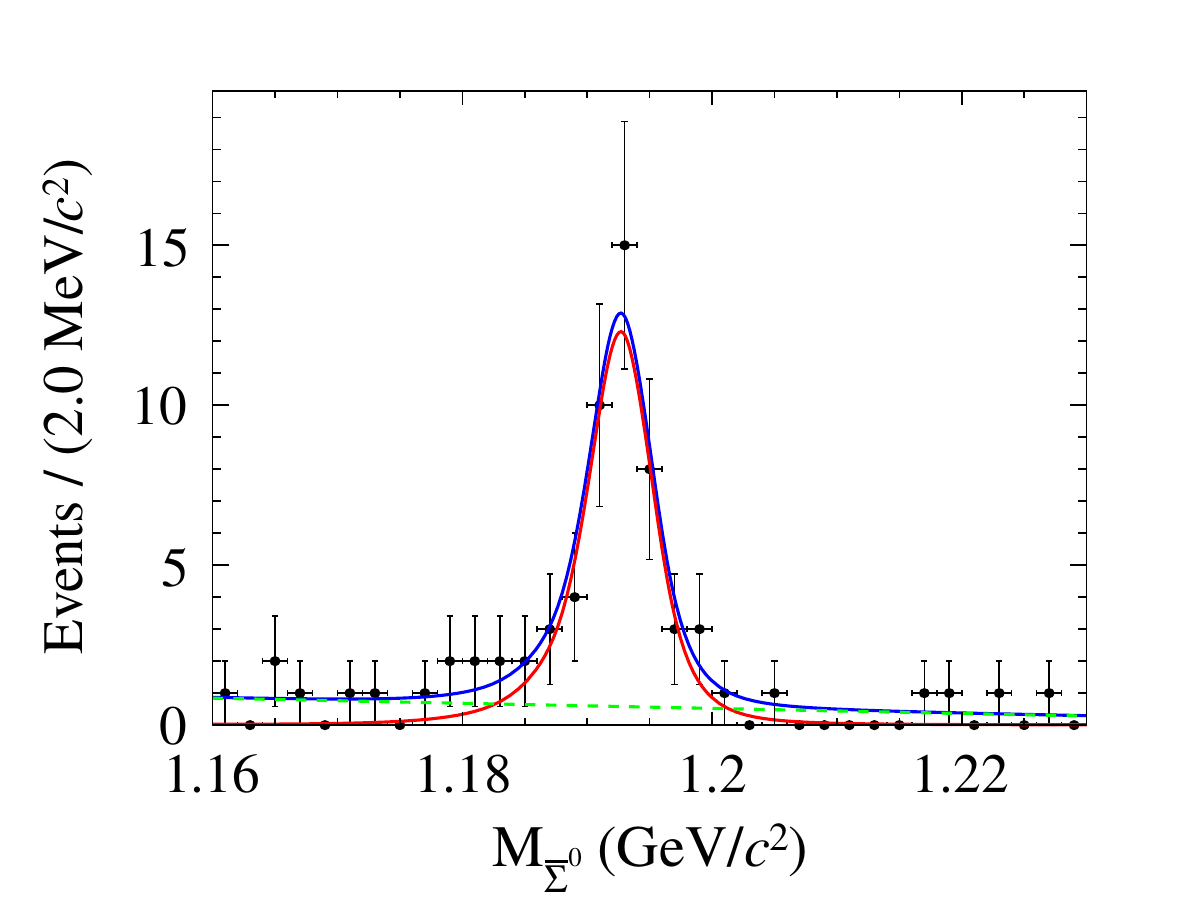}}
      \put(20,65){\color{blue}\scalebox{0.4}{$\sqrt{s}=4.416$ GeV}}
    \end{picture}
  \end{minipage}
  \begin{minipage}[b]{0.24\linewidth}
    \begin{picture}(50,80)
      \put(0,0){\includegraphics[width=\linewidth]{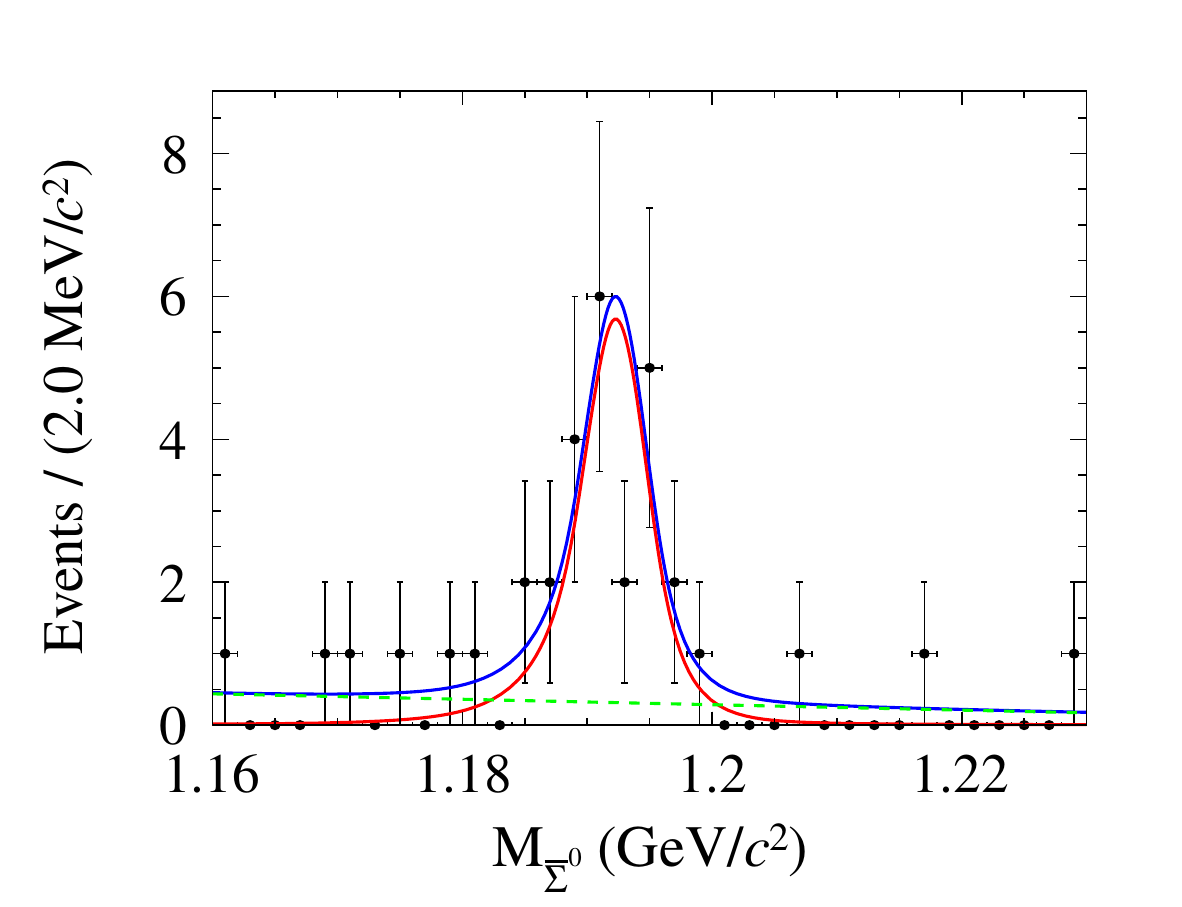}}
      \put(20,65){\color{blue}\scalebox{0.4}{$\sqrt{s}=4.436$ GeV}}
    \end{picture}
  \end{minipage}

  \begin{minipage}[b]{0.24\linewidth}
    \begin{picture}(50,80)
      \put(0,0){\includegraphics[width=\linewidth]{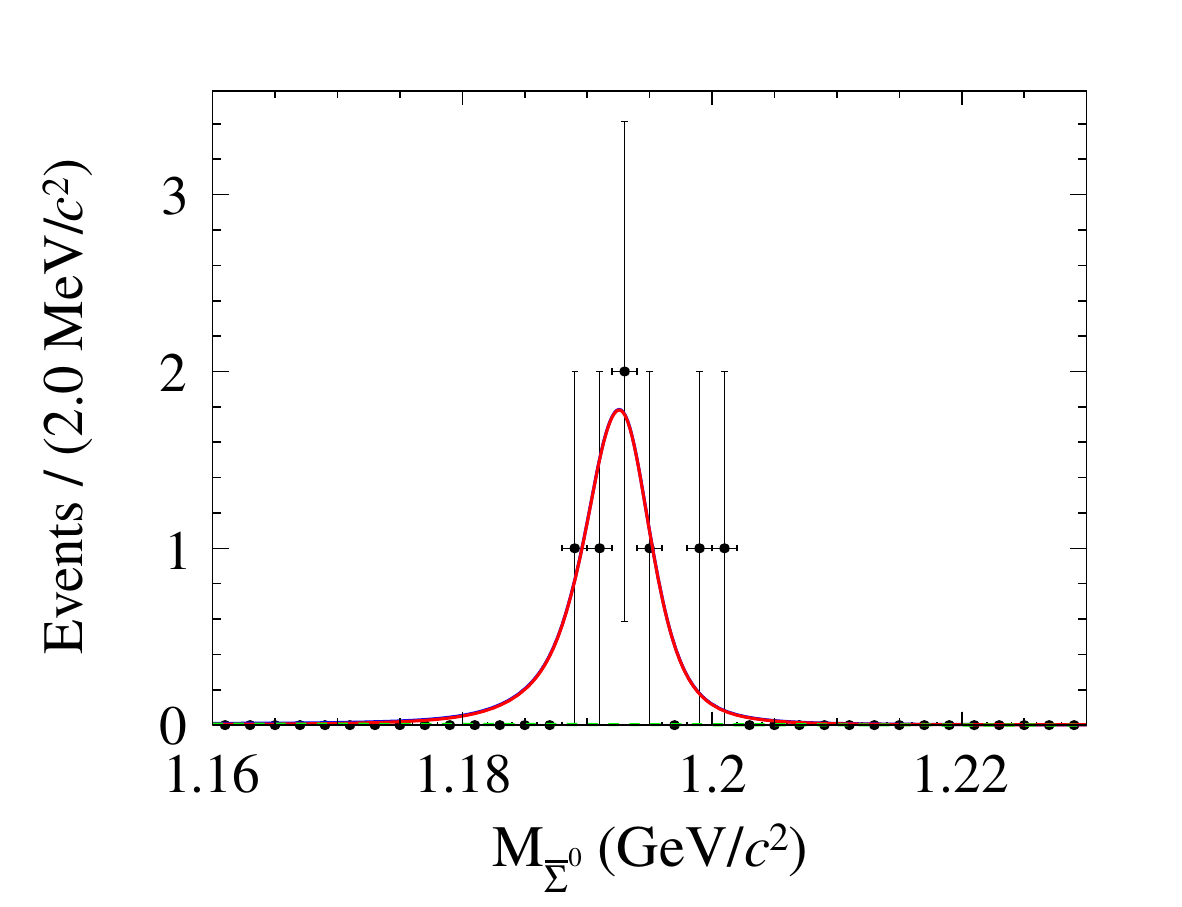}}
      \put(20,65){\color{blue}\scalebox{0.4}{$\sqrt{s}=4.467$ GeV}}
    \end{picture}
  \end{minipage}
  \begin{minipage}[b]{0.24\linewidth}
    \begin{picture}(50,80)
      \put(0,0){\includegraphics[width=\linewidth]{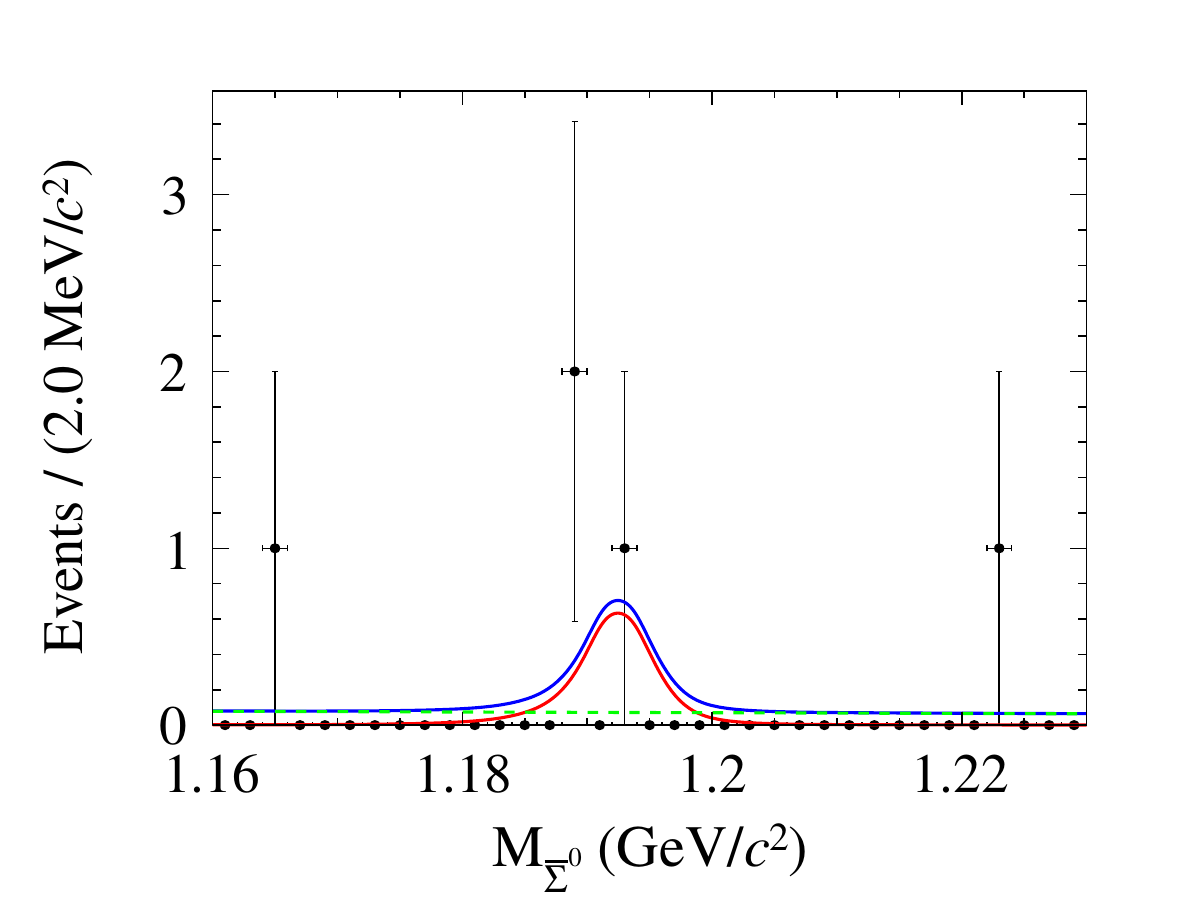}}
      \put(20,65){\color{blue}\scalebox{0.4}{$\sqrt{s}=4.527$ GeV}}
    \end{picture}
  \end{minipage}
  \begin{minipage}[b]{0.24\linewidth}
    \begin{picture}(50,80)
      \put(0,0){\includegraphics[width=\linewidth]{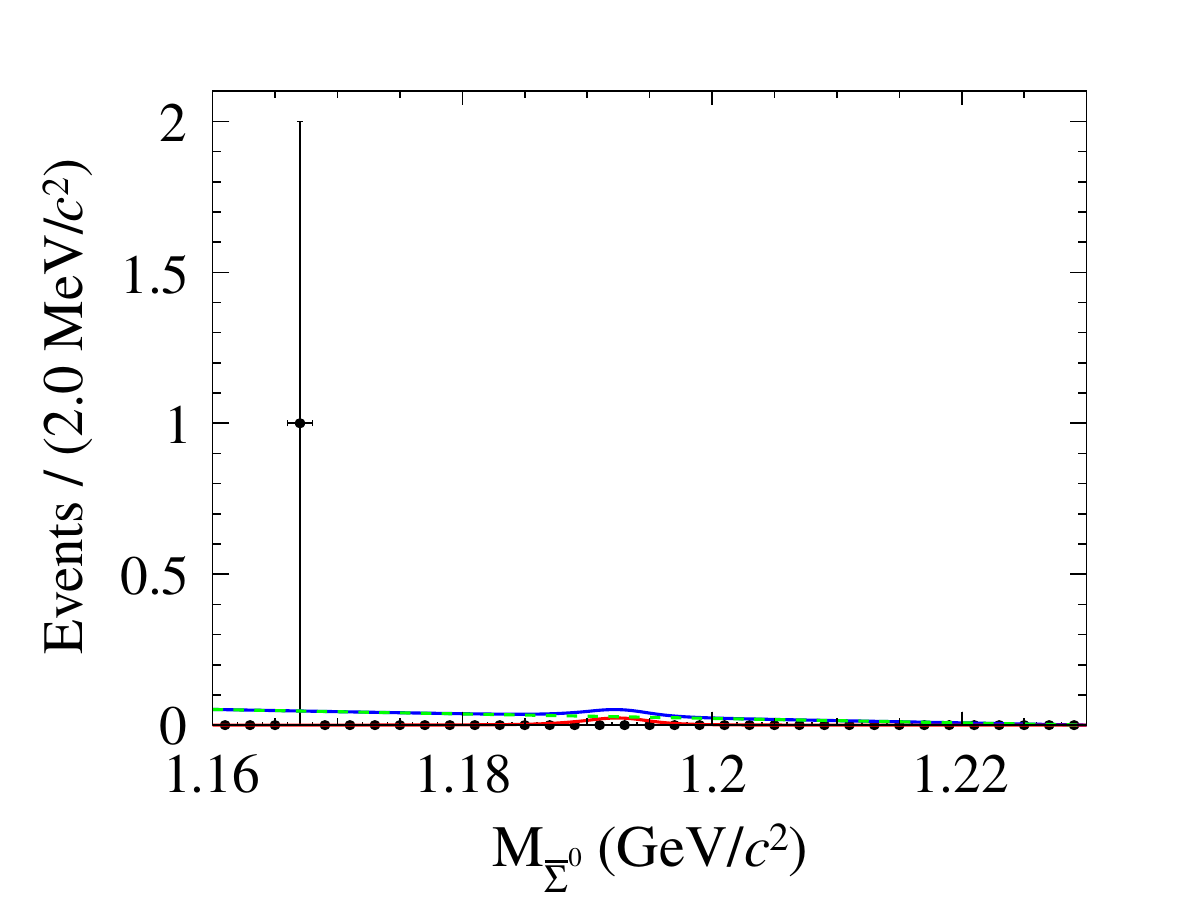}}
      \put(20,65){\color{blue}\scalebox{0.4}{$\sqrt{s}=4.574$ GeV}}
    \end{picture}
  \end{minipage}
  \begin{minipage}[b]{0.24\linewidth}
    \begin{picture}(50,80)
      \put(0,0){\includegraphics[width=\linewidth]{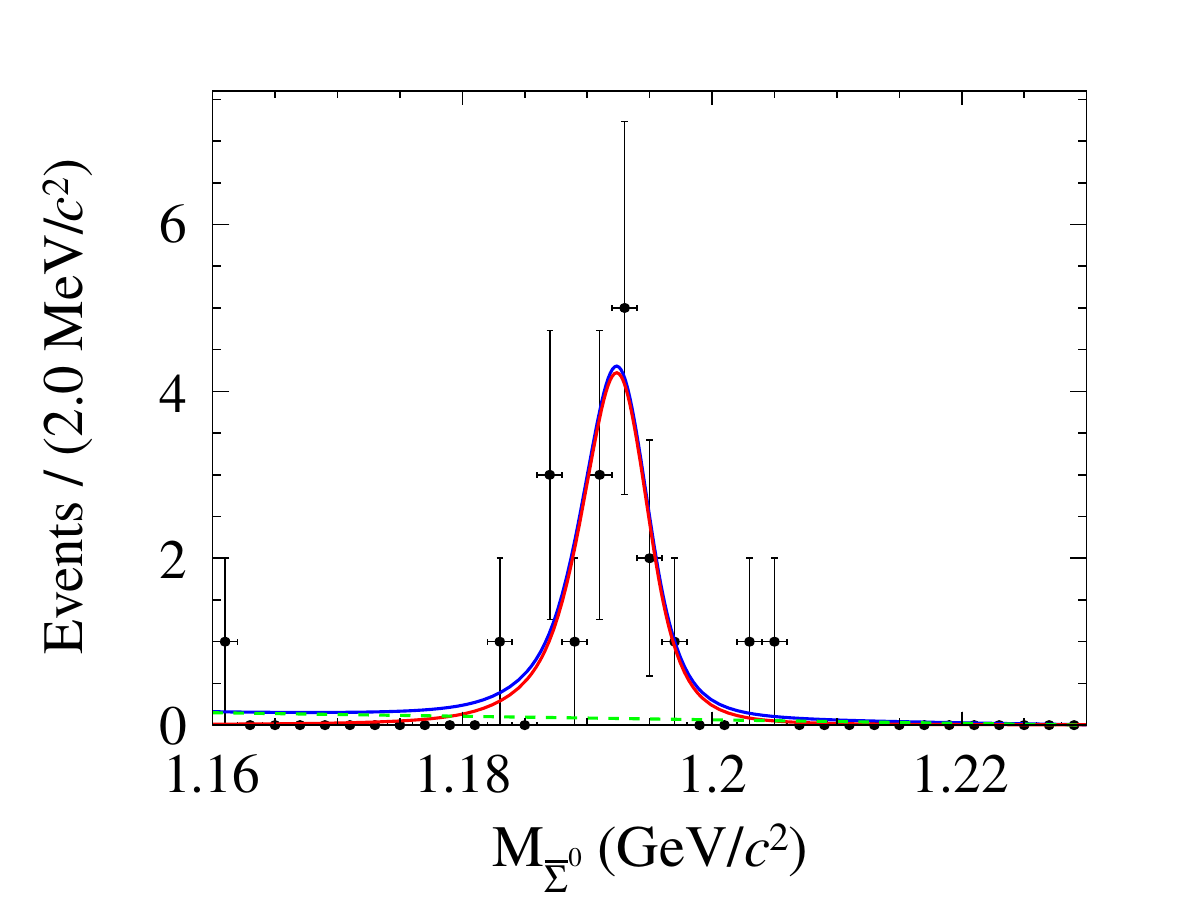}}
      \put(20,65){\color{blue}\scalebox{0.4}{$\sqrt{s}=4.600$ GeV}}
    \end{picture}
  \end{minipage}

  \begin{minipage}[b]{0.24\linewidth}
    \begin{picture}(50,80)
      \put(0,0){\includegraphics[width=\linewidth]{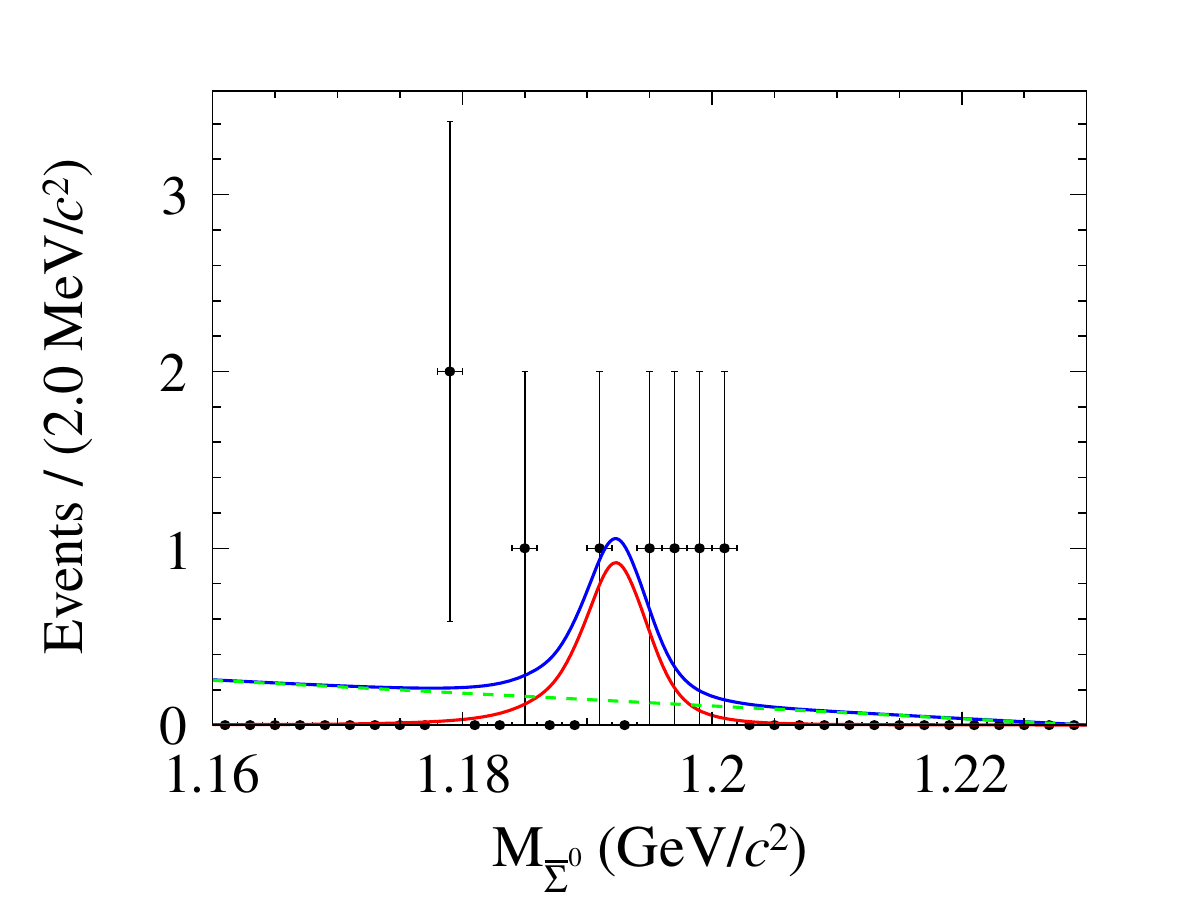}}
      \put(20,65){\color{blue}\scalebox{0.4}{$\sqrt{s}=4.612$ GeV}}
    \end{picture}
  \end{minipage}
  \begin{minipage}[b]{0.24\linewidth}
    \begin{picture}(50,80)
      \put(0,0){\includegraphics[width=\linewidth]{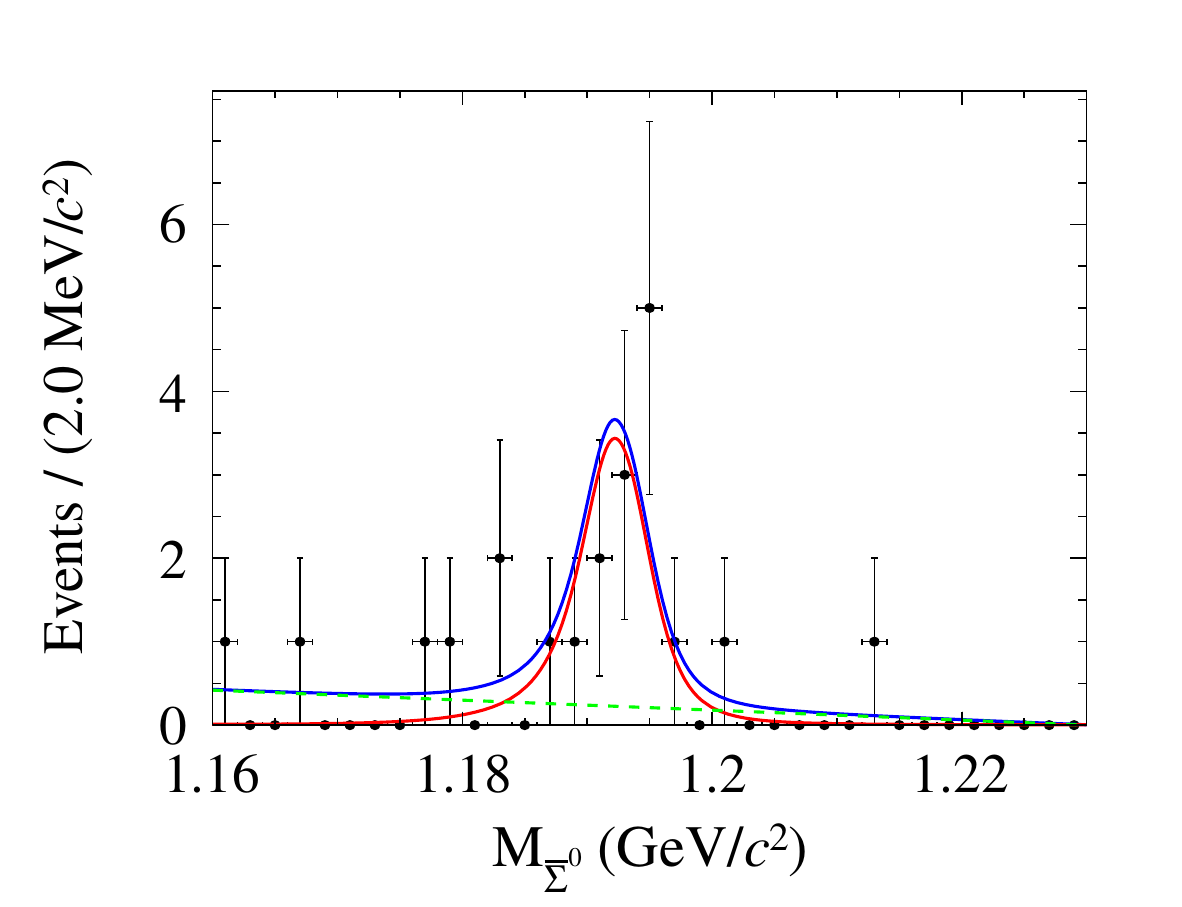}}
      \put(20,65){\color{blue}\scalebox{0.4}{$\sqrt{s}=4.628$ GeV}}
    \end{picture}
  \end{minipage}
  \begin{minipage}[b]{0.24\linewidth}
    \begin{picture}(50,80)
      \put(0,0){\includegraphics[width=\linewidth]{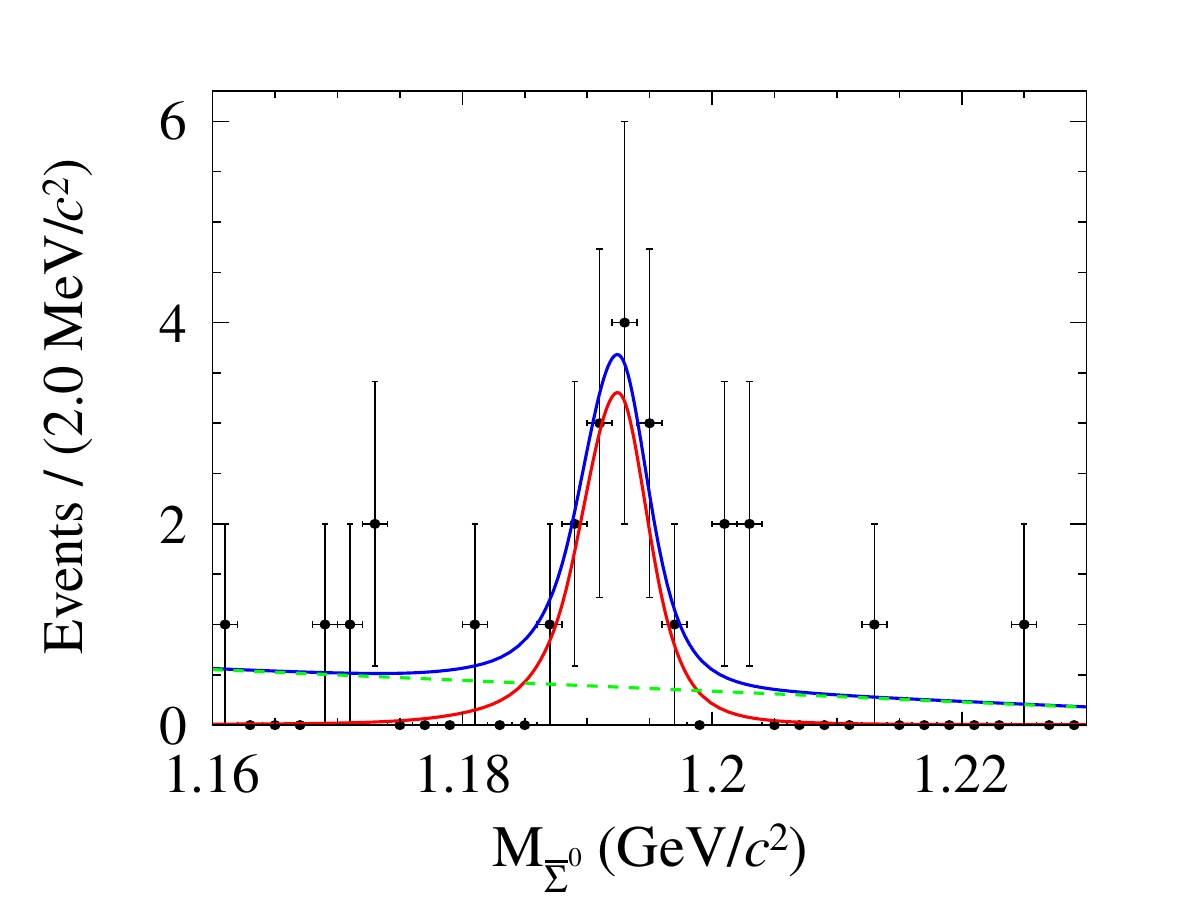}}
      \put(20,65){\color{blue}\scalebox{0.4}{$\sqrt{s}=4.641$ GeV}}
    \end{picture}
  \end{minipage}
  \begin{minipage}[b]{0.24\linewidth}
    \begin{picture}(50,80)
      \put(0,0){\includegraphics[width=\linewidth]{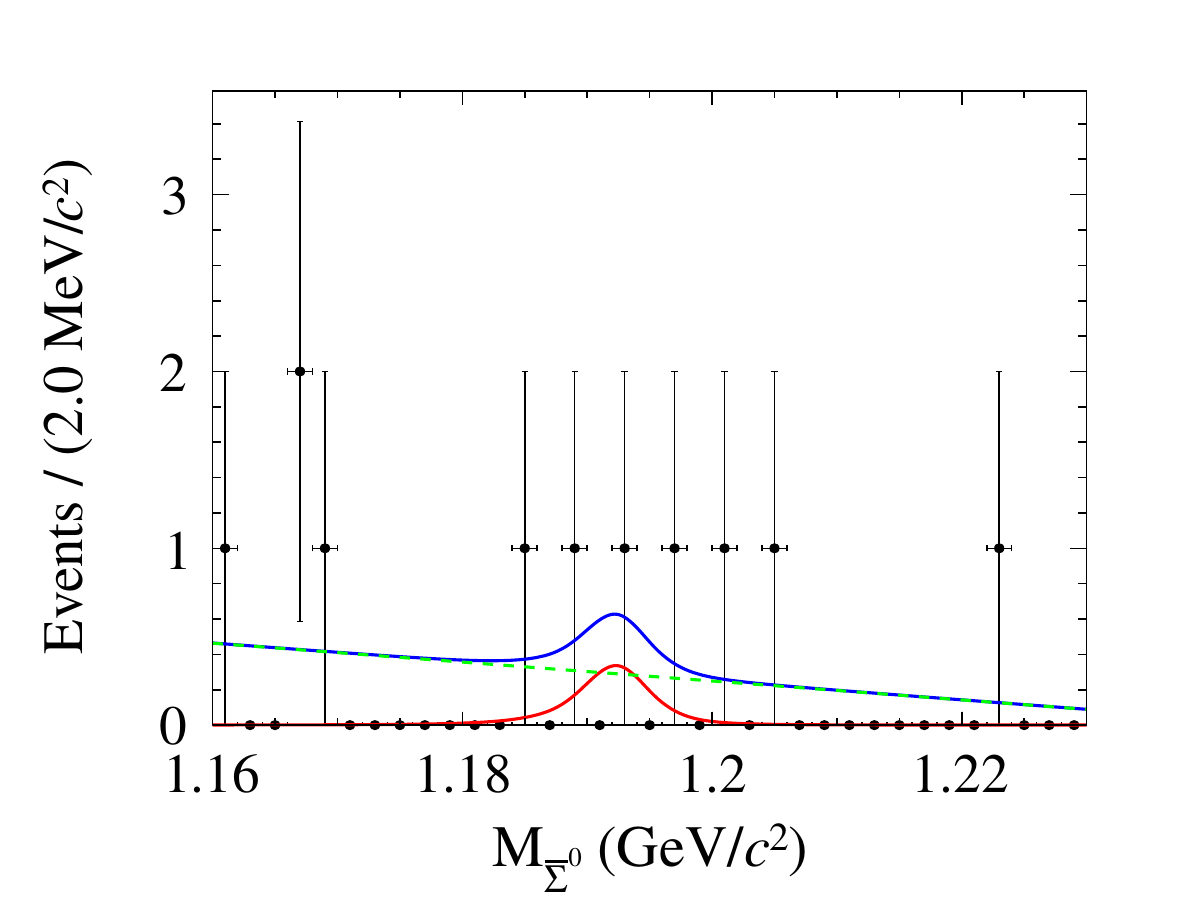}}
      \put(20,65){\color{blue}\scalebox{0.4}{$\sqrt{s}=4.661$ GeV}}
    \end{picture}
  \end{minipage}

  \begin{minipage}[b]{0.24\linewidth}
    \begin{picture}(50,80)
      \put(0,0){\includegraphics[width=\linewidth]{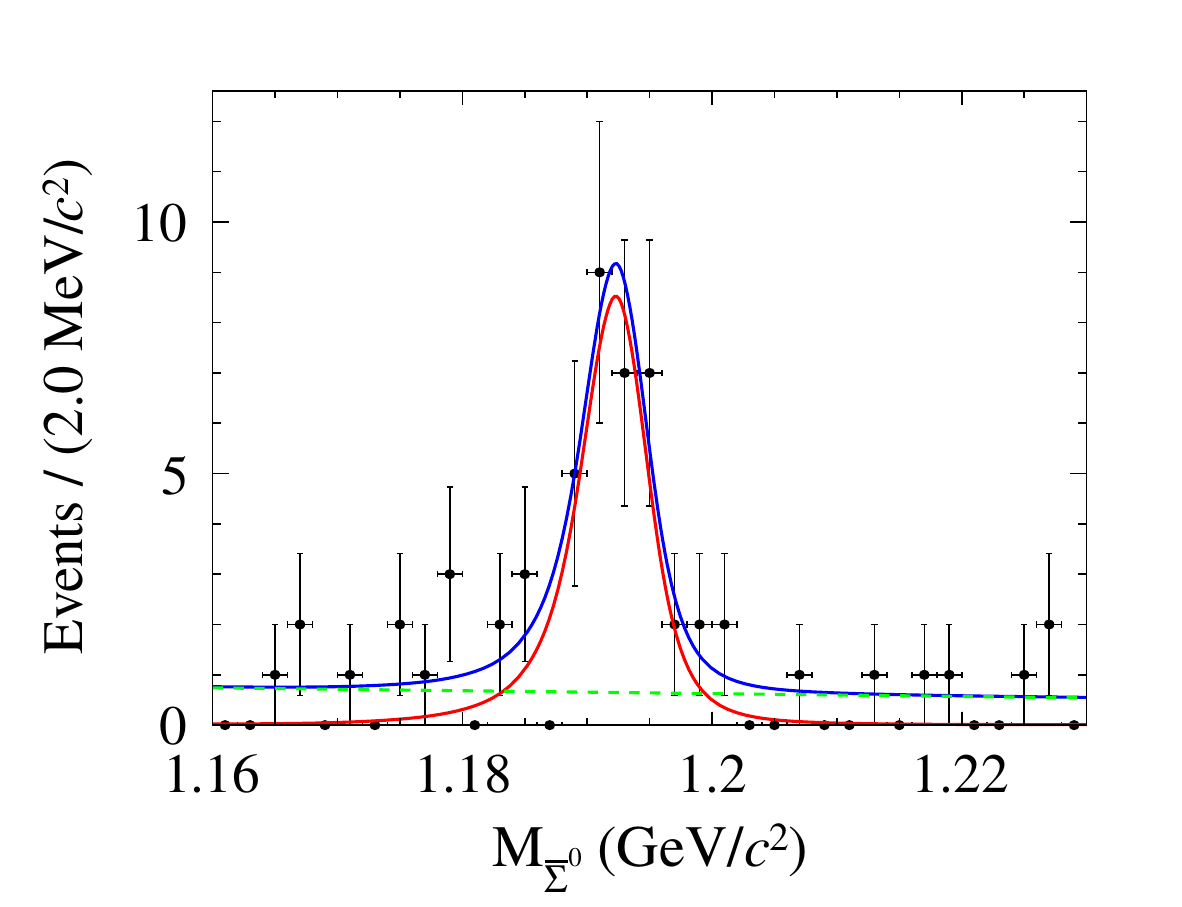}}
      \put(20,65){\color{blue}\scalebox{0.4}{$\sqrt{s}=4.682$ GeV}}
    \end{picture}
  \end{minipage}
  \begin{minipage}[b]{0.24\linewidth}
    \begin{picture}(50,80)
      \put(0,0){\includegraphics[width=\linewidth]{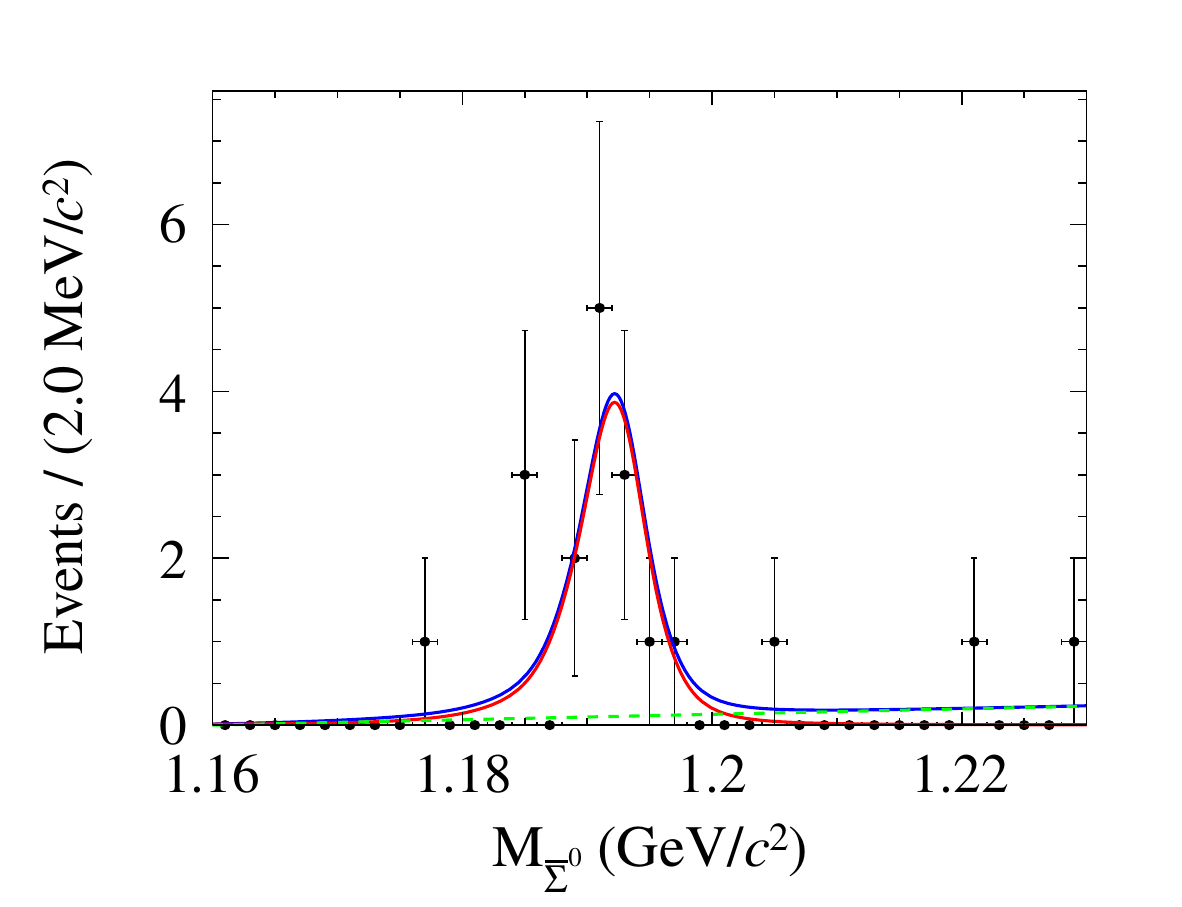}}
      \put(20,65){\color{blue}\scalebox{0.4}{$\sqrt{s}=4.699$ GeV}}
    \end{picture}
  \end{minipage}
  \begin{minipage}[b]{0.24\linewidth}
    \begin{picture}(50,80)
      \put(0,0){\includegraphics[width=\linewidth]{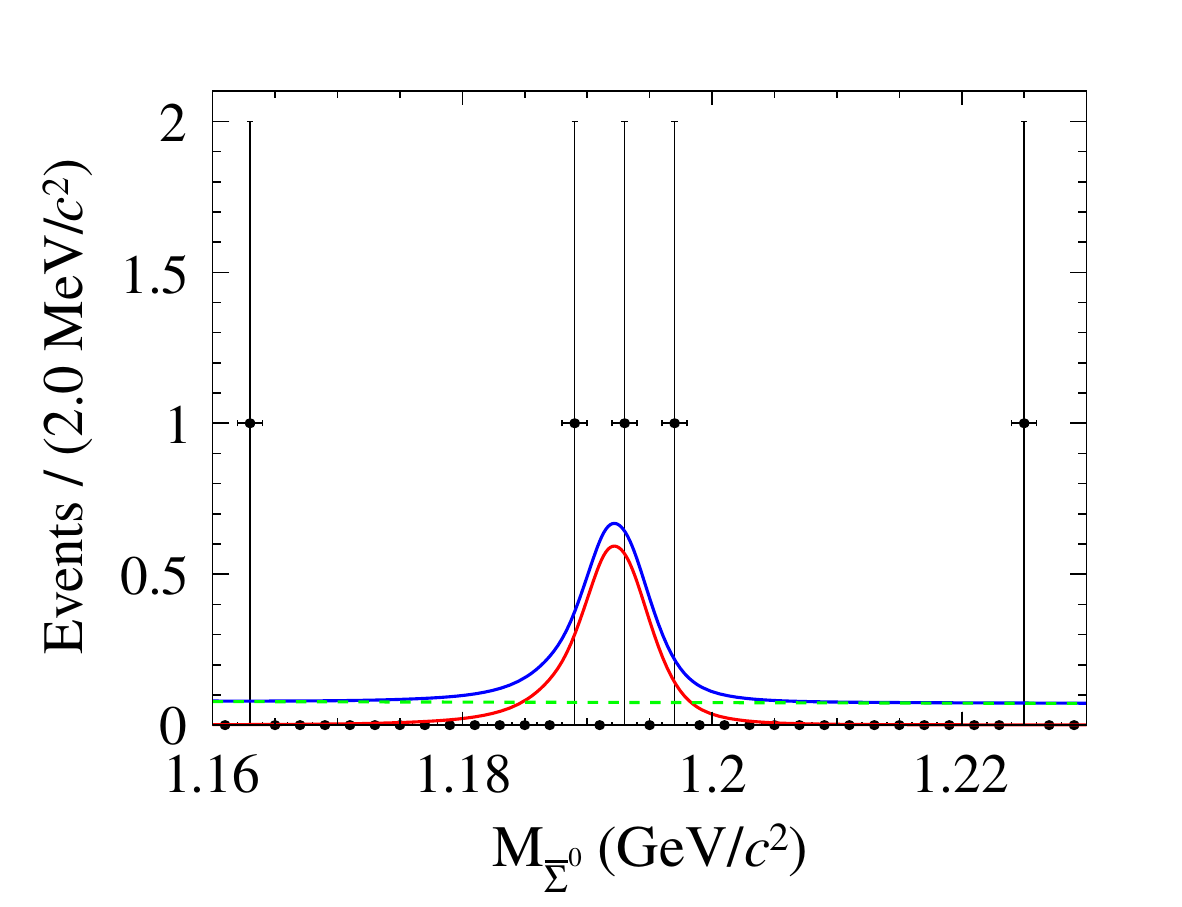}}
      \put(20,65){\color{blue}\scalebox{0.4}{$\sqrt{s}=4.740$ GeV}}
    \end{picture}
  \end{minipage}
  \begin{minipage}[b]{0.24\linewidth}
    \begin{picture}(50,80)
      \put(0,0){\includegraphics[width=\linewidth]{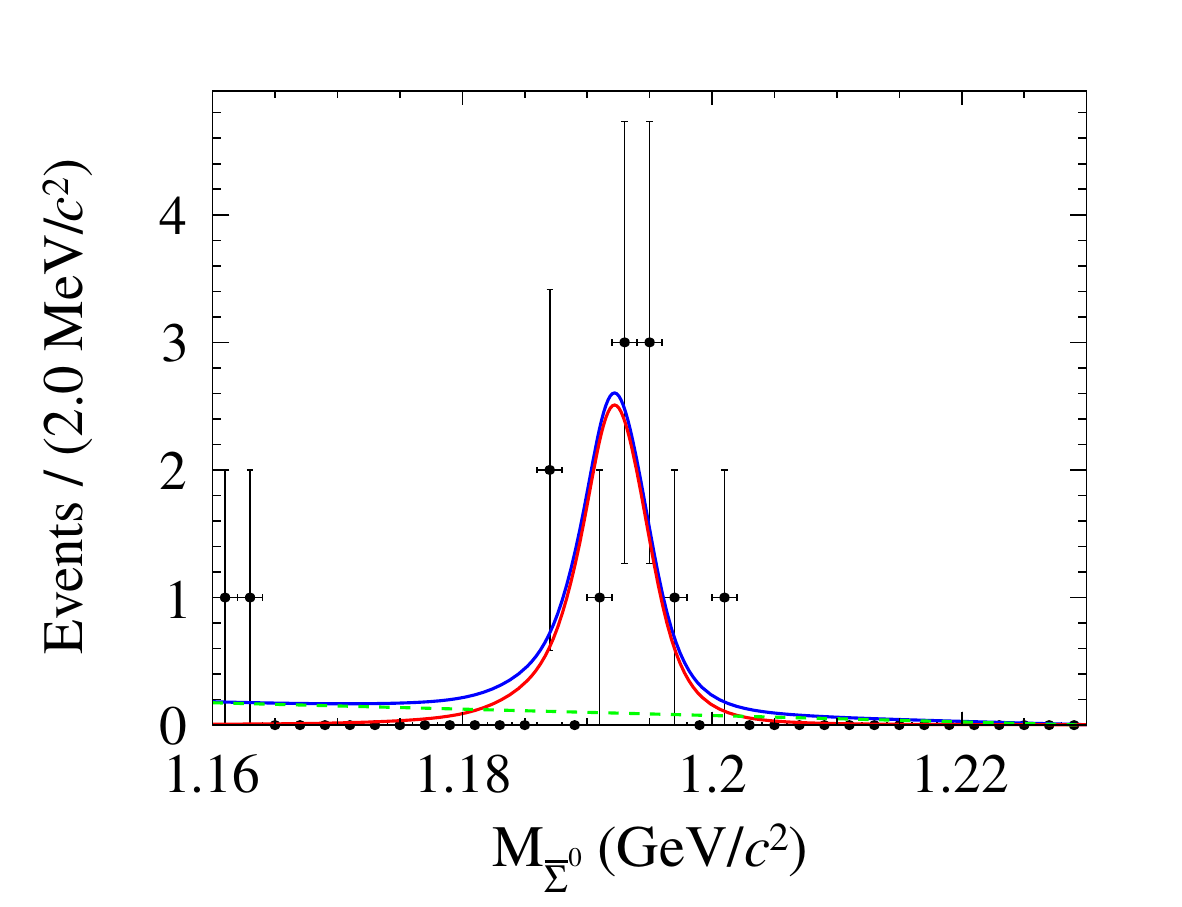}}
      \put(20,65){\color{blue}\scalebox{0.4}{$\sqrt{s}=4.750$ GeV}}
    \end{picture}
  \end{minipage}

  \begin{minipage}[b]{0.24\linewidth}
    \begin{picture}(50,80)
      \put(0,0){\includegraphics[width=\linewidth]{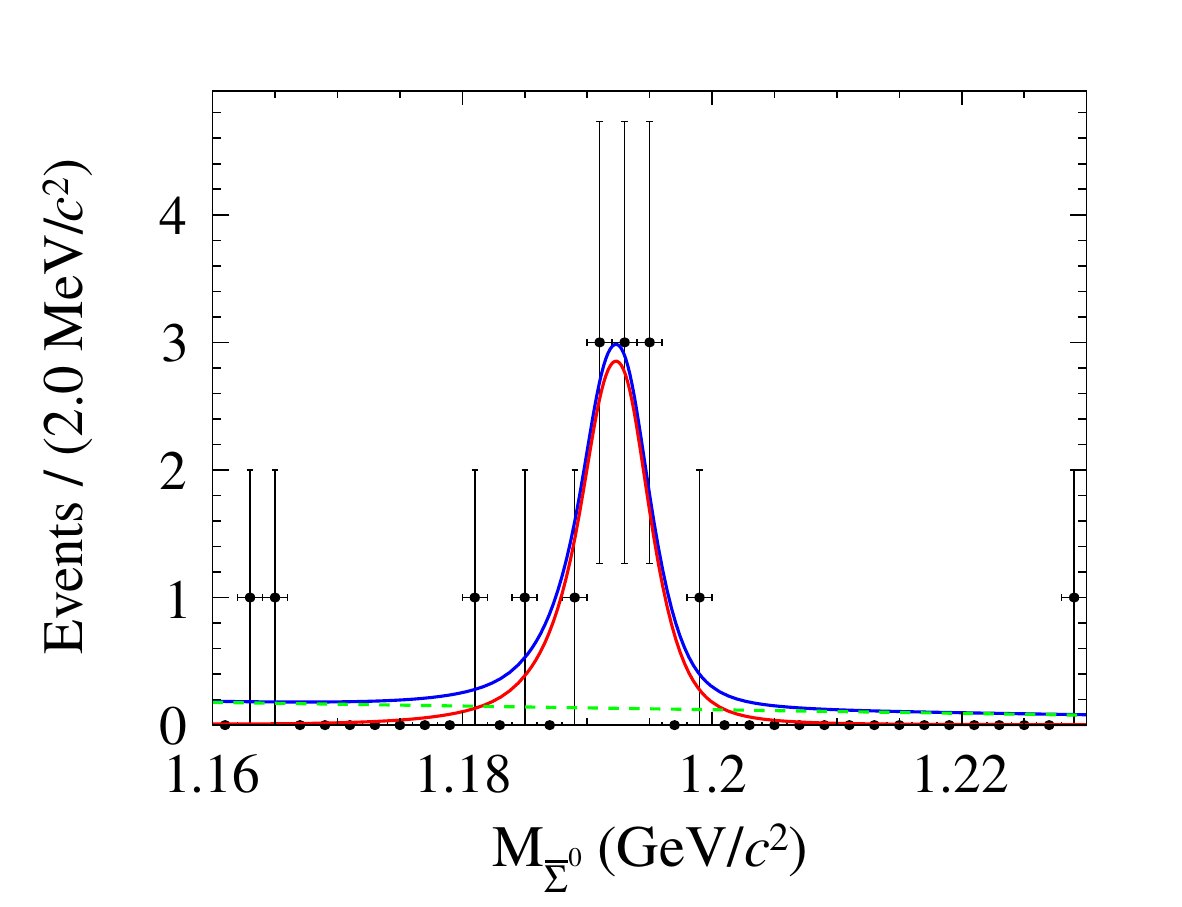}}
      \put(20,65){\color{blue}\scalebox{0.4}{$\sqrt{s}=4.781$ GeV}}
    \end{picture}
  \end{minipage}
  \begin{minipage}[b]{0.24\linewidth}
    \begin{picture}(50,80)
      \put(0,0){\includegraphics[width=\linewidth]{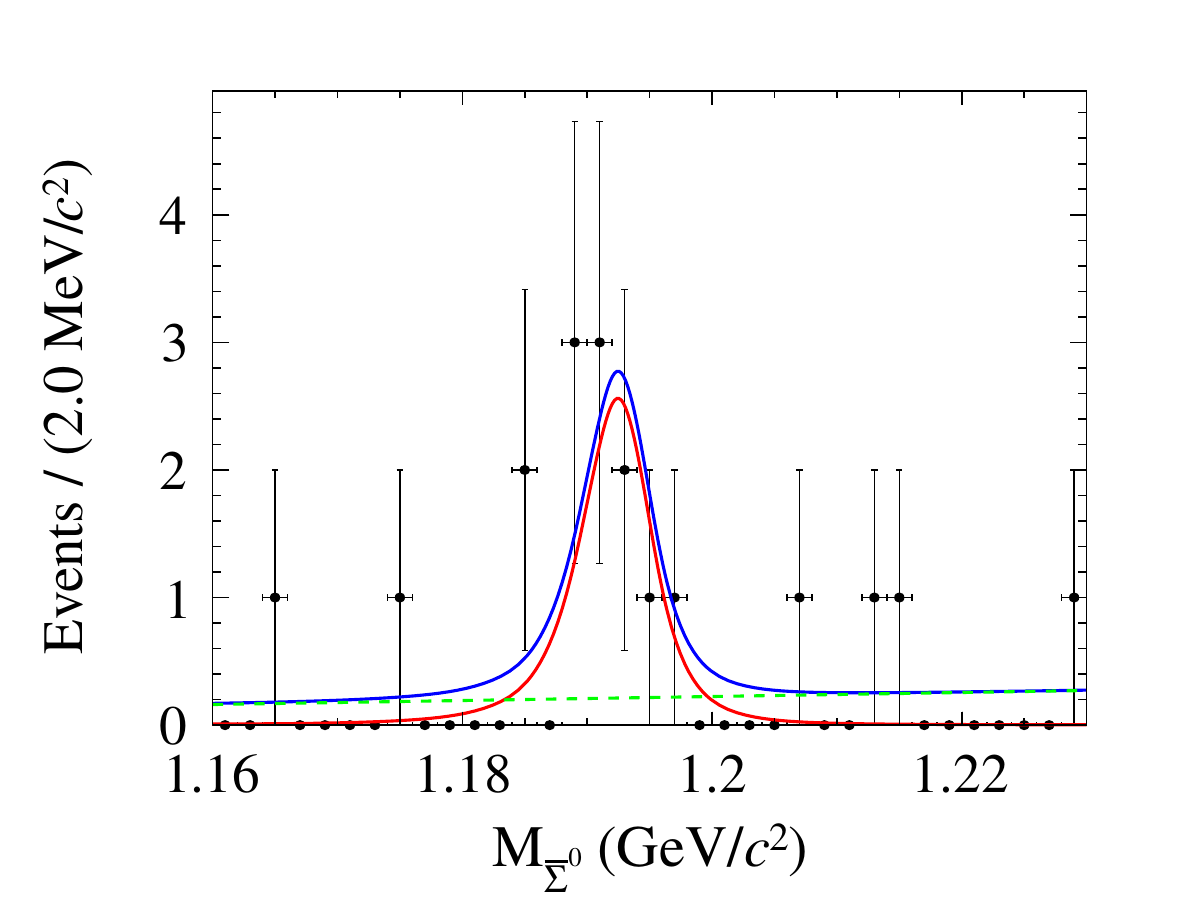}}
      \put(20,65){\color{blue}\scalebox{0.4}{$\sqrt{s}=4.843$ GeV}}
    \end{picture}
  \end{minipage}
  \begin{minipage}[b]{0.24\linewidth}
    \begin{picture}(50,80)
      \put(0,0){\includegraphics[width=\linewidth]{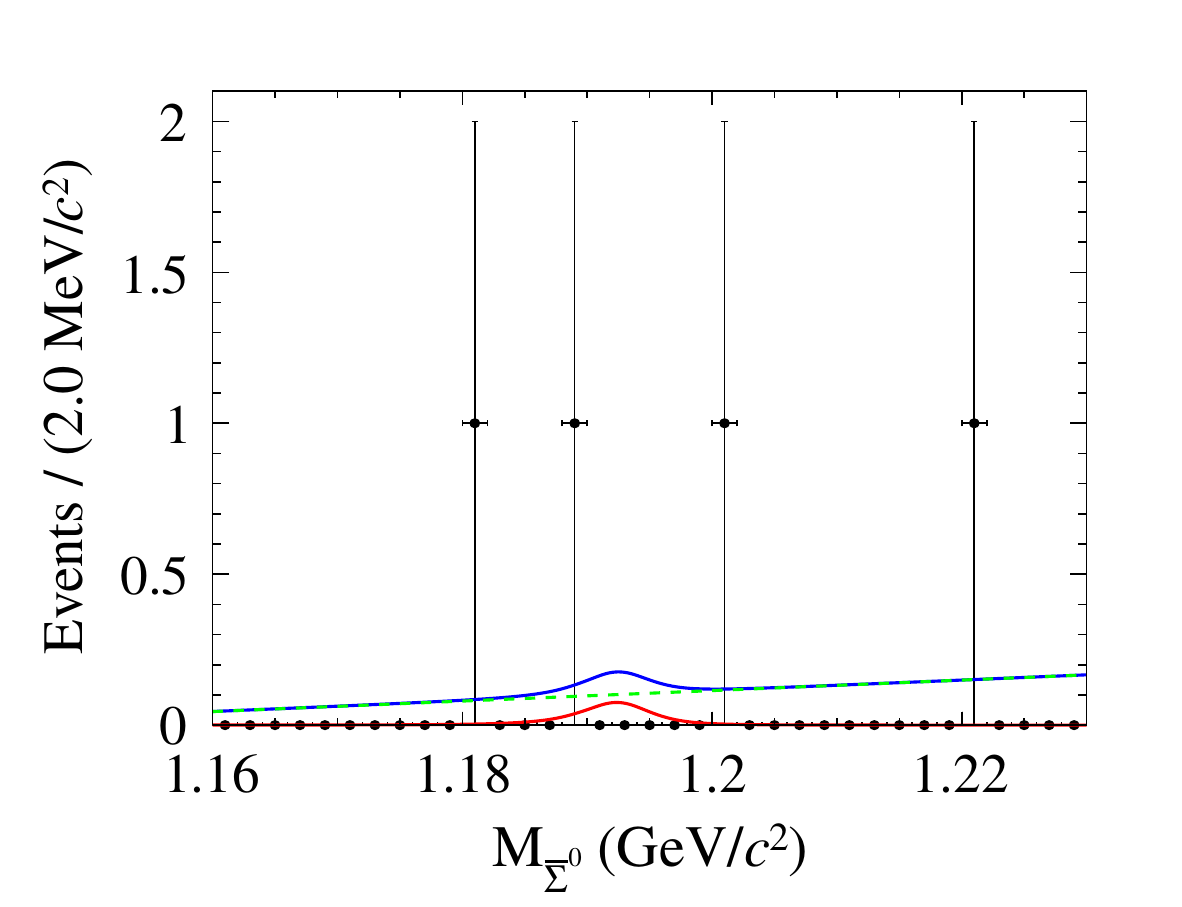}}
      \put(20,65){\color{blue}\scalebox{0.4}{$\sqrt{s}=4.918$ GeV}}
    \end{picture}
  \end{minipage}
  \begin{minipage}[b]{0.24\linewidth}
    \begin{picture}(50,80)
      \put(0,0){\includegraphics[width=\linewidth]{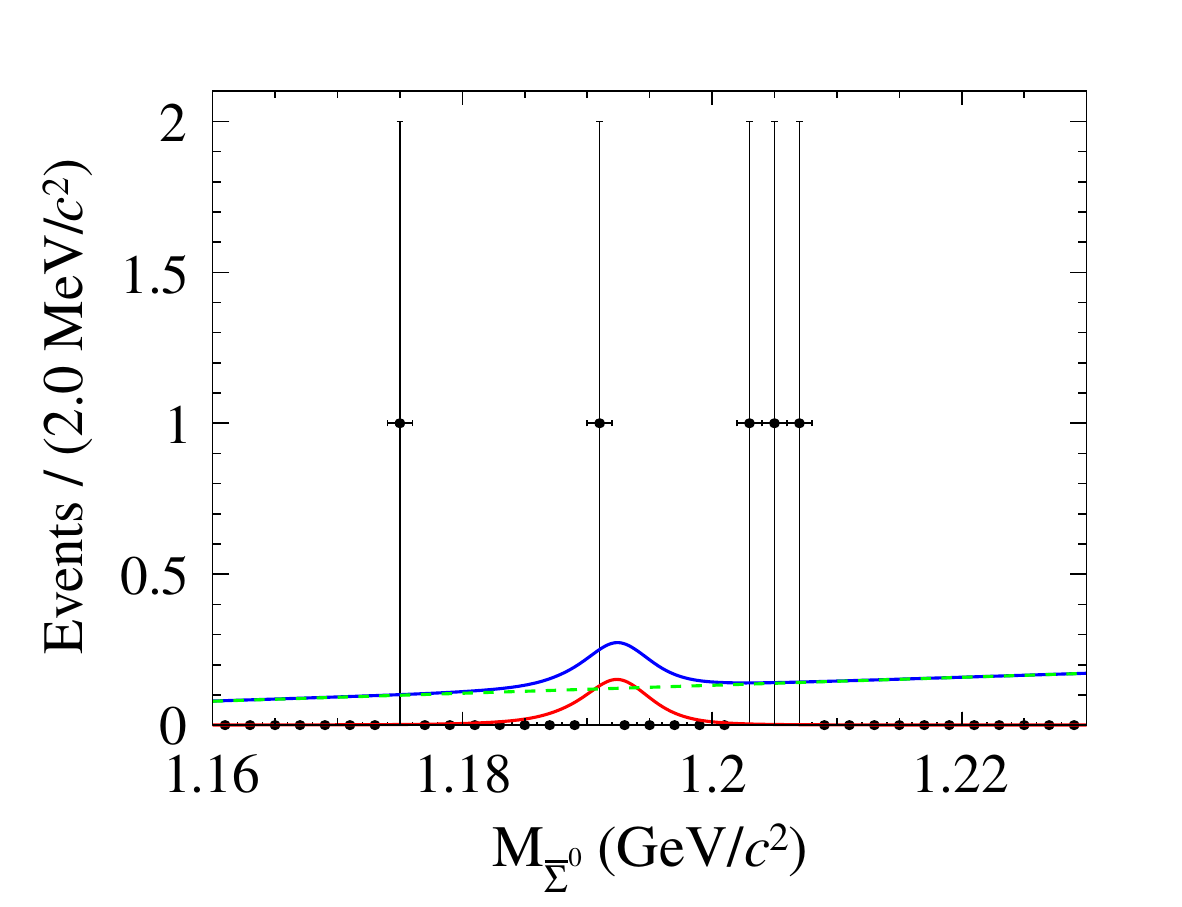}}
      \put(20,65){\color{blue}\scalebox{0.4}{$\sqrt{s}=4.951$ GeV}}
    \end{picture}
  \end{minipage}
  \caption{Fits to the $M_{\bar{\Sigma}^{0}}$ distributions at $\sqrt{s}$ between 4.27 and 4.95 GeV. The dots with error bars represent the data, the blue solid lines show the fit results, the red solid lines correspond to the signal shapes, and the green dashed lines indicate the background contributions.}
  \label{fitMsigma2}
\end{figure}

\noindent The signal yields ($N_{\rm obs}$) for the $e^{+}e^{-} \to pK^{-}\bar{\Sigma}^{0}$ + c.c. process at each energy point are determined by performing an unbinned extended maximum-likelihood fit to the $M_{\bar{\Sigma}^{0}}$ spectrum in the mass range from 1.16 to 1.23 GeV/$c^2$, as shown in Figs.~\ref{fitMsigma1} and~\ref{fitMsigma2}. In this fit, the signal shape is taken from the MC simulation convolved with a Gaussian function, accounting for the mass resolution difference between data and MC simulation. A first or second-order polynomial function is utilized to model the background. However, the significance at some of the energy points is less than $3.0\sigma$, as assessed by comparing the change in likelihood $L_{0}(N_{\rm obs})$ with and without the signal contribution in the fits. 
The upper limits of signal yields at the 90\% C.L. for these energy points are determined with the Bayesian method~\cite{Zhu:2008ca}. 
The additive systematic uncertainties associated with the fitting procedure are incorporated by convolving $L_{0}(N_{\rm obs})$ with a Gaussian function, whose width corresponds to the total additive uncertainty on $N_{\rm obs}$. 
The combined likelihood $L(N_{\rm obs})$ accounts for both statistical fluctuations and additive systematic effects.
The upper limit on the signal yield ($N_\mathrm{UL}$) at the 90\% C.L. is determined from the condition
$\int_{0}^{N_\mathrm{UL}} L {\rm{d}}N_\mathrm{obs}/\int_{0}^{\infty} L {\rm{d}}N_\mathrm{obs} = 0.9$. To incorporate the multiplicative systematic uncertainties, the probability distribution of the signal yield obtained from the maximum-likelihood fit is convolved with a Gaussian function whose mean is equal to the nominal detection efficiency and whose width is $\sigma_{\epsilon}$, where $\sigma_{\epsilon}$ is the product of the detection efficiency and the multiplicative systematic uncertainties in percent.


\subsection{Determination of the Born cross section} 
\noindent 
The Born cross section~($\sigma^B$) for $e^{+}e^{-} \to pK^{-}\bar{\Sigma}^{0}$ + c.c. at a given CM energy is calculated by
\begin{linenomath*}
\begin{equation}
\sigma^{B} = \frac{N_{\rm{obs}}}{\mathcal{L} \cdot \frac{1}{|1-\Pi|^{2}} \cdot (1 + \delta) \cdot \epsilon \cdot \mathcal{B}(\bar{\Sigma}^{0} \to \gamma\bar{\Lambda}) \cdot  \mathcal{B}(\bar{\Lambda} \to \bar{p}\pi^+)},
\end{equation}
\end{linenomath*}
where $N_{\rm{obs}}$ is the number of observed signal events with the inclusion of charge conjugate modes, $\mathcal{L}$ corresponds to the integrated luminosity, $\frac{1}{|1-\Pi|^2}$ is the vacuum polarization (VP) correction factor, $(1 + \delta)$ is the ISR correction factor, $\epsilon$ is the detection efficiency averaged over the two charge-conjugate modes, and $\mathcal{B}(\bar{\Sigma}^{0} \to \gamma\bar{\Lambda})$ and $\mathcal{B}(\bar{\Lambda} \to \bar{p}\pi^+)$ are the branching fractions taken from the PDG~\cite{PDG}. The VP correction factor is calculated with the Fortran package provided by Fred Jegerlehner~\cite{Jegerlehner:2011ti}. Note that the efficiencies and ISR correction factors are determined through an iterative procedure to accurately measure the Born cross section as proposed in Ref.~\cite{Sun:2020ehv}. 

Initially, the cross section is measured without any ISR correction. Using this initial line shape, signal MC samples are regenerated to obtain revised values of the efficiencies and ISR correction factors, and the Born cross sections are updated subsequently. 
The Born cross sections are calculated via an iterative method, which is considered to have converged when the $(1+\delta) \cdot \epsilon$ difference between the last two iterations is less than 0.1\%.
The values of the efficiency, ISR correction factor, and Born cross section are obtained through this iterative process. The numerical results for each CM energy point are summarized in table~\ref{bcs1}.
\setlength{\tabcolsep}{9pt}
\begin{longtable}[l]{>{\footnotesize}c >{\footnotesize}c >{\footnotesize}c  >{\footnotesize}c >{\footnotesize}c >{\footnotesize}c >{\footnotesize}c >{\footnotesize}c}
\caption{The Born cross sections~($\sigma^B$) for $e^{+}e^{-} \to pK^{-}\bar{\Sigma}^{0}$ + c.c. at 56 energy points between 3.51 and 4.95~GeV, where the first uncertainties are statistical and the second are systematic; $\sqrt{s}$ is the $e^{+}e^{-}$ CM energy, $\mathcal{L}$ is the integrated luminosity of each dataset~\cite{Ablikim:2013ntc,BESIII:2015qfd,BESIII:2022dxl,BESIII:2022ulv,BESIII:2024lks}, $\frac{1}{|1-\Pi|^{2}}$ is the VP correction factor, $(1+\delta)$ is the ISR correction factor$, \epsilon$ is the detection efficiency; $N_{\text{obs}}$ is the signal yield (the upper limits of signal yields including systematic uncertainty at the 90\%~C.L. at these energy points); $\mathcal{S}$ is the statistical significance. 
}
\label{bcs1}
\endfirsthead
\multicolumn{8}{c}{} \\
\hline \hline
$\sqrt{s}$ (GeV)  & $\mathcal{L}$~($\rm{pb}^{-1}$) & $\frac{1}{{\left|1-\Pi \right|}^{2}}$ & $(1+\delta)$&  $\epsilon$ & $N_{\text{obs}}~(N_{\text{UL}})$ & $\sigma^{B}(\sigma^{B}_\text{UL})~(\rm {fb})$ & $\mathcal{S}$~$(\sigma)$  \\
\hline
\endhead
\hline
\endfoot
\bottomrule
\endlastfoot
\hline \hline
$\sqrt{s}$ (GeV)  & $\mathcal{L}$~($\rm{pb}^{-1}$) & $\frac{1}{{\left|1-\Pi \right|}^{2}}$ & $(1+\delta)$&  $\epsilon$ & $N_{\text{obs}}~(N_{\text{UL}})$ & $\sigma^{B}(\sigma^{B}_\text{UL})~(\rm {fb})$ & $\mathcal{S}$~$(\sigma)$  \\
\hline
3.508 &  404.7 &  1.04  &  0.97  & 0.17  & $126.9^{+12.7}_{-11.9}$    &     $2798.4^{+280.6}_{-264.0} \pm 95.1$   &12.8      \\
3.514 &  40.9 &  1.04  &  0.98  & 0.17  & $8.2^{+3.7}_{-3.0}$    &     $1752.6^{+796.9}_{-641.3} \pm 59.6$   &1.3            \\
3.553 &  42.2 &  1.04  &  1.09  & 0.17  & $8.7^{+3.4}_{-2.7}$    &     $1663.5^{+643.5}_{-515.5} \pm 56.6$   &1.2            \\
3.554 &  129.4 &  1.04  &  1.10  & 0.18  & $29.7^{+6.5}_{-5.8}$    &     $1731.9^{+377.2}_{-337.1} \pm 58.9$   &5.8          \\
3.582 &  85.7 &  1.04  &  1.11  & 0.16  & $14.7^{+5.0}_{-4.4}$    &     $1423.3^{+484.2}_{-430.6} \pm 48.4$   &3.7           \\
3.650 &  410.0 &  1.02  &  1.15  & 0.18  & $78.3^{+9.6}_{-8.9}$    &     $1446.1^{+177.8}_{-165.3} \pm 49.2$   &9.1          \\
3.670 &  83.6 &  0.99  &  1.17  & 0.16  & $13.1^{+4.4}_{-3.8}$    &     $1319.3^{+446.8}_{-378.6} \pm 44.9$   &3.9           \\
3.768 &  415.8 &  1.05  &  1.24  & 0.16  & $42.8^{+7.7}_{-7.0}$    &     $756.6^{+136.5}_{-124.7} \pm 25.7$   &6.3           \\
3.773 &  20274.8 &  1.06  &  1.25  & 0.16  & $3117.0^{+64.6}_{-63.6}$    &     $1147.9^{+25.4}_{-25.1} \pm 39.0$   &30     \\
3.780 &  410.0 &  1.06  &  1.25  & 0.16  & $63.6^{+9.2}_{-8.5}$    &     $1130.1^{+163.7}_{-151.4} \pm 38.4$   &8.7          \\
3.808 &  50.5 &  1.06  &  1.15  & 0.16  & $7.2^{+3.6}_{-2.8}$    &     $1130.5^{+568.5}_{-443.4} \pm 38.4$   &2.7            \\
3.867 &  108.9 &  1.05  &  1.20  & 0.15  & $13.1^{+4.7}_{-4.0}$    &     $968.8^{+345.8}_{-296.5} \pm 32.9$   &3.9           \\
3.871 &  110.3 &  1.05  &  1.20  & 0.16  & $10.1^{+3.7}_{-3.1}$    &     $719.7^{+265.6}_{-217.5} \pm 24.5$   &3.2           \\
3.896 &  52.6 &  1.05  &  1.22  & 0.16  & $1.9^{+1.8}_{-1.1}$    &     $272.2^{+259.4}_{-163.6} \pm 9.3$   &0.8              \\
4.008 &  482.0 &  1.04  &  1.24  & 0.15  & $35.3^{+7.4}_{-6.6}$    &     $622.0^{+129.8}_{-116.0} \pm 21.1$   &5.7           \\
4.085 &  52.9 &  1.05  &  1.44  & 0.13  & $2.7^{+2.1}_{-1.5}$    &     $406.0^{+312.4}_{-214.8} \pm 12.6$   &1.7             \\
4.128 &  401.5 &  1.05  &  1.67  & 0.12  & $31.3^{+6.4}_{-5.7}$    &     $566.6^{+115.3}_{-103.0} \pm 17.6$   &6.0           \\
4.157 &  408.7 &  1.05  &  1.73  & 0.12  & $22.8^{+5.9}_{-5.3}$    &     $399.3^{+103.3}_{-92.1} \pm 12.4$   &4.5            \\
4.178 &  3194.5 &  1.05  &  1.56  & 0.12  & $224.3^{+16.8}_{-16.1}$    &     $540.0^{+40.7}_{-39.1} \pm 16.7$   &15.7        \\
4.189 &  526.7 &  1.06  &  1.59  & 0.12  & $35.0^{+6.8}_{-6.1}$    &     $503.4^{+98.2}_{-87.9} \pm 15.6$   &7.1             \\
4.199 &  526.0 &  1.06  &  1.59  & 0.13  & $31.1^{+6.9}_{-6.2}$    &     $438.0^{+97.8}_{-87.9} \pm 13.6$   &5.5             \\
4.208 &  572.0 &  1.06  &  1.60  & 0.13  & $27.2^{+6.4}_{-5.7}$    &     $335.4^{+78.8}_{-70.2} \pm 10.4$   &6.0             \\
4.217 &  569.2 &  1.06  &  1.61  & 0.12  & $26.9^{+6.2}_{-5.6}$    &     $351.0^{+81.4}_{-72.4} \pm 10.9$   &5.5             \\
4.226 &  1100.9 &  1.06  &  1.62  & 0.12  & $78.1^{+10.0}_{-9.2}$    &     $538.1^{+68.7}_{-63.8} \pm 16.7$   &10.2          \\
4.236 &  530.3 &  1.06  &  1.63  & 0.12  & $27.2^{+6.2}_{-5.5}$    &     $382.0^{+87.1}_{-77.6} \pm 11.8$   &5.0             \\
4.242 &  55.9 &  1.06  &  1.65  & 0.12  & $5.5^{+2.8}_{-2.2}$    &     $721.1^{+371.9}_{-286.8} \pm 22.4$   &0.8             \\
4.244 &  538.1 &  1.06  &  1.64  & 0.12  & $35.6^{+7.0}_{-6.3}$    &     $501.6^{+98.9}_{-89.3} \pm 15.6$   &6.1             \\
4.258 &  828.4 &  1.05  &  1.70  & 0.11  & $34.0^{+7.0}_{-6.3}$    &     $313.9^{+65.0}_{-58.5} \pm 9.7$   &5.6              \\
4.267 &  531.1 &  1.05  &  1.68  & 0.12  & $26.1^{+5.9}_{-5.2}$    &     $372.9^{+83.7}_{-73.9} \pm 11.6$   &5.1             \\
4.278 &  175.7 &  1.05  &  1.71  & 0.11  & $13.5^{+4.5}_{-3.8}$    &     $583.5^{+193.1}_{-162.9} \pm 18.1$   &4.1           \\
4.288 &  502.4 &  1.05  &  1.94  & 0.10  & $34.2^{+6.8}_{-6.1}$    &     $497.3^{+99.4}_{-89.3} \pm 15.4$   &6.1             \\
4.308 &  45.1 &  1.05  &  1.74  & 0.11  & $0.0^{+0.9}_{-0.0}$    &     $1.7^{+151.9}_{-1.7} \pm 0.1$   &0.1                  \\
4.312 &  501.2 &  1.05  &  1.98  & 0.10  & $23.9^{+5.6}_{-4.9}$    &     $364.9^{+86.3}_{-75.6} \pm 11.3$   &5.3             \\
4.337 &  505.0 &  1.05  &  2.05  & 0.10  & $21.5^{+5.5}_{-4.8}$    &     $316.4^{+81.6}_{-71.2} \pm 9.8$   &4.5              \\
4.358 &  543.9 &  1.05  &  1.87  & 0.11  & $14.1^{+4.5}_{-3.9}$    &     $186.9^{+60.0}_{-51.4} \pm 5.8$   &3.6              \\
4.377 &  522.7 &  1.05  &  2.19  & 0.09  & $24.7^{+5.7}_{-5.0}$    &     $351.1^{+80.9}_{-71.4} \pm 10.9$   &6.0             \\
4.387 &  55.6 &  1.05  &  1.95  & 0.10  & $2.3^{+2.6}_{-1.9}$    &     $310.7^{+340.6}_{-255.4} \pm 9.6$   &0.4              \\
4.396 &  507.8 &  1.05  &  2.21  & 0.09  & $18.6^{+5.0}_{-4.3}$    &     $266.4^{+71.2}_{-61.4} \pm 8.3$   &4.5              \\
4.416 &  1090.7 &  1.05  &  2.00  & 0.10  & $47.3^{+8.0}_{-7.3}$    &     $320.0^{+54.5}_{-49.6} \pm 9.9$   &7.0             \\
4.436 &  569.9 &  1.05  &  2.27  & 0.09  & $22.3^{+5.7}_{-5.1}$    &     $279.7^{+72.2}_{-63.5} \pm 8.7$   &4.8              \\
4.467 &  111.1 &  1.05  &  2.16  & 0.09  & $7.0^{+2.6}_{-2.6}$    &     $477.7^{+180.5}_{-180.5} \pm 14.8$   &2.4            \\
4.527 &  112.1 &  1.05  &  2.22  & 0.09  & $2.5^{+2.2}_{-1.5}$    &     $167.0^{+142.6}_{-100.2} \pm 5.2$   &1.6             \\
4.574 &  48.9 &  1.05  &  2.33  & 0.08  & $0.0^{+0.6}_{-0.0}$    &     $1.5^{+92.9}_{-1.5} \pm 0.0$   &0.1                   \\
4.600 &  586.9 &  1.05  &  2.92  & 0.07  & $16.4^{+4.7}_{-4.1}$    &     $202.9^{+58.6}_{-50.4} \pm 6.3$   &4.7              \\
4.612 &  103.7 &  1.05  &  2.81  & 0.07  & $4.5^{+3.6}_{-3.1}$    &     $338.1^{+269.0}_{-233.2} \pm 10.5$   &1.2            \\
4.628 &  521.5 &  1.05  &  2.92  & 0.06  & $13.6^{+4.6}_{-3.9}$    &     $208.0^{+70.9}_{-59.7} \pm 6.4$   &3.7              \\
4.641 &  551.6 &  1.05  &  2.96  & 0.06  & $13.3^{+4.7}_{-4.0}$    &     $193.0^{+68.5}_{-58.1} \pm 6.0$   &3.7              \\
4.661 &  529.4 &  1.05  &  3.02  & 0.06  & $1.3^{+2.5}_{-1.3}$    &     $20.0^{+38.0}_{-20.0} \pm 0.6$   &0.6                \\
4.682 &  1667.4 &  1.05  &  3.10  & 0.06  & $33.6^{+7.1}_{-6.4}$    &     $164.6^{+34.7}_{-31.5} \pm 5.1$   &5.5             \\
4.699 &  535.5 &  1.05  &  3.10  & 0.06  & $15.0^{+4.5}_{-3.8}$    &     $225.3^{+67.3}_{-57.6} \pm 7.0$   &3.8              \\
4.740 &  163.9 &  1.05  &  3.25  & 0.06  & $2.4^{+2.1}_{-1.5}$    &     $113.3^{+102.9}_{-72.0} \pm 3.5$   &1.3              \\
4.750 &  366.6 &  1.05  &  3.34  & 0.06  & $9.9^{+3.8}_{-3.1}$    &     $210.9^{+80.2}_{-66.4} \pm 6.5$   &3.0               \\
4.781 &  511.5 &  1.06  &  3.52  & 0.05  & $11.5^{+4.0}_{-3.4}$    &     $172.3^{+60.5}_{-50.6} \pm 5.3$   &3.5              \\
4.843 &  525.2 &  1.06  &  3.67  & 0.05  & $10.4^{+4.1}_{-3.4}$    &     $157.2^{+61.0}_{-50.7} \pm 4.9$   &1.5              \\
4.918 &  207.8 &  1.06  &  4.28  & 0.04  & $0.3^{+1.9}_{-0.3}$    &     $11.4^{+76.4}_{-11.4} \pm 0.4$   &0.3                \\
4.951 &  159.3 &  1.06  &  4.73  & 0.04  & $0.6^{+1.6}_{-0.6}$    &     $32.7^{+87.8}_{-32.7} \pm 1.0$   &0.5                \\

\hline
\end{longtable}

\section{Partial wave analysis} 
\subsection{Datasets and background treatment}
\noindent To investigate potential $N^{*}$, $\Lambda^{*}$ and $\Sigma^{*}$ resonances, a partial wave analysis~(PWA) of $e^+e^- \to pK^{-}\bar{\Sigma}^{0}$ + c.c. is performed at 3.773 GeV. 
The PWA employs the helicity-amplitude formalism~\cite{Richman:1984gh,Chung:1971ri} as implemented in the open-source framework TF-PWA~\cite{TFPWA}. The basic procedure follows Ref.~\cite{BESIII:2026glf}. 
To improve the signal purity, the invariant mass of $\gamma\bar{\Lambda}$ is required to be within $1.177 < M_{\gamma\bar{\Lambda}} < 1.207~\text{GeV}/c^2$ with 3 684 remaining events. The $\bar{\Sigma}^{0}$ sideband (defined as $M_{\gamma\bar{\Lambda}}$ within $(1.154, 1.167)~\text{GeV}/c^2$ or $(1.216, 1.229)~\text{GeV}/c^2$) is used to describe the mis-combination background and the yield is estimated to be $384.0\pm19.6$. The $\bar{\Sigma}^{0}$ sideband events are normalized with a factor of 1.0 and added to the data with negative weights. Decay amplitudes are constructed from sequential helicity amplitudes and relativistic Breit–Wigner propagators for intermediate resonances; non-resonant~($N\!R$) terms are set to unity. Blatt–Weisskopf barrier factors~\cite{Chung:1993da} are included for the resonant but not for the $N\!R$ contributions.
\subsection{Nominal fit results}
\begin{figure}[htbp]
    \centering
\includegraphics[width=0.48\linewidth]{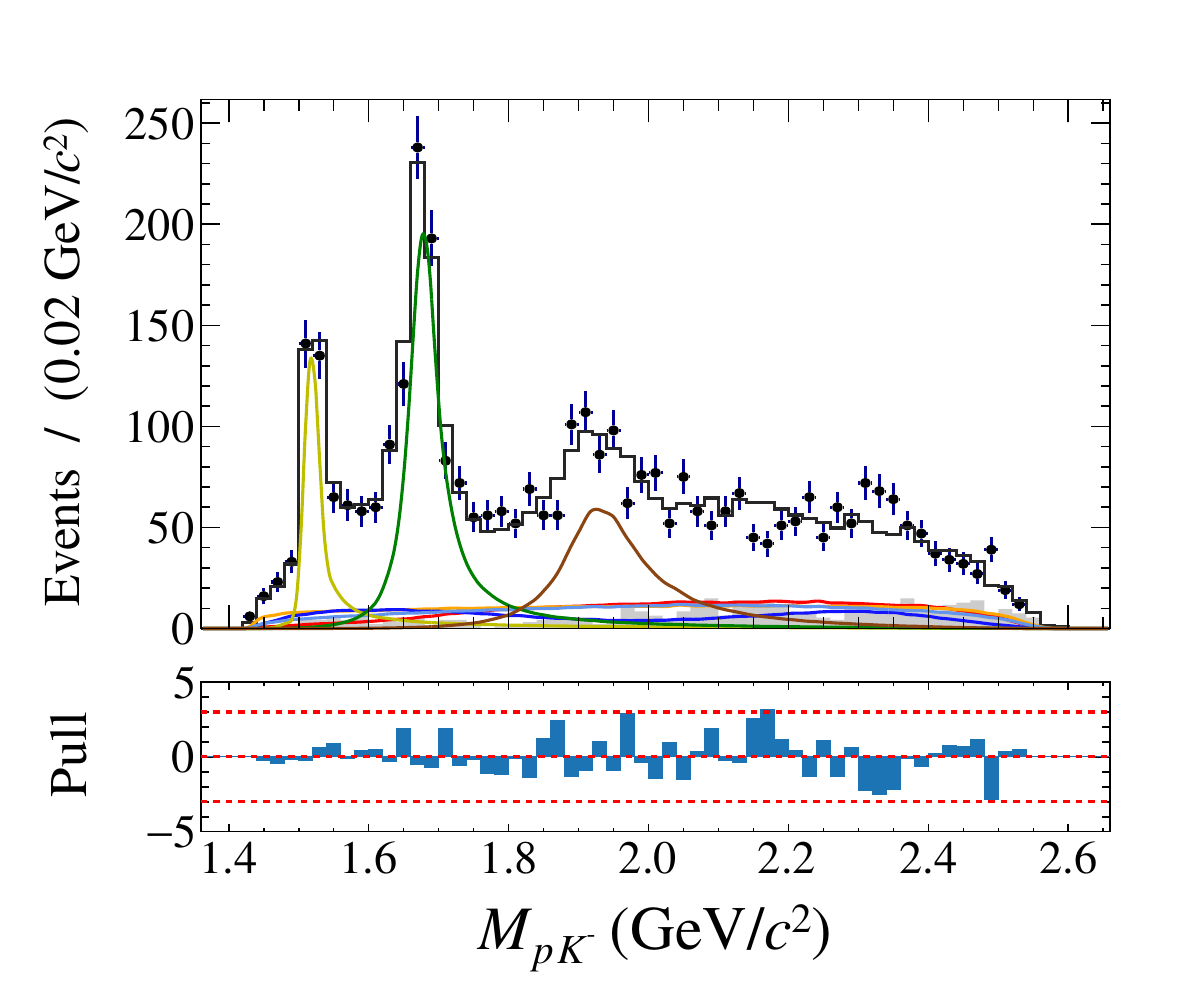}
\includegraphics[width=0.48\linewidth]{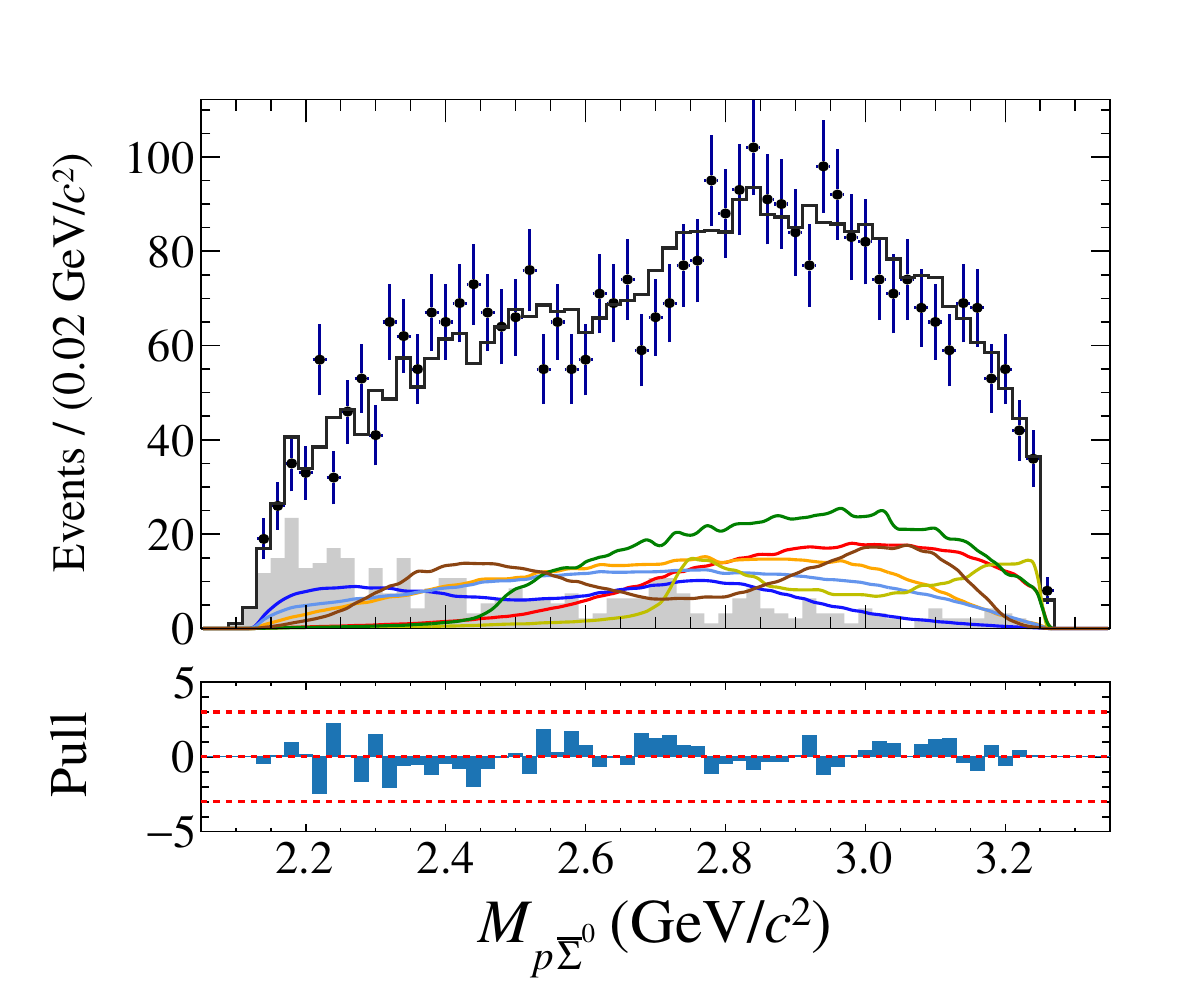}
\includegraphics[width=0.48\linewidth]{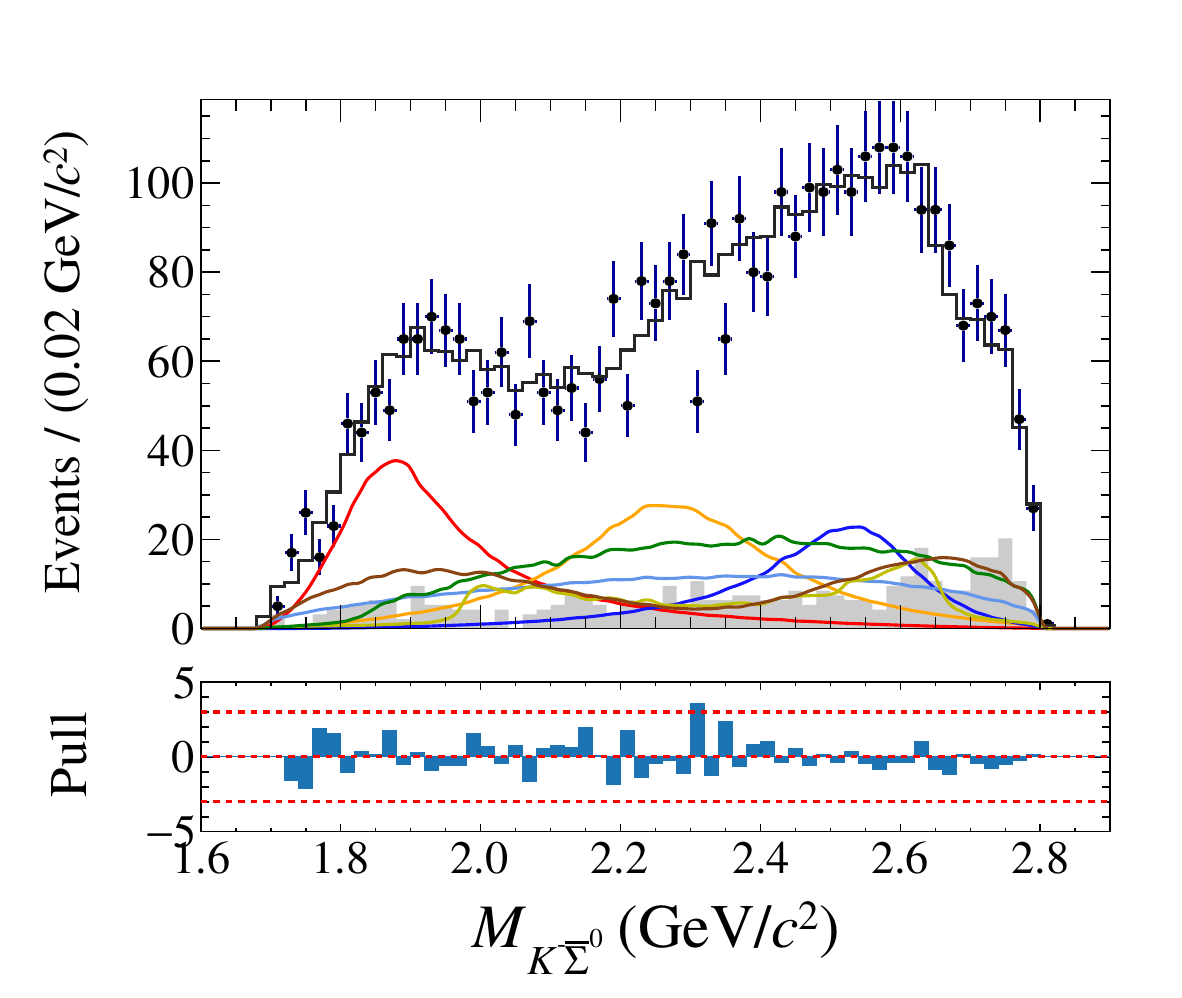}
\includegraphics[width=0.48\linewidth]{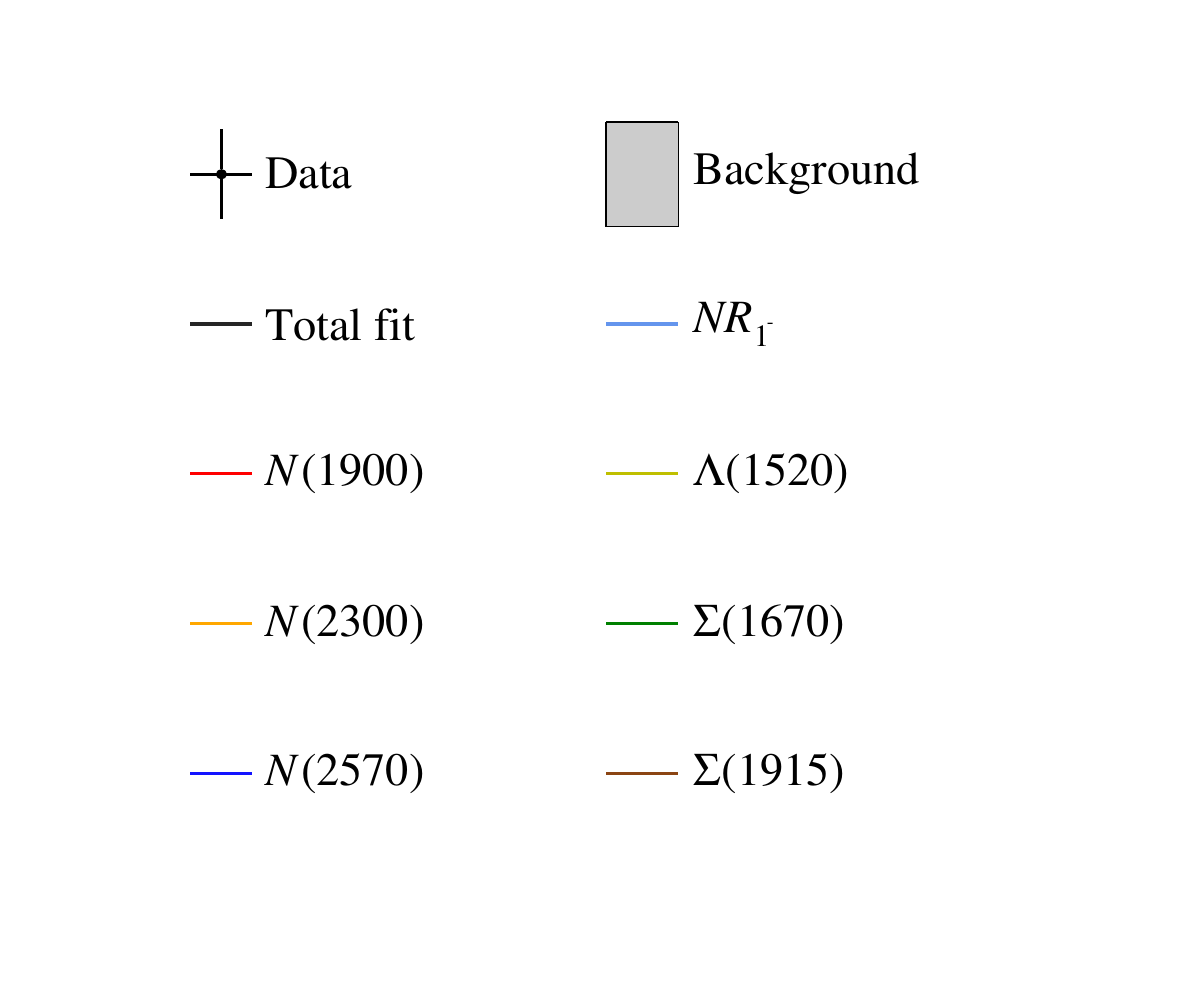}
    \caption{Distributions of $M_{pK^{-}}$, $M_{p\bar{\Sigma}^{0}}$ and $M_{K^{-}\bar{\Sigma}^{0}}$. Black dots with error bars represent the data sample. The grey-hatched histogram shows the background estimated from the $\bar{\Sigma}^{0}$ sideband. The colored histograms display the contributions from the individual resonant states and the sum of all these contributions together. }
    \label{pwa_fig}
\end{figure}

\begin{figure}[htbp]
    \centering
\includegraphics[width=0.48\linewidth]{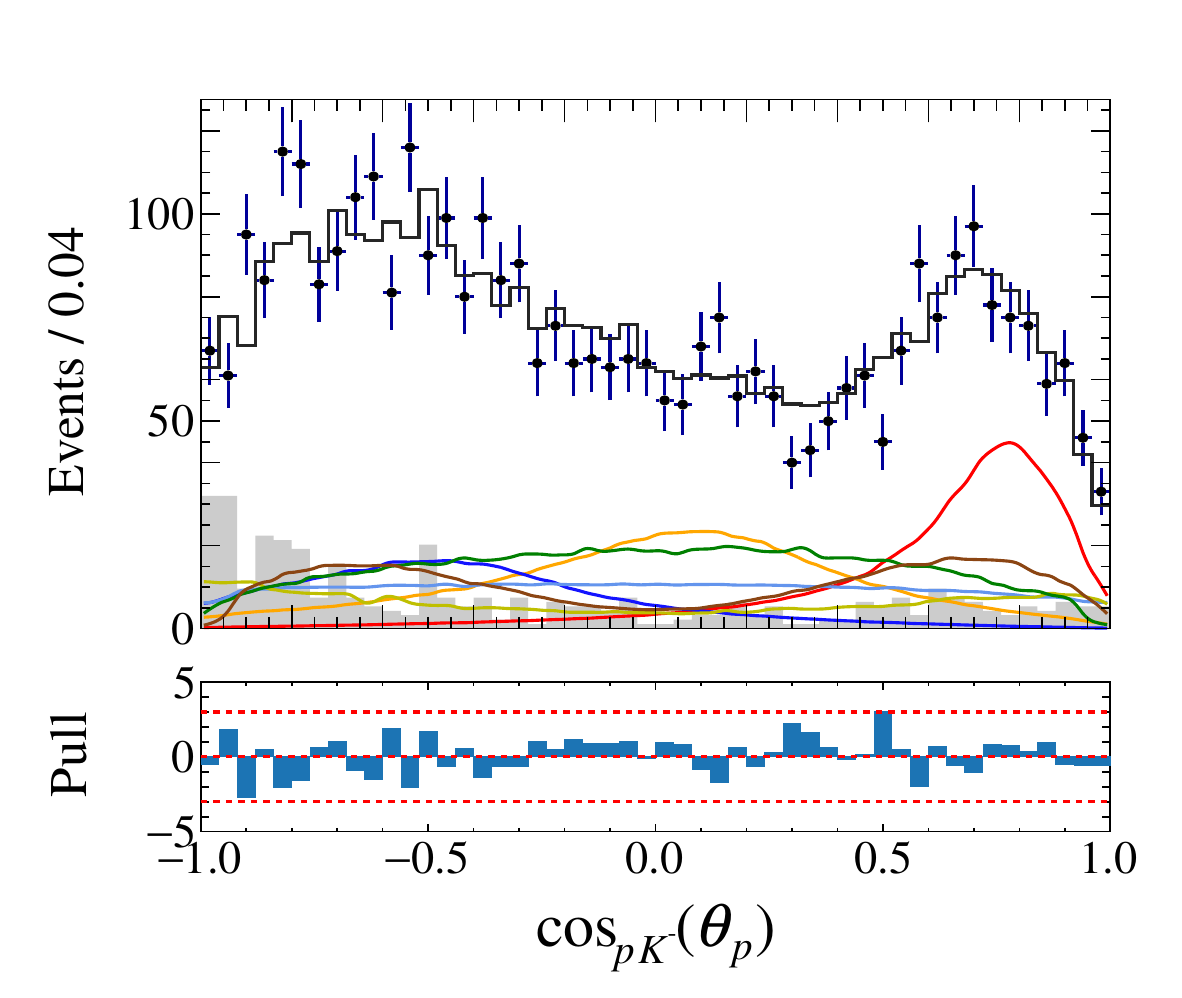}
\includegraphics[width=0.48\linewidth]{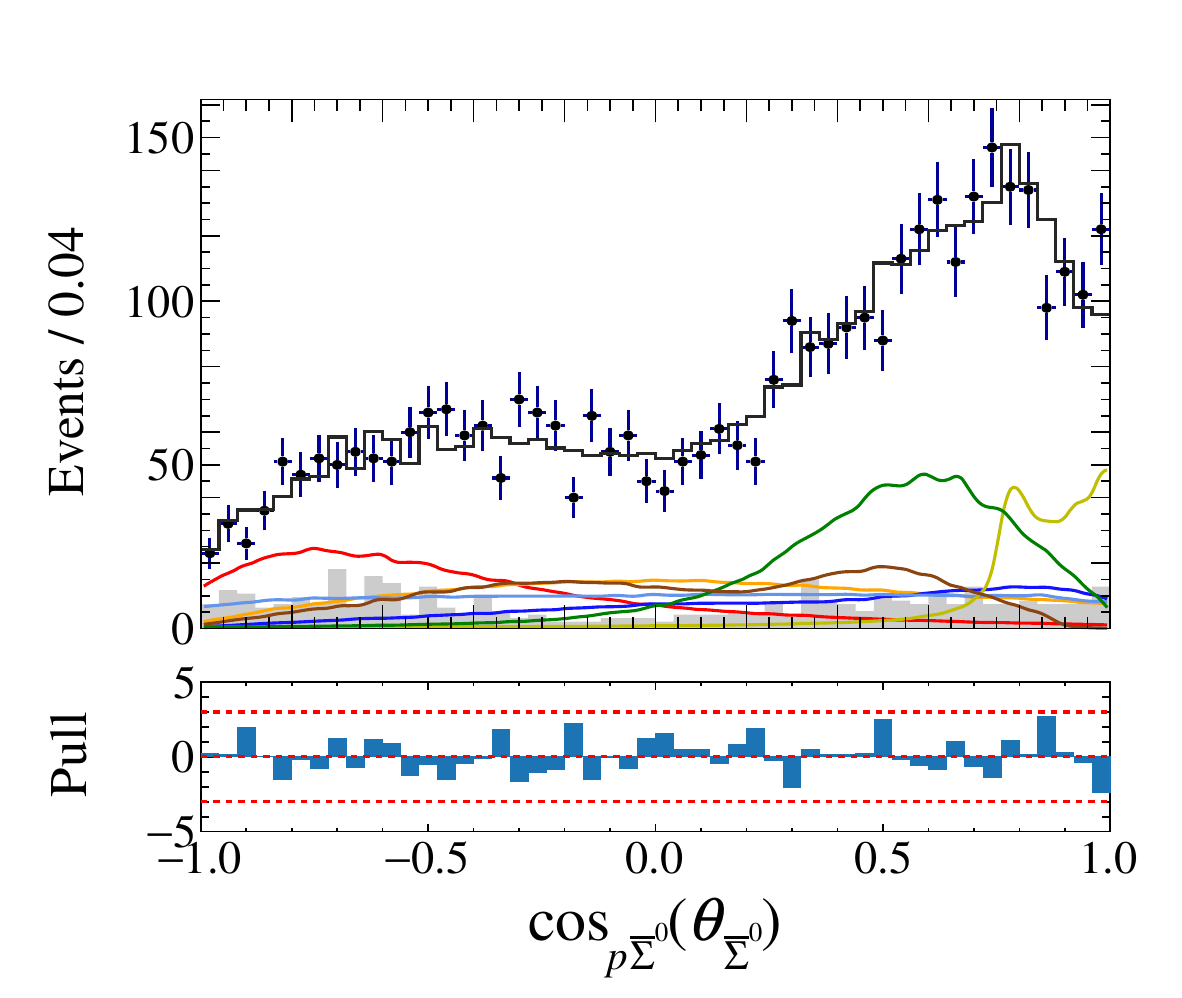}
\includegraphics[width=0.48\linewidth]{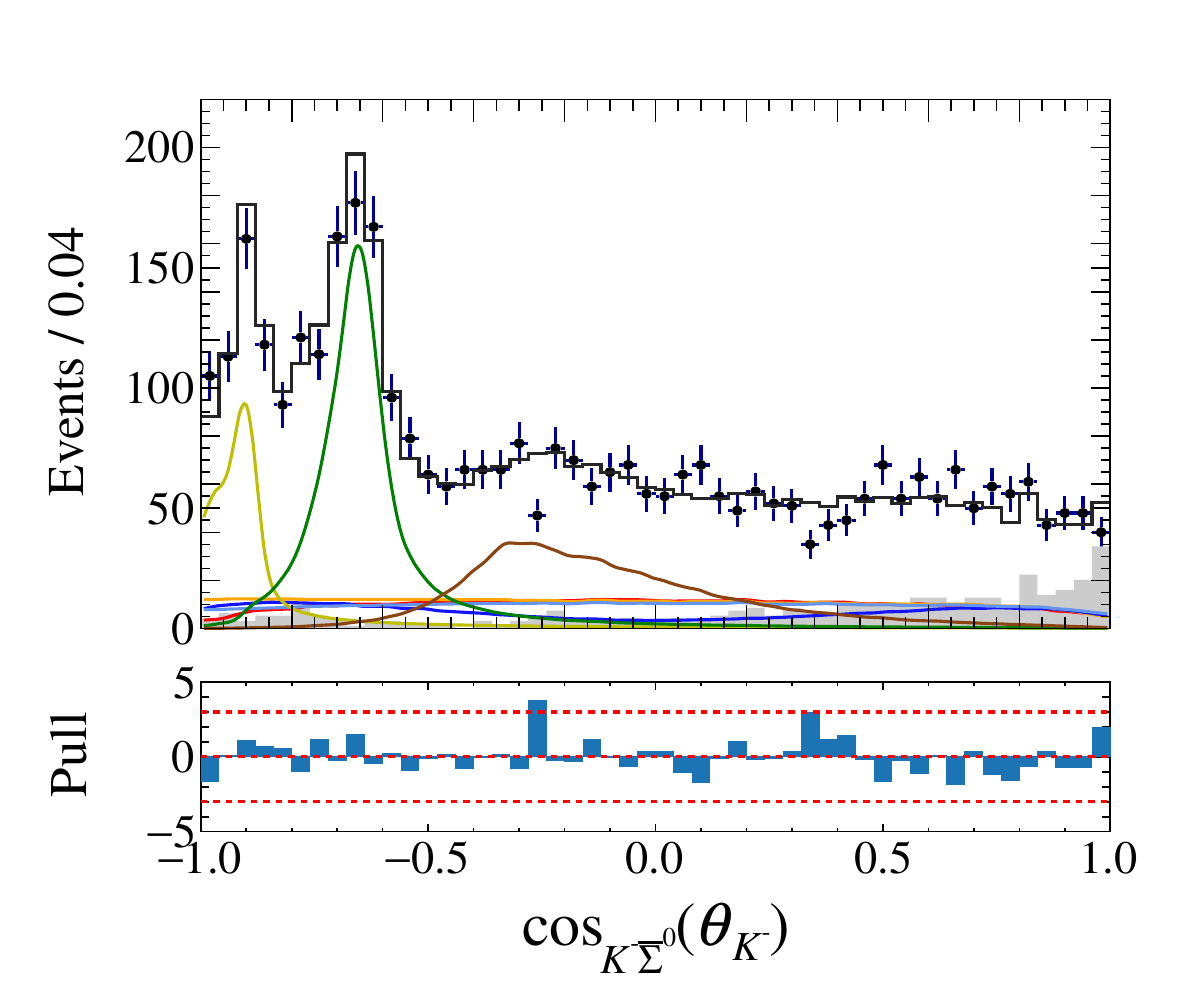}
\includegraphics[width=0.48\linewidth]{legend_combined.pdf}
    \caption{Distributions of ${\rm{cos}}_{pK^{-}}(\theta_{p})$, ${\rm{cos}}_{p\bar{\Sigma}^{0}}(\theta_{\bar{\Sigma}^{0}})$ and ${\rm{cos}}_{K^{-}\bar{\Sigma}^{0}}(\theta_{K^{-}})$. Black dots with error bars represent the data sample. The grey-hatched histogram shows the background estimated from the $\bar{\Sigma}^{0}$ sideband. The colored histograms display the contributions from the individual resonant states and the sum of all these contributions together.}
    \label{pwa_figangle}
\end{figure}

\noindent The baseline solution follows Ref.~\cite{BESIII:2026glf} with the same final-state particles. To determine the nominal fit hypothesis of the PWA, the statistical significance of each potential component is evaluated based on the change of the negative log likelihood value when it is included, taking into account the change of the number of degrees of freedom ($N_{\text{n.d.f.}}$).
The considered components include excited nucleons, $\Lambda$ and $\Sigma$ states (considered as established by the PDG~\cite{PDG} with a status of at least three-star of existence and a spin not more than $5/2$), and an $\mathcal{S}$-wave non-resonant component~$N\!R_{1^{-}}$ in the $M_{p\bar{\Sigma}^{0}}$ distribution. In addition, two extra resonances $N(2300)$ and $N(2570)$ are considered as potential components in the $M_{K^{-}\bar{\Sigma}^{0}}$ distribution. The results show that the contributions are from $N(1900)$, $N(2300)$, $N(2570)$, $\Lambda(1520)$, $\Sigma(1670)$, $\Sigma(1915)$ and the $N\!R_{1^{-}}$, all of which are found to have statistical significances greater than 5$\sigma$.
In the fit, the mass and width of the $\Lambda(1520)$ are fixed to the values from the PDG~\cite{PDG}, while the masses and widths of other resonances are free parameters. The invariant mass and helicity angle distributions are presented in Figs.~\ref{pwa_fig} and~\ref{pwa_figangle}. 
\section{Systematic uncertainty}
\label{Systematic uncertainty}
\subsection{Born cross-section measurement}
\noindent Systematic uncertainties on the cross sections measurement for $e^{+}e^{-} \to pK^{-}\bar{\Sigma}^{0}$ + c.c. originate from integrated luminosity, tracking and PID of $p$ and $K$ from IP, photon reconstruction, $\Lambda$ reconstruction, kinematic fit, the branching fraction of the intermediate state, the physical model, the signal and background shapes, fit range and input line shape for the cross section. 
\subsubsection{Luminosity}
\noindent
The luminosities at each energy point are measured using the Bhabha scattering process with an uncertainty of about $1.0\%$~\cite{BESIII:2015qfd} below $4.0$~GeV, $0.7\%$~\cite{BESIII:2022dxl} from $4.0$ to $4.6$~GeV, and $0.6\%$~\cite{BESIII:2022ulv} above $4.6$~GeV. 
\subsubsection{Tracking}
\noindent
The uncertainties of tracking for the $p$ and $K$ are determined by the control samples of $J/\psi \to p\bar{p}\pi^+\pi^-$ and $J/\psi \to K_{s}^{0}K^{-}\pi^{+}$ events~\cite{Tracking}. Two-dimensional tracking efficiencies as a function of the transverse momentum and $\cos\theta$ are evaluated for both the data and MC simulation. To correct the MC efficiencies to those of the data, the corresponding two-dimensional correction factors and their uncertainties are derived. To estimate its impact on the systematic uncertainty, the MC simulation events are then weighted to determine the detection efficiency. The difference in the detection efficiency is taken as the systematic uncertainty. 
\subsubsection{PID}
\noindent
The uncertainty of PID is explored using as control samples the $J/\psi \to K^{*}\bar{K}$~\cite{BESIII:2011ysp} and $J/\psi \to p\bar{p}\pi^{+}\pi^{-}$ events~\cite{BESIII:2011wmh}, and the difference in efficiency between the data and MC simulation is assigned as the corresponding systematic uncertainty.

\subsubsection{Photon reconstruction}
\noindent
The systematic uncertainty of the photon reconstruction is estimated to be $0.5\%$ for each photon by analyzing the ISR process $e^{+}e^{-} \to \gamma \mu^{+}\mu^{-}$. 
\subsubsection{$\Lambda$ reconstruction}
\noindent
The systematic uncertainty associated with  $\Lambda$ reconstruction includes the contributions from tracking and PID efficiencies of the $p$ and $\pi^{-}$ candidates, as well as the $\Lambda$  mass-window and  decay-length requirements. It is determined using the control sample $\psi(3686)\to\Lambda\bar{\Lambda}$, as in Ref.~\cite{BESIII:2017kqw}, and a systematic uncertainty of $1.1\%$  is assigned. 
\subsubsection{Kinematic fit}
\noindent
The systematic uncertainty associated with the kinematic fit is assigned by comparing the results with and without the track helix parameter corrections.  Half of the difference in the detection efficiency is taken as the systematic uncertainty~\cite{4C}.
\subsubsection{Branching fractions}
\noindent
The uncertainties in the branching fractions of the intermediate decays $\bar{\Sigma}^{0}\to\gamma\bar{\Lambda}$ and $\bar{\Lambda}\to \bar{p}\pi^{+}$ are obtained from the PDG~\cite{PDG}. 
\subsubsection{Physical model}
\noindent
The systematic uncertainty associated with the physical model is evaluated from the difference in detection efficiency between the PHSP model and the angular distribution obtained from a PWA of $e^{+}e^{-} \to pK^{-}\bar{\Sigma}^{0}$ + c.c. at $\sqrt{s}$ = 3.773 GeV. The relative difference in efficiency, 1.4\%, is assigned as the corresponding systematic uncertainty. 

\subsubsection{Fit method}
\noindent
The systematic uncertainties associated with the fit to the $M_{\bar{\Sigma}^{0}}$ spectrum arise from the signal shape, background shape and fit range. They are evaluated using the combined data sample from all center-of-mass energy points to reduce the impact of statistical fluctuations. The contribution from the signal shape is evaluated by replacing the nominal signal shape, given by the MC shape convolved with a Gaussian function, with the unconvolved MC shape, and the yield difference, 1.0\%, is taken as the systematic uncertainty. 
The contribution from the background shape is estimated through an alternative fit with a higher-order polynomial function; we take the yield difference, 0.6\%, as the systematic uncertainty.  
The systematic uncertainty associated with the fit range is evaluated by varying the lower and upper boundaries of the $M_{\bar{\Sigma}^{0}}$ fit range independently by 0.01 or 0.02 GeV/$c^2$, and the resulting 0.7\% change in the signal yield is taken as the corresponding systematic uncertainty.
\subsubsection{Input line shape}
\noindent
The systematic uncertainty associated with the input cross-section line shape used to determine the product of the ISR correction factor and the detection efficiency, $(1+\delta)\cdot\epsilon$, consists of two components. One component arises from the statistical uncertainty of the input cross-section line shape and is evaluated by varying the central value at each energy point within its $\pm1\sigma$ statistical uncertainty. The $(1+\delta)\cdot\epsilon$ values for each CM energy point are then recalculated. This process is repeated 200 times, and a Gaussian function is used to fit the $(1+\delta)\cdot\epsilon$ distribution. The width of the Gaussian function is taken as the corresponding systematic uncertainty. 
The other component arises from the parameterization of the line shape and is estimated by using an alternative input cross-section line shape described by a resonance plus a power-law (PL) function. The $(1+\delta)\cdot\epsilon$ value for each of CM energy point is then recalculated, and the largest change is taken as the systematic uncertainty. The total systematic uncertainty for the input line shape is 0.3\% by adding both contributions in quadrature.

\subsubsection{Total systematic uncertainty} 
The various sources of systematic uncertainty on the Born cross-section measurement are summarized in table~\ref{sysT}. Assuming all sources are independent, the total systematic uncertainty on the cross-section measurement is determined by adding them in quadrature.

\begin{table}[hbtp]
\footnotesize
\centering
\caption{Relative systematic uncertainties (\%) in the cross-section measurement. The sources of systematic uncertainty include the luminosity, tracking, PID,  photon reconstruction ($\gamma_{\rm REC}$), $\bar{\Lambda}$ reconstruction ($\Lambda_{\rm REC}$), the 4C kinematic fit (4C), branching fractions ($\mathcal{B}$), the physical model (PM), the signal shape (SS), the background shape (BS), the fit range (FR) and input line shape (ILS).}
\setlength{\tabcolsep}{5pt}
\label{sysT}
\resizebox{15.cm}{!}{
\renewcommand{\arraystretch}{1.2}
\begin{tabular}{cccccccccccccc}
\hline \hline
$\sqrt{s}$ (GeV)   & Lum. &Tracking & PID & $\gamma_{\rm REC}$  &$\Lambda_{\rm REC}$ &4C&$\mathcal{B}$ &PM &SS &BS &FR &ILS &Total \\
\hline
From 3.510 to 3.871 & 1.0  &0.3     &2.0  &0.5  &1.1   & 0.3 &0.8 &1.4 &1.0 &0.6 &0.7  &0.3 &3.4 \\
\hline
From 4.009 to 4.440 & 0.7  &0.3     &2.0  &0.5  &1.1   & 0.3 &0.8  &1.4 &1.0 &0.6 &0.7 &0.3 &3.3 \\
\hline
From 4.600 to 4.950 & 0.6  &0.3     &2.0  &0.5  &1.1   & 0.3 &0.8   &1.4 &1.0 &0.6 &0.7 &0.3 &3.3 \\
\hline \hline 
\end{tabular}}
\end{table}

\subsection{PWA}
\noindent The systematic uncertainties in the measurements of the resonance masses, widths, and fit fractions (FFs) obtained from the PWA mainly arise from the inclusion of additional possible components, the background estimation, the orbital angular momentum assigned to the non-resonant component, and the choice of the Blatt--Weisskopf barrier radius. All sources are listed in Tables~\ref{sys_ff},~\ref{sys_mass}, and~\ref{sys_width}, and discussed in detail below. Assuming all sources are independent, the total systematic uncertainty is determined by adding them in quadrature.

\subsubsection{Additional possible components}
\noindent
To assess the impact of possible additional components on the PWA results, the PWA is repeated by adding one extra resonance at a time ($K_2(2250)$, $K_3(2320)$, $N(1895)$, $N(2100)$, $N(2120)$, $N(1875)$, $\Sigma(1660)$, and $\Sigma(1910)$). The largest changes in the fitted masses, widths, and fit fractions of the resonances are taken as the corresponding systematic uncertainties.
\subsubsection{Background estimation}
\noindent
The uncertainty associated to the background estimation is evaluated by varying the background  yield within $\pm 1\sigma$ of its statistical uncertainty. The largest differences in each fitted parameter are taken as the corresponding systematic uncertainties.
\subsubsection{Orbital angular momentum of $N\!R$}
\noindent
The uncertainty associated with the orbital angular momentum assigned to the $NR$ component is evaluated by varying its orbital angular momentum from 0 to 1 or 2. The largest changes in the fitted masses, widths, and FFs are taken as the corresponding systematic uncertainties.
\subsubsection{Radius of the Blatt-Weisskopf barrier factor}
\noindent
In the PWA fit, the radius $d$ in the Blatt-Weisskopf barrier factor~\cite{Chung:1993da} is set to $d = 0.73~\text{fm} \approx 3.7~\text{GeV}^{-1}$ following Ref.~\cite{BESIII:2019dme};  the uncertainty is estimated by varying the radius within the range $1.0 \sim 5.0~\text{GeV}^{-1}$~\cite{PDG}. The largest changes of the masses, widths, and FFs are taken as the corresponding systematic uncertainties.
\begin{table}[htbp]
\footnotesize
\centering
\caption{Systematic uncertainty sources for the fit fraction (in \%). Here, EX-RES denotes the contributions from additional resonances,  BKG the background, $N\!R$ the orbital angular momentum of the $N\!R$ component, and  BW-BF  the Blatt-Weisskopf barrier factor.}
\begin{tabular}{lcccccc}
\hline \hline
Resonance & EX-RES & BKG & $N\!R$& BW-BF & Total \\
\hline 
\( N(1900) \) & 13.3 & 17.5 & 6.1 & 12.8 & 26.2 \\
\( N(2300) \) & 34.3 & 12.6 & 20.5 & 19.6 & 46.3 \\
\( N(2570) \) & 25.1 & 17.7 & 18.3 & 21.0 & 41.5 \\
\( \Lambda(1520) \) & 11.4 & 15.6 & 11.7 & 10.8 & 25.0 \\
\( \Sigma(1670) \) & 11.4 & 21.6 & 13.3 & 19.5 & 34.0 \\
\( \Sigma(1915) \) & 32.6 & 34.1 & 35.6 & 34.7 & 68.5 \\
\hline \hline
\label{sys_ff}
\end{tabular}
\end{table}

\begin{table}[htbp]
\footnotesize
\centering
\caption{Systematic uncertainty sources for the masses (in MeV/c$^2$). Here, EX-RES denotes the contributions from additional resonances,  BKG the background, $N\!R$ the orbital angular momentum of the $N\!R$ component, and  BW-BF  the Blatt-Weisskopf barrier factor.}
\begin{tabular}{lccccc}
\hline \hline
Resonance & EX-RES & BKG  & $N\!R$ & BW-BF  & Total  \\
\hline 
\( N(1900) \) & 40.0 & 28.9 & 17.8 & 12.3 & 53.8 \\
\( N(2300) \) & 15.6 & 23.2 & 14.3 & 5.1 & 31.8 \\
\( N(2570) \) & 22.4 & 16.1 & 14.8 & 15.2 & 34.8 \\
\( \Sigma(1670) \) & 3.0 & 1.5 & 2.8 & 2.8 & 5.2 \\
\( \Sigma(1915) \) & 26.1 & 21.1 & 19.8 & 23.6 & 45.6 \\
\hline \hline
\label{sys_mass}
\end{tabular}
\end{table}

\begin{table}[htbp]
\footnotesize
\centering
\caption{Systematic uncertainty sources for the widths (in MeV). Here, EX-RES denotes the contributions from additional resonances,  BKG the background, $N\!R$ the orbital angular momentum of the $N\!R$ component, and  BW-BF  the Blatt-Weisskopf barrier factor.}
\begin{tabular}{lccccc}
\hline \hline
Resonance & EX-RES & BKG & $N\!R$ & BW-BF  & Total \\
\hline 
\( N(1900) \) & 64.7 & 55.1 & 68.5 & 60.8 & 124.9 \\
\( N(2300) \) & 11.2 & 11.7 & 13.7 & 8.1 & 22.7 \\
\( N(2570) \) & 6.1 & 11.4 & 18.2 & 2.1 & 22.4 \\
\( \Sigma(1670) \) & 6.8 & 7.1 & 2.8 & 2.8 & 10.6 \\
\( \Sigma(1915) \) & 23.5 & 24.3 & 32.3 & 23.6 & 52.4 \\
\hline \hline
\label{sys_width}
\end{tabular}
\end{table}

\section{Fit to the dressed cross section}

\noindent Potential resonances or charmonium(-like) states contributing to the $e^{+}e^{-} \to pK^{-}\bar{\Sigma}^{0}$ + c.c. process are investigated by fitting the line shape of the dressed cross section, obtained as $\sigma^{\text{dressed}} = \sigma^B \cdot\frac{1}{|1 - \Pi|^2}$, which incorporates  the VP effect. The fit is performed using the least-squares method, which minimizes the $\chi^2$ function defined by
\begin{linenomath*}
\begin{equation}
    \chi^2 = \sum_i \left(\frac{x_i - fk_i}{\sigma_i}\right)^2 + \frac{(1 - f)^2} {\sigma_f^2},
\end{equation}
\end{linenomath*}
where $f$ is a normalization factor used to account for correlated contributions, which is compatible with 1 in the fit; $\sigma_f$ is the relative systematic uncertainty associated with $f$; the quantity $x_i$ denotes the measured cross section, $k_i$ the expected cross section, and $\sigma_i$ the uncorrelated uncertainty at the $i$th CM energy. To incorporate asymmetric uncertainties in the fit, the positive uncertainty is used when the fit is above the data point, and the negative uncertainty when it falls below.
The systematic uncertainties associated with the luminosity, branching fraction, detection efficiency, $\Lambda$ reconstruction, kinematic fit , physical model and input line shape are assumed to be fully correlated among the CM energies, while the other systematic uncertainties are assumed to be uncorrelated.

\begin{figure}[htpb]
    \centering
    \includegraphics[width=0.47\linewidth]{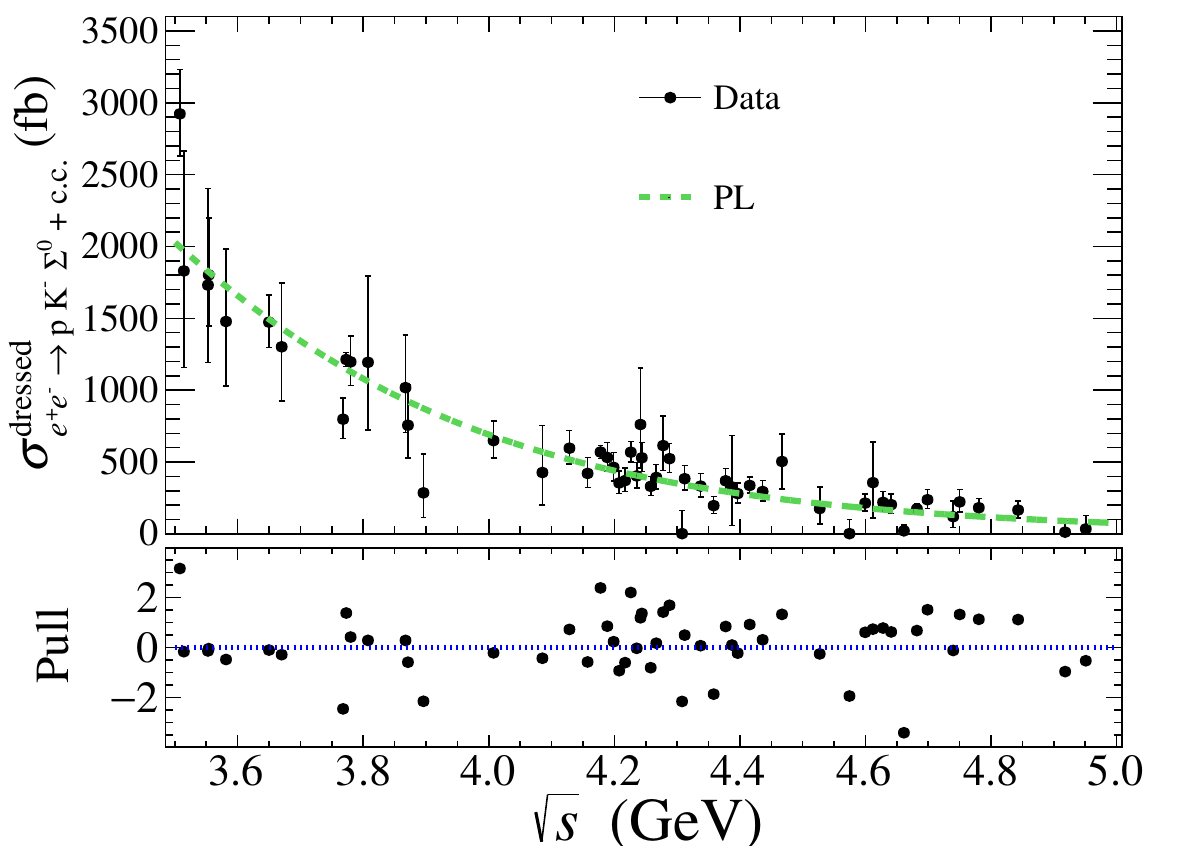}
    \includegraphics[width=0.47\linewidth]{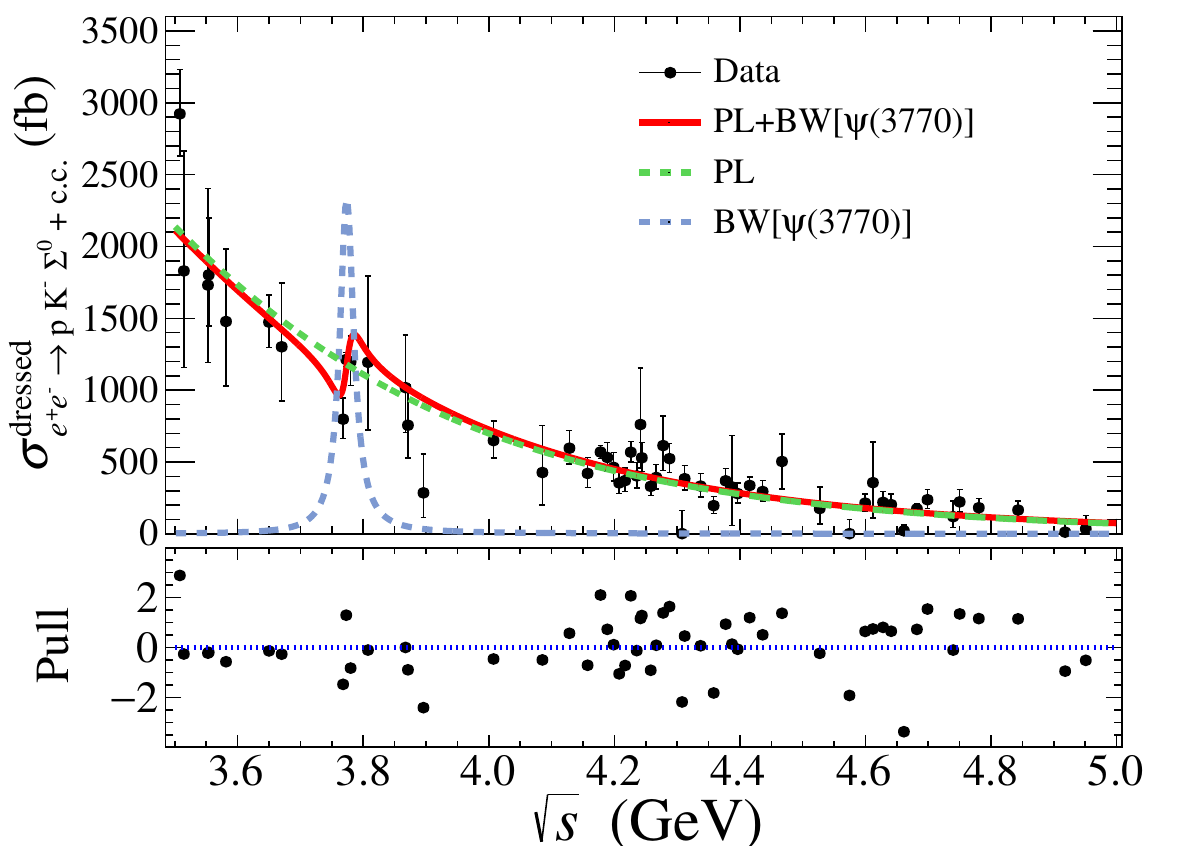}
    \caption{Fit results for the dressed cross sections at the CM energy from 3.51 to 4.95~GeV, with the assumptions of a PL function only (left) and a PL function plus the $\psi(3770)$ resonance (right). The dots with error bars are the measured dressed cross sections including the statistical and systematic uncertainties. The green dashed line corresponds to the fit curve with only the PL function, while the blue dashed line represents the resonance signal.}
\label{R3773}
\end{figure}

\begin{figure}[htbp]
    \centering
    \includegraphics[width=0.47\linewidth]{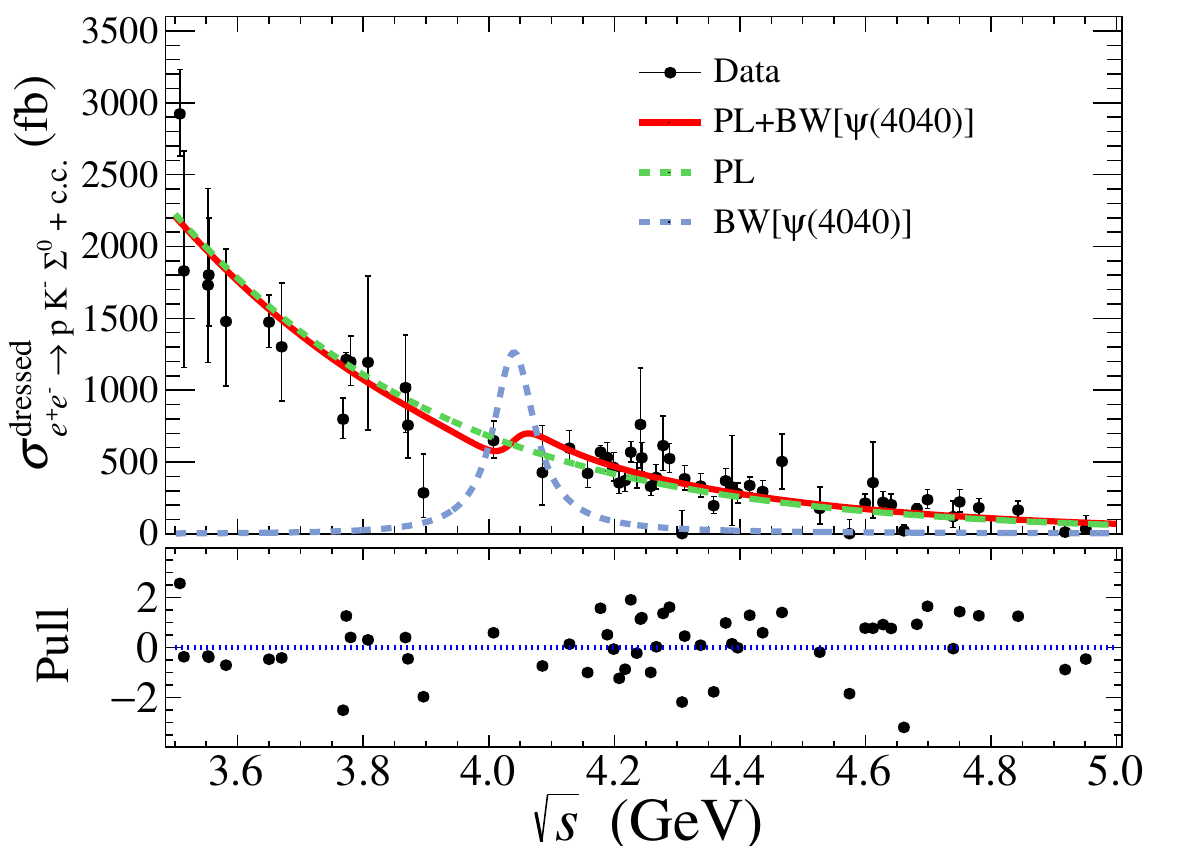}
    \includegraphics[width=0.47\linewidth]{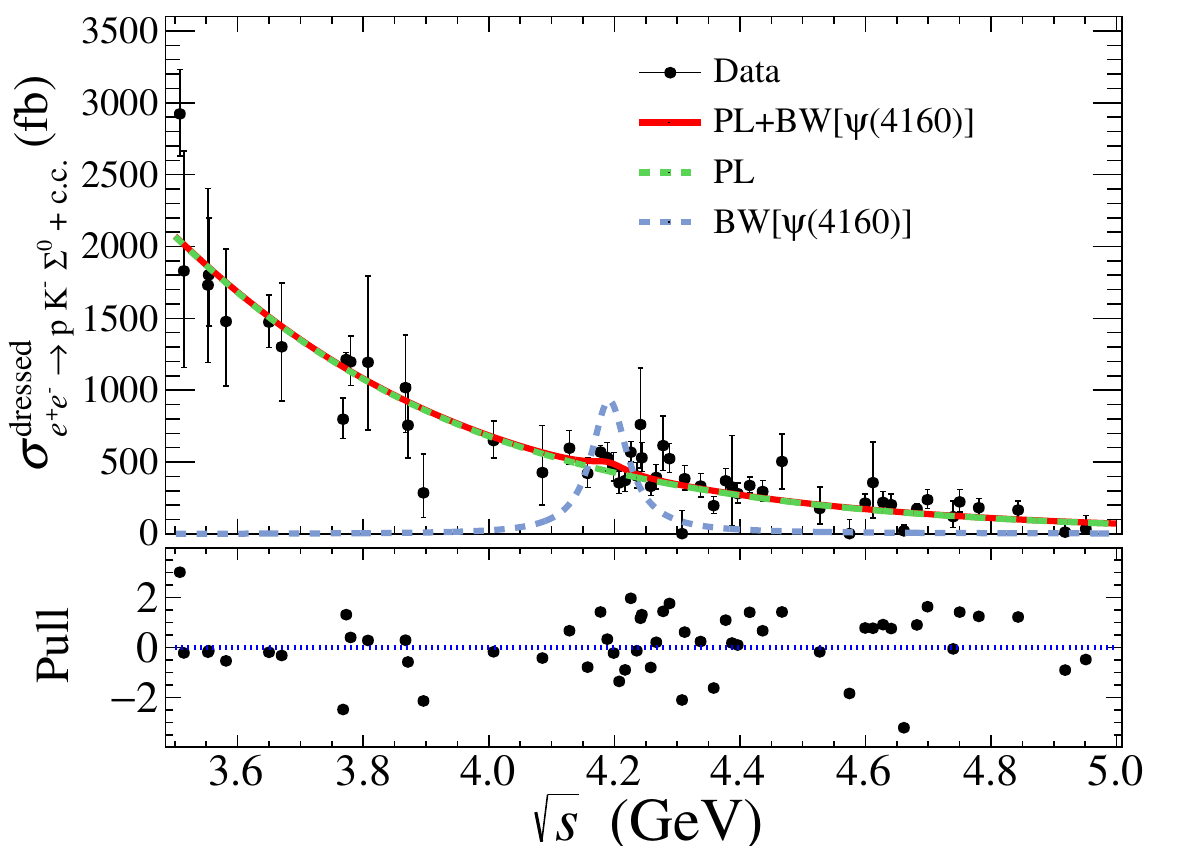}
    \includegraphics[width=0.47\linewidth]{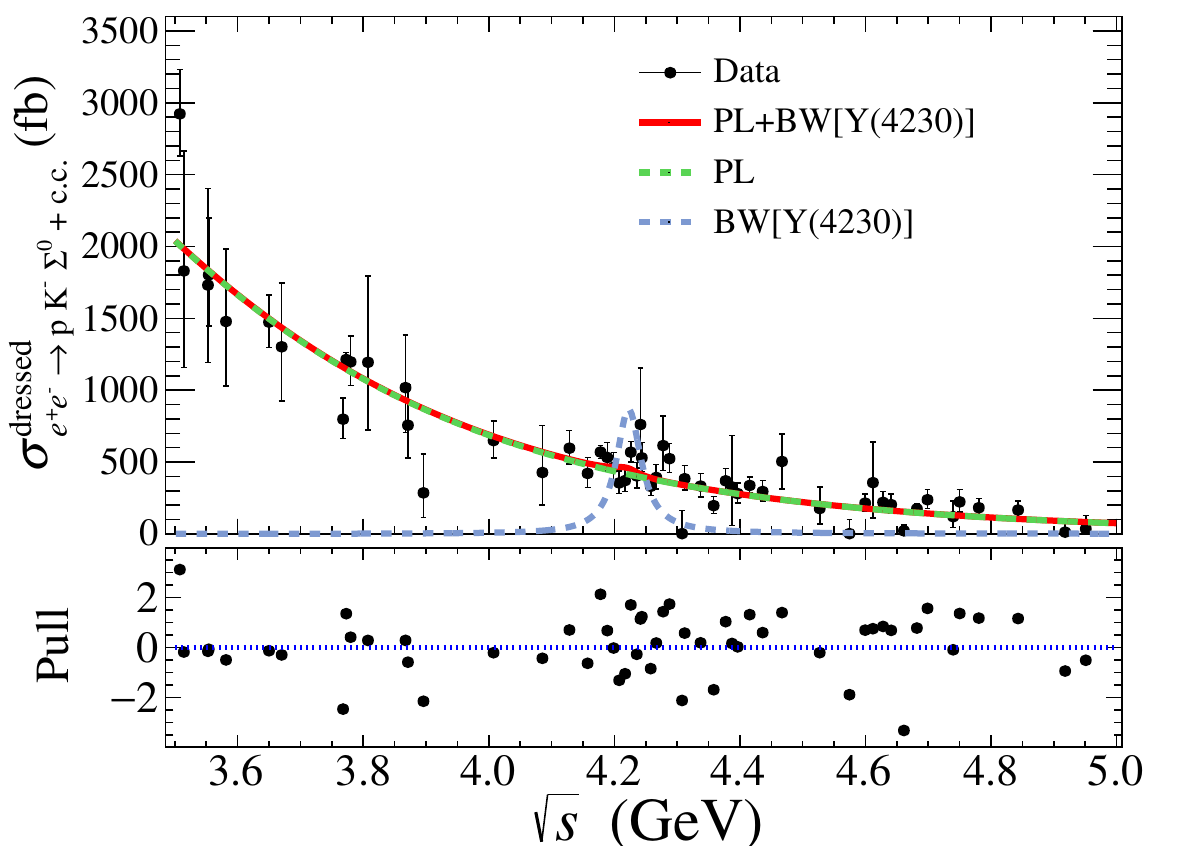}
    \includegraphics[width=0.47\linewidth]{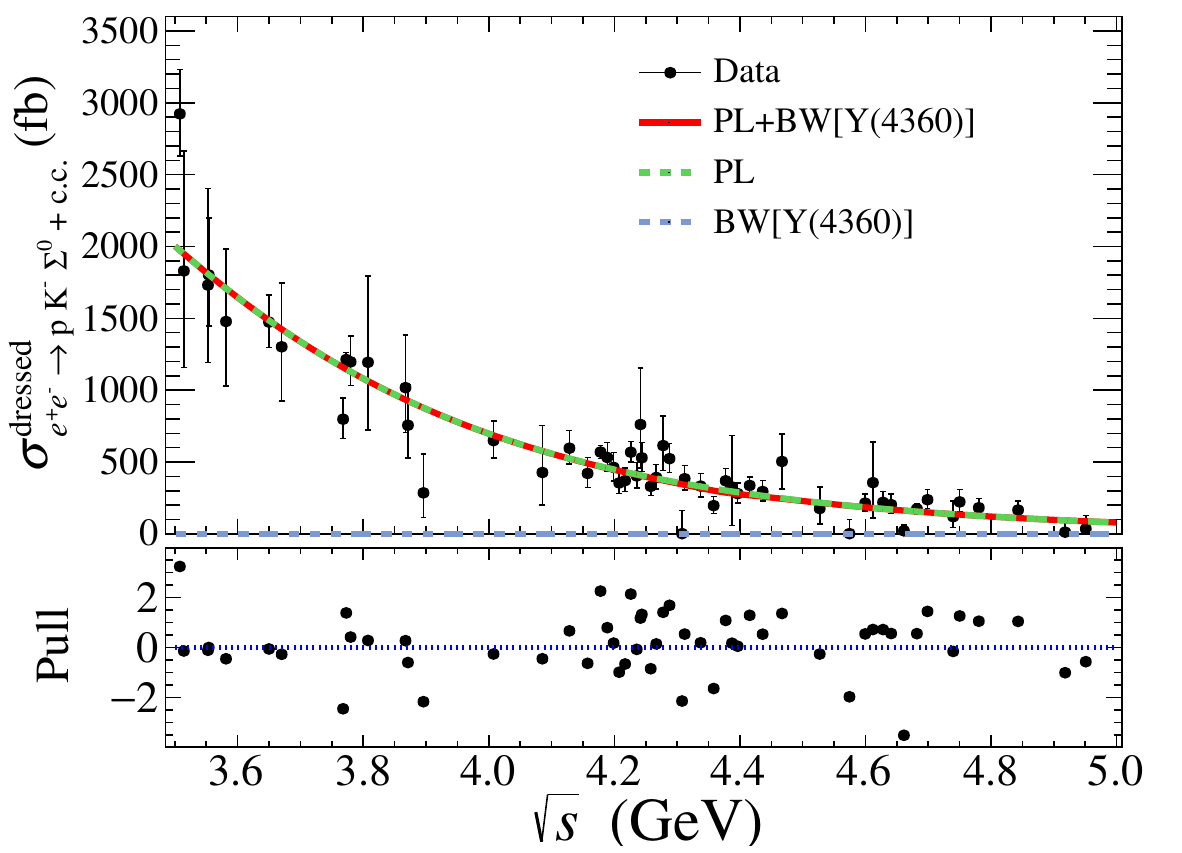}
    \includegraphics[width=0.47\linewidth]{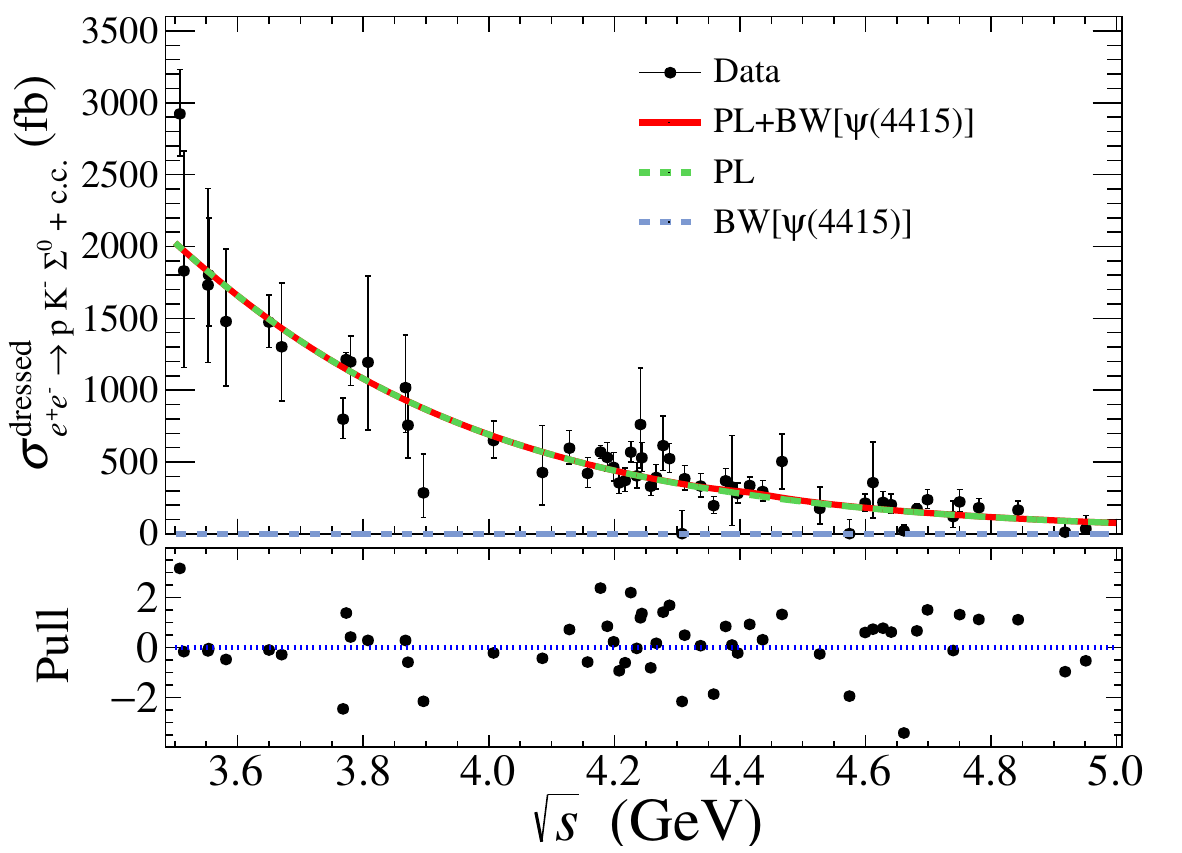}
    \includegraphics[width=0.47\linewidth]{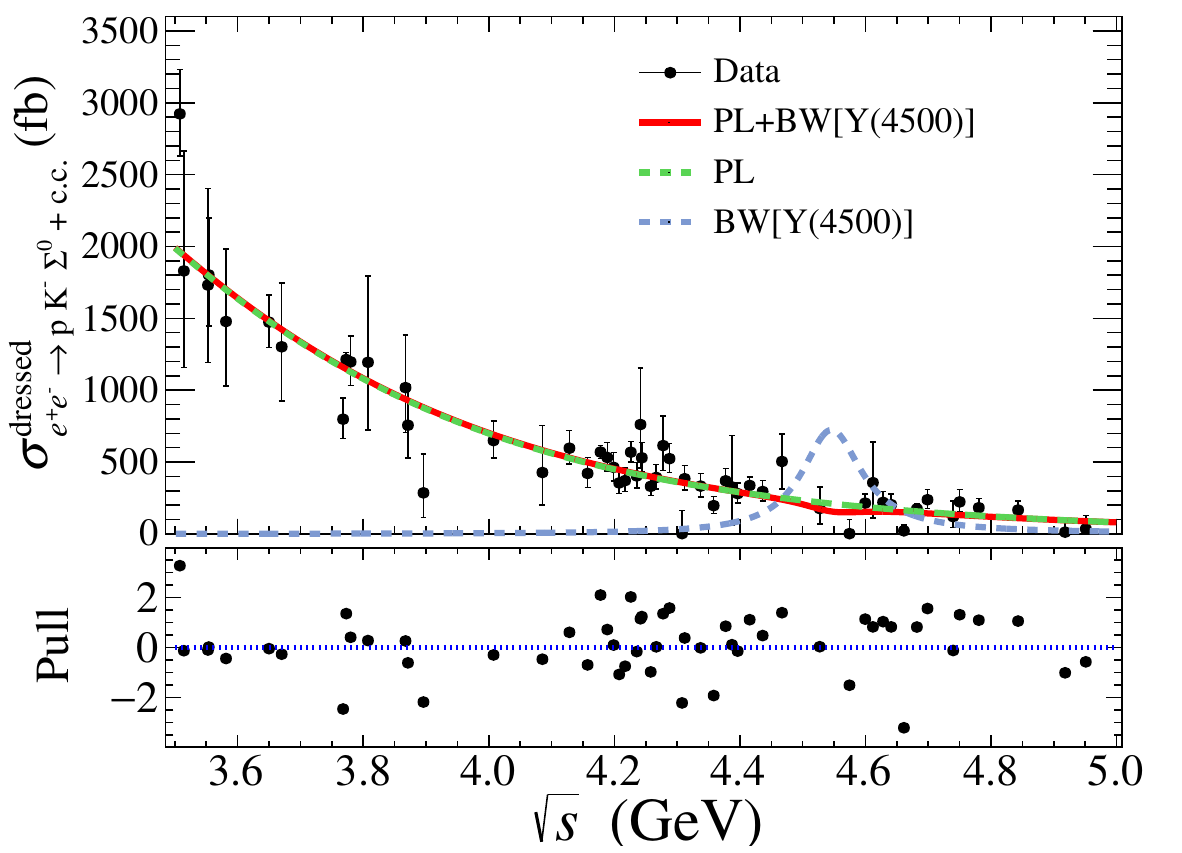}
    \includegraphics[width=0.47\linewidth]{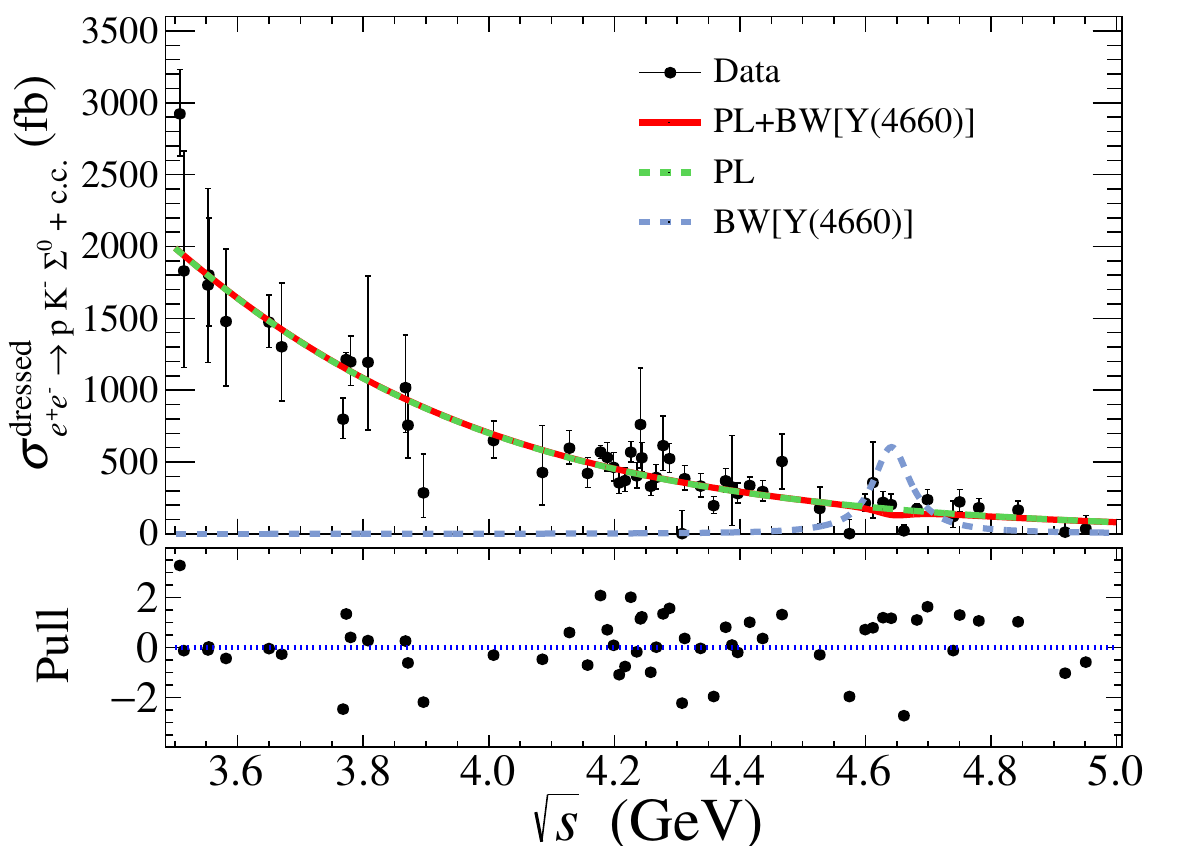}
    \includegraphics[width=0.47\linewidth]{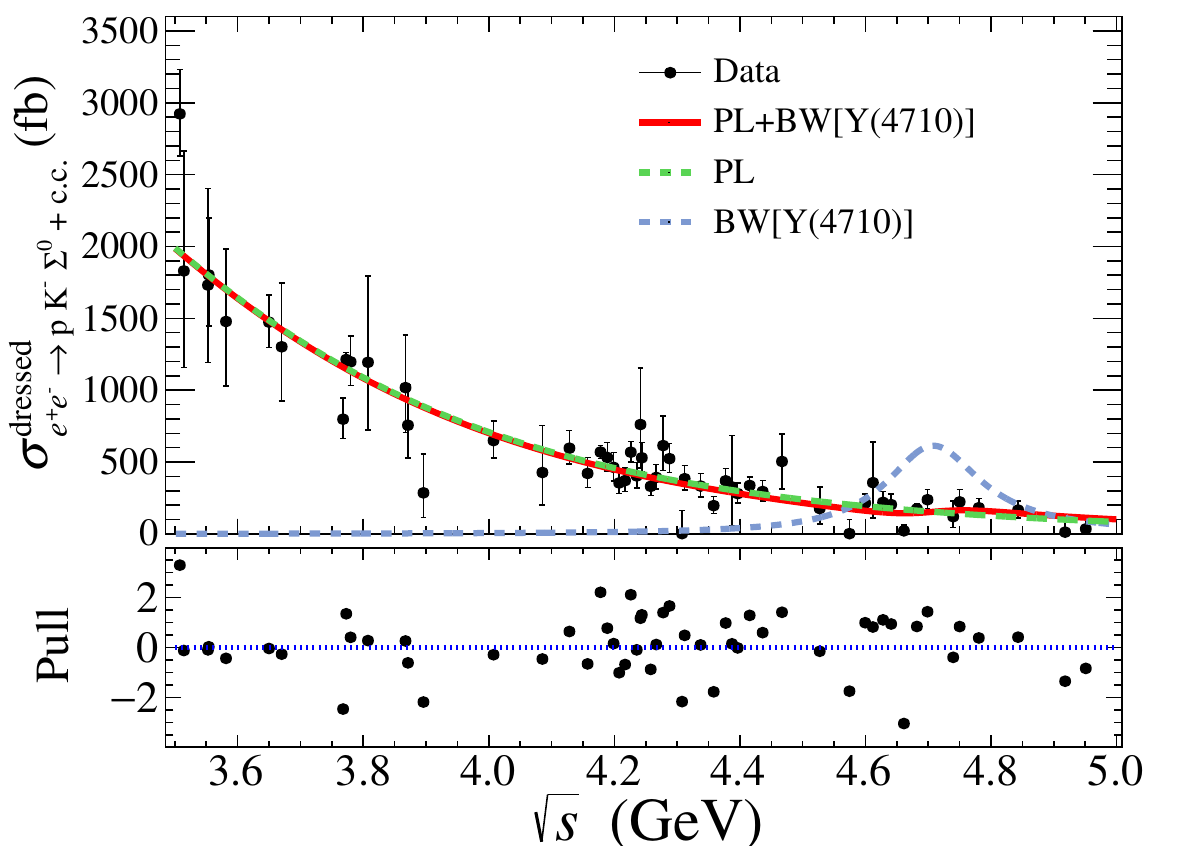}
    \caption{Fit results for the dressed cross sections at the CM energy from 3.51 to 4.95~GeV, with the assumptions of a PL function plus a resonance ($\psi(4040)$, $\psi(4160)$, $Y(4230)$, $Y(4360)$, $\psi(4415)$, $Y(4500)$, $Y(4660)$ and $Y(4710)$). The dots with error bars are the measured dressed cross sections including the statistical and systematic uncertainties. The green dashed line corresponds to the fit curve with only the PL function, while the blue dashed line represents the resonance signal.}
\label{Rother}
\end{figure}

Assuming the cross-section line shape of $e^{+}e^{-} \to pK^{-}\bar{\Sigma}^{0}$ + c.c. includes a resonance [i.e., $\psi(3770)$, $\psi(4040)$, $\psi(4160)$, $Y(4230)$, $Y(4360)$, $\psi(4415)$, $Y(4500)$, $Y(4660)$ and $Y(4710)$] plus a continuum contribution, a fit to the dressed cross sections is applied with the coherent sum of a PL function plus a Breit-Wigner (BW) function,
\begin{linenomath*}
\begin{equation}
\label{333}
\sigma_{\text{dressed}}(\sqrt{s}) = {\left|c_0 \frac {\sqrt{P(\sqrt{s})}}{ (\sqrt{s})^n} + e^{i\Phi} {\rm{BW}}(\sqrt{s}) \sqrt{\frac{P(\sqrt{s})}{P(M)}}\right|}^2,
\end{equation}
\end{linenomath*}
where $\Phi$ represents the relative phase between the BW function,
\begin{linenomath*}
\begin{equation}
{\text{BW}}(\sqrt{s})=\frac{\sqrt{12\pi\Gamma_{ee}\mathcal{B}\Gamma}}{s - M^2 + iM\Gamma},
\end{equation}
\end{linenomath*}
and the PL function. The $c_0$ and $n$ are free fit parameters, $\sqrt{P(\sqrt{s})}$ is the three-body PHSP factor. 
Figure~\ref{R3773} shows the fits to the dressed cross sections under the $\psi(3770)$ resonance assumption. The fits to other possible resonances are shown in Fig.~\ref{Rother}.
The mass $M$ and total width $\Gamma$ of the assumed resonance are fixed to the individual PDG values~\cite{PDG}, and $\Gamma_{ee}\mathcal{B}$ is the product of the electronic partial width and the branching fraction for the assumed resonance decaying into the $pK^{-}\bar{\Sigma}^{0}$ final state. The significance is found to be less than 3$\sigma$ for all assumed resonances. 
Multiple solutions arise from the quadratic form of Eq.~\ref{333}, obtained by scanning the parameter space~\cite{Bai:2019jrb}, and the values of $\Gamma_{ee}\mathcal{B}$ for the assumed charmonium(-like) states decaying into the $pK^{-}\bar{\Sigma}^{0}$ final state, including their upper limits at the 90\% C.L. with a Bayesian approach~\cite{Zhu:2008ca}, are provided. The fit to the cross sections with the PL function alone, i.e. without additional resonance, gives a goodness-of-fit of $\chi^{2}/\mathrm{n.d.f.} = 83.8/(56-2)$, and the fit results under different resonance hypotheses are summarized in table~\ref{tab:multisolution}.

\begin{table}[htbp]
\centering
\caption{Fitted resonance parameters for $\Gamma_{ee}\mathcal{B}$ ($10^{-3}$~eV) and $\phi$ (rad) with two solutions (I, II). The fit procedure includes both statistical and systematic uncertainties. Here, the $\mathcal{S}$ indicates the statistical significance, the $\chi^2$/n.d.f. indicates the fit quality for each model.}
\resizebox{15.cm}{!}{
\renewcommand{\arraystretch}{1.2}
\begin{tabular}{c|c|c|c|c|c|c}
\hline
\hline
\multirow{2}{*}{Resonance}  & \multicolumn{2}{|c|}{$\Gamma_{ee}\mathcal{B}$ ($10^{-3}$ eV)}& \multicolumn{2}{|c|}{$\phi$ (rad)} & \multirow{2}{*}{$\mathcal{S}~(\sigma)$} & \multirow{2}{*}{$\chi^2/{\rm n.d.f}$}\\
\cline{2-5}
& \uppercase\expandafter{\romannumeral1} & \uppercase\expandafter{\romannumeral2}&  \uppercase\expandafter{\romannumeral1} & \uppercase\expandafter{\romannumeral2} &   & \\
\hline
$\psi(3770)$ &$123.8^{+3.2}_{-0.1}$~($<132$) & $1.4^{+1.1}_{-0.9}$ & $-1.4\pm0.4$ &$-0.3\pm0.3$       & 1.9 & $79.0/(56-4)$    \\

$\psi(4040)$ & $225.6^{+22.1}_{-22.8}$~($<243$) & $1.8^{+1.6}_{-0.1}$ & $-1.6\pm0.2$  &$0.8\pm0.5$  & 2.0 & $78.5/(56-4)$     \\

$\psi(4160)$ &$155.2^{+4.9}_{-4.9}$~($<167$)  & $0.2^{+0.2}_{-0.1}$ & $-1.6\pm0.2$  &$1.6\pm0.3$   & 1.5 & $80.8/(56-4)$     \\

$Y(4230)$ &$101.9^{+7.9}_{-7.6}$~($<110$) & $0.0^{+0.1}_{-0.0}$ &$-1.6\pm0.2$ &$1.8\pm0.5$   & 0.9 & $82.4/(56-4)$     \\

$Y(4360)$ &$179.7^{+8.3}_{-7.9}$~($<199$)  & $0.0^{+0.1}_{-0.0}$ &$-1.7\pm0.2$  &$1.5\pm1.1$  & 0.3 & $83.8/(56-4)$    \\

$\psi(4415)$ &$160.7^{+8.2}_{-7.7}$~($<180$)   & $0.0^{+0.1}_{-0.0}$ &$-1.7\pm0.3$  &$1.3\pm0.7$  & 0.5 & $83.4/(56-4)$   \\

$Y(4500)$ &$118.1^{+15.9}_{-13.7}$~($<153$)   &$0.7^{+1.5}_{-0.6}$   &$-1.8\pm0.3$  &$-1.7\pm0.3$  & 0.7 & $82.1/(56-4)$    \\

$Y(4660)$ &$64.6^{+5.2}_{-5.1}$~($<76$)   & $0.3^{+0.5}_{-0.3}$ &$-1.6\pm0.5$ &$1.8\pm0.7$ & 1.3 & $81.3/(56-4)$   \\

$Y(4710)$ &$169.5^{+10.8}_{-11.1}$~($<195$)   & $1.8^{+1.5}_{-1.3}$   &$-1.6\pm0.4$  &$-1.8\pm0.2$  & 0.9 & $81.2/(56-4)$    \\
\hline
\hline
\end{tabular}
}
\label{tab:multisolution}
\end{table}

\section{Summary}

\noindent In summary, using $e^+e^-$ collision data collected with the BESIII detector at the BEPCII collider corresponding to an integrated luminosity of $44.6~\text{fb}^{-1}$, we present the first measurement of Born cross sections for the reaction $e^{+}e^{-} \to pK^{-}\bar{\Sigma}^{0}$ + c.c. at fifty-six center-of-mass energies ranging from $3.51$ to $4.95~\text{GeV}$. A fit to the dressed cross sections of $e^{+}e^{-} \to pK^{-}\bar{\Sigma}^{0}$ + c.c. is performed under the assumption of a charmonium(-like) resonance [i.e., $\psi(3770)$, $\psi(4040)$, $\psi(4160)$, $Y(4230)$, $Y(4360)$, $\psi(4415)$, $Y(4500)$, $Y(4660)$ and $Y(4710)$] plus a continuum contribution. No significant evidence from any
of these charmonium(-like) states is observed. 
In addition, the products of branching fractions and electronic partial widths for these charmonium(-like) states decaying into the $pK^{-}\bar{\Sigma}^{0}$ final state, as well as their upper limits at the 90\% C.L., are provided.
Furthermore, we perform a PWA of $e^{+}e^{-} \to pK^{-}\bar{\Sigma}^{0}$ + c.c. at $\sqrt{s}$ = 3.773 GeV, where the largest data sample is available. The components considered in the PWA include excited $\Lambda^{*}$, $\Sigma^{*}$ and $N^{*}$ states, namely $\Lambda(1520)$, $\Sigma(1670)$, $\Sigma(1915)$, $N(1900)$, $N(2300)$ and $N(2570)$. The resonant parameters and the fit fractions for each component are summarized in table~\ref{pwa_result}, where the first uncertainties are statistical and the second systematic.

These results provide important input for understanding the decays of vector charmonium(-like) states into baryonic final states and for further investigating the production mechanism of charmonium(-like) states above the open-charm threshold.
The PWA results indicate that the measured masses and widths of the excited states are consistent with the PDG~\cite{PDG}. Although $N(2300)$ and $N(2570)$ are listed as two-star states in the PDG~\cite{PDG},  they are found to be among the dominant contributions to $e^{+}e^{-} \to pK^{-}\bar{\Sigma}^{0}$ + c.c. at $\sqrt{s}$ = 3.773 GeV, providing additional evidence for their existence. 
Owing to the limited statistical precision, the current results in table~\ref{pwa_result} differ from the recent theoretical prediction~\cite{FewBodySystems}. In addition, the fit fractions of the intermediate resonant components differ from those measured in $\psi(3686) \to pK^{-}\bar{\Sigma}^{0}+c.c.$~\cite{BESIII:2026glf}. Further studies with larger data samples and additional decay channels are needed to improve our understanding of the underlying dynamics and baryon spectroscopy. 

\begin{table}[htbp]
\centering
\caption{Spin-parity ($J^{P}$), masses, widths, fit fractions (FFs), and statistical significances ($\mathcal{S}$) of the resonances obtained from the nominal PWA fit, together with the corresponding PDG values~\cite{PDG}. The first uncertainties are statistical and the second are systematic.}
\label{pwa_result}
\resizebox{15.2cm}{!}{
\renewcommand{\arraystretch}{1.2}
\begin{tabular}{c|c|c|c|c|c|c|c}
\hline\hline
\multirow{2}{*}{Resonance} & \multirow{2}{*}{$J^{P}$} & \multicolumn{2}{c|}{Mass (MeV/$c^2$)} & \multicolumn{2}{c|}{Width (MeV)} & \multirow{2}{*}{FF (\%)} & \multirow{2}{*}{$\mathcal{S}$ ($\sigma$)} \\
\cline{3-6}
& & This work & PDG~\cite{PDG} & This work & PDG~\cite{PDG} & & \\
\hline
$N(1900)$ & $3/2^{-}$ & $1912.2 \pm 12.7 \pm 53.8$ & $1890\sim1950$~($\approx 1920$) & $204.6 \pm 30.1 \pm 124.9$ & $100\sim320$~($\approx 200$) & $15.1 \pm 3.3 \pm 4.0$ & 5.7\\
$N(2300)$ & $1/2^{-}$ & $2294.9 \pm 14.3 \pm 31.8$ & $2300^{+40~+109}_{-30~-0}$ & $348.9 \pm 14.9 \pm 22.7$ & $340\pm30^{~+110}_{~-58}$ & $13.4 \pm 4.7 \pm 6.2$ & 4.7\\
$N(2570)$ & $5/2^{+}$ & $2574.7 \pm 12.4 \pm 34.8$ & $2570^{+19~+34}_{-10~-10}$ & $247.3 \pm 14.1 \pm 22.4$ & $250^{+14~+69}_{-24~-21}$ & $15.2 \pm 4.3 \pm 6.3$ &5.2\\
$\Sigma(1670)$ & $3/2^{-}$  & $1679.4 \pm 1.6 \pm 5.2$ &$1665\sim1685$~($\approx 1675$) & $41.1 \pm 4.2 \pm 10.6$ & $40\sim100$~($\approx 70$) & $21.1 \pm 3.5 \pm 7.2$ &13.1\\
$\Sigma(1915)$ & $5/2^{+}$  & $1923.5 \pm 7.3 \pm 45.6$ & $1900\sim1930$~($\approx 1915$) & $103.3 \pm 20.8 \pm 52.4$ & $80\sim160$~($\approx 120$) & $11.0 \pm 2.5 \pm 7.5$ &5.7\\
$\Lambda(1520)$ & $3/2^{-}$ & $...$ & $1519.42\pm 0.19$ & $...$ & $15.73\pm 0.26$ & $10.3 \pm 4.4 \pm 2.6$ &8.4\\
\hline\hline
\end{tabular}}
\end{table}

\acknowledgments
\noindent 
The BESIII Collaboration thanks the staff of BEPCII (https://cstr.cn/31109.02.BEPC) and the IHEP computing center for their strong support. This work is supported in part by National Key R\&D Program of China under Contracts Nos. 2025YFA1613900, 2023YFA1606000, 2023YFA1606704, 2025YFA1613900; National Natural Science Foundation of China (NSFC) under Contracts Nos. 
12247101,
11635010, 11935015, 11935016, 11935018, 12025502, 12035009, 12035013, 12061131003, 12192260, 12192261, 12192262, 12192263, 12192264, 12192265, 12221005, 12225509, 12235017, 12342502, 12361141819, 12535005; 
he Fundamental Research Funds for the Central Universities No.
lzujbky-2025-ytA05, No. lzujbky-2025-it06, No. lzujbky-2024-jdzx06;
the Natural Science Foundation of Gansu Province No. 22JR5RA389, No. 25JRRA799;
the "111 Center" under Grant No. B20063;
the Chinese Academy of Sciences (CAS) Large-Scale Scientific Facility Program; the Strategic Priority Research Program of Chinese Academy of Sciences under Contract No. XDA0480600; CAS under Contract No. YSBR-101; 100 Talents Program of CAS; The Institute of Nuclear and Particle Physics (INPAC) and Shanghai Key Laboratory for Particle Physics and Cosmology; Agencia Nacional de Investigación y Desarrollo de Chile (ANID), Chile under Contract No. ANID CCTVal CIA250027; ERC under Contract No. 758462; German Research Foundation DFG under Contract No. FOR5327; Istituto Nazionale di Fisica Nucleare, Italy; Knut and Alice Wallenberg Foundation under Contracts Nos. 2021.0174, 2021.0299, 2023.0315; Ministry of Development of Turkey under Contract No. DPT2006K-120470; National Research Foundation of Korea under Contract No. NRF-2022R1A2C1092335; National Science and Technology fund of Mongolia; Polish National Science Centre under Contract No. 2024/53/B/ST2/00975; STFC (United Kingdom); Swedish Research Council under Contract No. 2019.04595; U. S. Department of Energy under Contract No. DE-FG02-05ER41374.

\newpage

\input{authorlist}

\end{document}

%% file: authorlist.tex
\noindent {\bf{The BESIII Collaboration}}
~\\

\noindent
M.~Ablikim$^{1}$\BESIIIorcid{0000-0002-3935-619X},
M.~N.~Achasov$^{4,c}$\BESIIIorcid{0000-0002-9400-8622},
P.~Adlarson$^{83}$\BESIIIorcid{0000-0001-6280-3851},
X.~C.~Ai$^{89}$\BESIIIorcid{0000-0003-3856-2415},
C.~S.~Akondi$^{31A,31B}$\BESIIIorcid{0000-0001-6303-5217},
R.~Aliberti$^{39}$\BESIIIorcid{0000-0003-3500-4012},
A.~Amoroso$^{82A,82C}$\BESIIIorcid{0000-0002-3095-8610},
Q.~An$^{78,65,\dagger}$,
Y.~H.~An$^{89}$\BESIIIorcid{0009-0008-3419-0849},
M.~S.~Anderson$^{39}$\BESIIIorcid{0009-0008-1550-2632}
% ...
\\
\vspace{0.2cm}
(BESIII Collaboration)\\
\vspace{0.2cm} {\it
$^{1}$ Institute of High Energy Physics, Beijing 100049, People's Republic of China\\
% ...
$^{4}$ Budker Institute of Nuclear Physics SB RAS (BINP), Novosibirsk 630090, Russia\\
% ...
$^{31}$ INFN Sezione di Ferrara, (A)INFN Sezione di Ferrara, I-44122, Ferrara, Italy; (B)University of Ferrara, I-44122, Ferrara, Italy\\
% ...
$^{39}$ Johannes Gutenberg University of Mainz, Johann-Joachim-Becher-Weg 45, D-55099 Mainz, Germany\\
% ...
$^{65}$ State Key Laboratory of Particle Detection and Electronics, Beijing 100049, Hefei 230026, People's Republic of China\\
% ...
$^{78}$ University of Science and Technology of China, Hefei 230026, People's Republic of China\\
% ...
$^{82}$ University of Turin and INFN, (A)University of Turin, I-10125, Turin, Italy; (B)University of Eastern Piedmont, I-15121, Alessandria, Italy; (C)INFN, I-10125, Turin, Italy\\
$^{83}$ Uppsala University, Box 516, SE-75120 Uppsala, Sweden\\
% ...
$^{89}$ Zhengzhou University, Zhengzhou 450001, People's Republic of China\\

\vspace{0.2cm}
$^{\dagger}$ Deceased\\
% ...
$^{c}$ Also at the Novosibirsk State University, Novosibirsk, 630090, Russia\\
% ...
}